\documentclass[fleqn,usenatbib]{mnras}

\usepackage{newtxtext,newtxmath}

\usepackage[T1]{fontenc}

\newcommand{\mZAMS}{\ensuremath{M_{\rm ZAMS}}}
\newcommand{\mCO}{\ensuremath{M_{\rm CO}}}
\newcommand{\msun}{\ensuremath{{\rm M}_\odot}}

\newcommand{\rsun}{\ensuremath{{\rm R}_\odot}}

\newcommand{\beq}{\begin{equation}}
\newcommand{\eeq}{\end{equation}}

\DeclareRobustCommand{\VAN}[3]{#2}
\let\VANthebibliography\thebibliography
\def\thebibliography{\DeclareRobustCommand{\VAN}[3]{##3}\VANthebibliography}

\usepackage{graphicx}	
\usepackage{amsmath}	

\title[Fate / SN progenitors in binary]{Fate of supernova progenitors in massive binary systems}

\author[T. Kinugawa et. al.]
{
Tomoya Kinugawa$^{(1)(2)(3)}$\thanks{E-mail: kinugawa@shinshu-u.ac.jp}, 
Shunsaku Horiuchi$^{(4)(5)}$\thanks{E-mail: horiuchi@vt.edu}, 
Tomoya Takiwaki$^{(6)}$, and  Kei Kotake$^{(7)(8)}$\\
\\
$^{1}$Faculty of Engineering, Shinshu University, 4-17-1, Wakasato, Nagano-shi, Nagano, 380-8553, Japan\\
$^{2}$Research Center for Aerospace System, Shinshu University,  4-17-1, Wakasato, Nagano-shi, Nagano, 380-8553, Japan\\
$^{3}$Research Center for the Early Universe, Graduate School of Science, University of Tokyo, 7-3-1 Hongo, Bunkyo-ku, Tokyo 113-0033, Japan\\
$^{4}$Center for Neutrino Physics, Department of Physics, Virginia Tech, Blacksburg, VA 24061, USA\\
$^{5}$Kavli IPMU (WPI), UTIAS, The University of Tokyo, Kashiwa, Chiba 277-8583, Japan\\
$^{6}$National Astronomical Observatory of Japan, 2-21-1 Osawa, Mitaka, Tokyo 181-8588, Japan\\
$^{7}$Research Institute of Stellar Explosive Phenomena, Fukuoka University, 8-19-1 Nanakuma, Jonan-ku, Fukuoka-shi, Fukuoka 814-0180, Japan\\
$^{8}$Department of Applied Physics, Faculty of Science, Fukuoka University, 8-19-1 Nanakuma, Jonan-ku, Fukuoka-shi, Fukuoka 814-0180, Japan
}

\date{Accepted XXX. Received YYY; in original form ZZZ}

\pubyear{2015}

\begin{document}
\label{firstpage}
\pagerange{\pageref{firstpage}--\pageref{lastpage}}
\maketitle

\begin{abstract}
How massive stars end their lives
depends on the core mass, core angular momentum, and hydrogen envelopes at death. 
However, these key physical facets of stellar evolution can be severely affected by binary interactions. In turn, the effectiveness of binary interactions itself varies greatly depending on the initial conditions of the binaries, making the situation much more complex. 
We investigate systematically how binary interactions influence core-collapse progenitors and their fates. 
Binary evolution simulations are performed to survey the parameter space of supernova progenitors in solar metallicity binary systems and to delineate major evolutionary paths. 
We first study fixed binary mass ratios ($q=M_2/M_1$ = 0.5, 0.7, and 0.9) to elucidate the impacts of initial mass and initial separation on the outcomes, treating separately Type Ibc supernova, Type II supernova, accretion induced collapse (AIC), rapidly rotating supernova (RSN), black hole formation, and gamma ray burst (GRB). 
We then conduct Binary Population Synthesis calculations for 12 models, varying the initial parameter distributions and binary evolution parameters, to estimate various supernova fractions. 
We obtain a Milky Way supernova rate $R_{\rm SN} = (1.14$--$1.57) \times10^{-2} \, {\rm yr}^{-1}$ which is consistent with observations. 
We find the rates of AIC, RSN, and GRB to be $\sim 1/100$ the rate of regular supernovae. Our estimated GRB rates are higher than the observed long GRB rate, but very close to the low luminosity GRB rate. 
Furthering binary modeling and improving the inputs one by one will enable more detailed studies of these and other transients associated with massive stars.
\end{abstract}




\section{Introduction} \label{sec:intro}
Stellar evolution holds paramount importance in astrophysics, providing a foundational framework for not only comprehending the behaviors of stars but also their profound influence on shaping the Universe. By studying how stars form, evolve, and eventually die, one can gain insights into a wide range of astrophysical phenomena, from the properties of individual stars to the formation and evolution of galaxies \citep{Conroy2013,Nomoto2013,Smith2014,Eldridge2022}.
In particular, supernovae (SNe) and gamma-ray bursts (GRBs) are
among the most powerful and intriguing transient phenomenon in the Universe, which are the explosions triggered by the death of massive stars \citep{Woosley2002,Woosley2006ARA,Smartt2009,Gehrels2009}.

In a seminal work on stellar evolution, \cite{Heger2003} illustrated that a star's mass and metallicity determine its eventual fate 
\citep[see also][for systematic studties]{Woosley1995,Woosley2002,Woosley2007,Sukhbold2016}.
A variety of models are provided and they are often used in SN modelling \citep[][]{Umeda2008,Umeda2012,Sukhbold2018,Chieffi2020}. 
With the ever-increasing power of supercomputers, studies of the systematic effects of the progenitor mass and metallicity have been explored in multi-dimensional numerical simulations of massive star collapse \citep{NakamuraKo2015,Summa2016,Vartanyan2023,Burrows2020}.
Among the many insights they have provided, of particular importance is that the anisotropy of the Si/O shell is enough to alter the course of dynamical evolution of the core-collapse SN \citep{Bollig2021}. Thus, three-dimensional simulations have been performed with limited timescale of the Si/O burning phase to explore more realistic profiles of the progenitors \citep{Couch2015,Muller2016b,MOcak2018,Yoshida2019,Yoshida2021a,Yadav2020}.

One commonly missing ingredient in previous studies of SN progenitors is rotation. 
This is despite the fact that massive stars initially have high angular momentum \citep{Wolff2006,Huang2010} and rotation affects the evolution of stars as strongly as mass and metallicity \citep[e.g.,][]{Maeder2009}; for example, centrifugal force, angular momentum transport, and rotation-induced mixing change the stellar structure \citep[see][and the refereces therein]{Langer2012}. Several evolutionary models have incorporated the impact of rotation \citep{Heger2000,Heger2005,Ekstrom2012,Georgy2012}.
\cite{Yoon2006} have helped to delineate the fate of the massive stars in the parameter space of mass, metallicity and rotation. At the extreme, a star can completely change its evolution when it undergoes chemically homogeneous evolution (CHE) due to efficient mixing induced by rotation \citep{Yoon2006,Woosley_Heger_2006,Aguilera-Dena2018}. Although CHE may not be common, more generally stars are deformed by rotation; axisymmetric equilibrium structures of rotating stars in two spatial dimensions have been obtained \citep[see][and the refereces therein]{Ogata2023}. 
It is imperative to also acknowledge the intricate connection of magnetic fields and rotation in stellar evolution \citep[see][and refereces therein]{Keszthelyi2023}.
A way to handle the shape of the magnetic fields was recently developed \citep{Takahashi2021}.
The effect of rotation and magnetic fields are eagerly studied by three dimensional simulations
in the phases right before the core collapse \citep{Varma2021,Yoshida2021b,Mcneill2022,Fields2022}.
The strength of the magnetic field and rotation of stellar cores are thought to even dictate the outcome of explosion, including explosion energies and morphologies, as well as occurrence of SNe versus GRBs \citep[e.g.,][]{Iwakami2014,Summa2018,Kuroda2020,Takiwaki2021,Obergaulinger2022,Varma2023,Matsumoto2022,Bugli2023,Shibagaki2023,Hsieh2023}. 

Therefore, it is not surprising that binary interactions have garnered substantial attention due to their impact on stellar rotation, involving factors such as tidal interactions in binary systems and mass transfer. Moreover, observational evidence suggests a high binary formation rate among high-mass stars \citep{Kobulnicky2007,Mason2009,Sana2012,Sana2013,Chiini2012,Kobulnicky2014,Moe2017}.
Efforts to incorporate binary effects into stellar evolution codes are ongoing \citep{Cantiello2007,Patton2020,Schneider2021,Laplace2021}, including even to SN modeling \citep{Vartanian2021}.
The most famous example of a SN from a binary system is perhaps SN1987A \citep[e.g.,][]{Menon2017,Urushibata2018,Ono2020,Utrobin2021,Nakamura2022}.
Observations of the explosion site are still ongoing \citep{Cigan2019,Lasson2023}. 
As another example, the SN that created Cassiopeia A may also have been a binary system \citep{Hirai2020}. Also, Betelgeuse may experience a stellar merger \citep{Chatzopoulos2020}. Finally, to make a double neutron star system, an ultra-stripped SN is considered necessary, motivating again binary evolution \citep{Tauris2017,Yoshida2017,Muller2018, Hijikawa2019}.

Nevertheless, while previous studies have explored the contribution of binaries to SNe in specific cases, a systematic understanding akin to the well-established scenario of single stars, as outlined by \cite{Heger2003} and \cite{Yoon2006}, remains lacking. 
To bridge this gap, population synthesis methods emerge as indispensable tools, enabling systematic exploration under a range of assumptions regarding stellar and binary physics, in particular wind mass loss, mass transfer, and common envelope treatments \citep{Hurley_2002,Belczynski_2002,Kinugawa2014, Zapartas2017, DeMarco2017,Stanway2018,Spera2019,Tanikawa2020,Breivik2020,Riley2022,Fragos2023}. 
The population synthesis method is a numerical calculation technique widely used in binary studies. It is employed to study the evolution and statistical properties of binary stellar systems considering various physics of binary interactions.
For example, population synthesis can estimate the gravitational wave sources from compact binary mergers \citep[e.g.,][]{Belczynski_2002,Dominik_2013,Kinugawa2020} and have predicted the massive stellar-mass binary black hole mergers \citep{Kinugawa2014,Kinugawa2016}.
The impact of the binary evolution on the Diffuse SN Neutrino Background (DSNB) have been studied \citep{Horiuchi2021}.
In such studies, the distribution of He or CO core mass is important \citep{Patton2022,Fragos2023}.

It is in these contexts that we investigate systematically how binary interactions influence SN progenitors. Using simplified models, we focus in particular on the final fates of massive stars: either a ``standard'' Type II SN, a stripped Ibc SN, a rapidly rotating SN, an accretion induced collapse (AIC), a collapse to black hole, or a GRB. This paper is organized as follows.
In Section \ref{sec:method}, we describe how to calculate the binary interactions, the stellar evolution, and how to determine the SNe type.
Section \ref{sec:model} shows our parameter survey and the binary population synthesis calculations for SNe.
In Section \ref{sec:summary}, we describe the summary of this paper and discussion comparison with previous studies and uncertainties.

\section{Methods} \label{sec:method}
We use the binary population synthesis code in \cite{Kinugawa2014} which is updated from the BSE code \citep{Hurley2000,Hurley_2002} in order to calculate binary evolution effects for SNe.
In this section, we describe the main binary effects which change stellar masses and the calculation methods.
The other binary interactions are described in Appendix \ref{sec:tidaletc}.

\subsection{Stability of mass transfer}
If the Roche lobe around a star is fulfilled, the material of the stellar surface is transferred to its companion through the L1 point.
The Roche lobe radius of the donor star is approximately described as \citep{Eggleton_1983}
\begin{equation}
R_{\rm L,1}\simeq \frac{0.49q_1^{2/3}}{0.6q_1^{2/3}+\ln(1+q_1^{1/3})}a \,,
\end{equation}
where $a$, $q_1=M_1/M_2$, $M_1$, and $M_2$ are the orbital separation, the mass ratio, the mass of the donor, and the mass of the accretor, respectively.

When the mass transfer is dynamically unstable, i.e., the orbit shrink too rapidly that the accretor star will plunge into the envelope of the donor star, the mass transfer become a common envelope phase.
On the other hand, when the mass transfer is dynamically stable, the mass transfer continue stably. This case is called the Roche lobe overflow. 
The dynamical stability of mass transfer is determined by how the stellar radius of the donor star is changed by the mass loss and by how the Roche lobe radius is changed by the mass transfer.

In order to consider the dynamical stability of the mass transfer, we use $\zeta_{\rm L}={d{\rm{log}}R_{\rm L,1}}/{d{\rm{log}}M_1}$ and $\zeta_{\rm ad}=({d{\rm{log}}R_{\rm 1}}/{d{\rm{log}}M_1})_{\rm ad}$. 
Here, $\zeta_\mathrm{L}$ is the response of the Roche lobe
radius $R_\mathrm{L,1}$ to the change in the mass of the donor star $M_1$ and
$\zeta_\mathrm{ad}$
is the response of the radius of the donor star $R_1$ to changes in the mass of the donor star within the dynamical timescale. 
When $\zeta_{\rm L}>\zeta_{\rm ad}$, the Roche lobe radius will be much smaller than the stellar radius by the mass transfer in the dynamical timescale.
In this case, the mass transfer becomes a common envelope.
On the other hand, when  $\zeta_{\rm L}>\zeta_{\rm ad}$, the donor stellar radius will be much smaller than the Roche lobe radius by the mass transfer in the dynamical timescale. In this case, the mass transfer is treated as the Roche lobe overflow. $\zeta_{\rm L}$ is a function of the mass ratio and the separation of the binary, described as \citep{Eggleton_1983, Eggleton2011}
\begin{align}\label{eq.stMT}
\zeta_{\rm L}&=\frac{d{\rm{log}}R_{\rm L,1}}{d{\rm{log}}M_1}\\
                  &=\frac{(0.33+0.13q_1)[1+ q_1)]+(1-\beta)(q_1^2-1)-\beta q_1}{1+q_1}.
\end{align}   
On the other hand, $\zeta_{\rm ad}$ strongly depends on the evolutionary stage of the donor star.
When the donor star is in the red giant phase, $\zeta_{\rm ad}$ is described as
\begin{equation}
    \zeta_{\rm ad}=-1+\frac{2}{3}\frac{M_1}{M_{\rm env,1}}. 
\end{equation}
When the donor star is in other evolutionary phases, $\zeta_{\rm ad}$ is 2.59, 6.85, 1.95 and 5.79 for the main sequence phase, the Hertzsprung gap phase \citep{Hjellming_1989}, the
naked-He main sequence phase and the naked-He giant star \citep{Ivanova_2002,Belczynski_2008}, respectively.

\subsection{Roche lobe overflow}
When the star starts mass transfer ($R_1>R_{\rm L,1}$) and the mass transfer is dynamically stable ($\zeta_{\rm ad}>\zeta_{\rm L}$),
the Roche lobe overflow occurs.
In order to calculate the mass transfer rate, we use the fitting formula by \cite{Hurley_2002},
\begin{equation}
\dot{M_1}=F(M_1)\left[{\rm{ln}} \left(\frac{R_{1}}{R_{\rm L,1}}\right)\right]^3~\rm{M_{\odot}~yr^{-1}} \,,
\label{eq:m1dotf}
\end{equation}
where
\begin{equation}
F(M_1)=3\times10^{-6}\left\{{\rm{min}}\left[\left(10\frac{M_1}{10~\msun}\right),5.0\right]\right\}^2 \,.
\end{equation}
As the radius of the donor changes by the Kelvin–Helmholtz (KH) timescale,
the maximum value of the mass transfer rate from the donor is given by
\begin{equation}
\dot{M}_{1,\rm{max}}=\frac{M_1}{\tau_{\rm{KH,1}}} \,,
\end{equation}
where $\tau_{\rm{KH,1}}$ is the KH timescale of the donor.

The accretion rate to the accretor is described as
\begin{equation}
    \dot{M}_2=-\beta\dot{M}_1 \,.
\end{equation}
\textcolor{blue}{where $\beta$ is the accretion parameter of the mass transfer.}
However, if the accretor is a white dwarf, a neutron star or a black hole, we consider
the mass accretion rate is limited by the Eddington accretion rate described by
\begin{equation}
\dot{M}_{\rm{Edd}}=-\frac{4\pi c R_2}{\kappa_{\rm T}}=2.08\times10^{-3}(1+X)^{-1}
\left(\frac{R_2}{\rsun}\right)~\rm{M_{\odot}~yr^{-1}} \,,
\end{equation}
where $R_2$, $\kappa_{\rm T} =0.2(1+X)\ \rm cm^2\ g^{-1}$, and $X=0.76$ are  the stellar radius of the accretor, the Thomson scattering opacity, and the hydrogen mass fraction, respectively. 

We calculate the spin angular momentum evolution of stars in a binary system during the RLOF. The spin angular momentum is carried from the donor to the accretor.
We estimate the spin angular momentum loss of the donor in this process with a thin shell approximation:
\begin{equation}
\frac{dJ_{\rm sp,1}}{dt}= \frac{2}{3}\dot{M}_1R_1^2\Omega_{\rm spin,1} \,,
\end{equation}
where $\Omega_{\rm spin,1}$ is the spin angular velocity of the donor.
For the spin angular momentum of the accretor, we consider whether the transferred mass accretes via an accretion disk or not. 
First, if there is no accretion disk, i.e., the secondary radius is larger than the critical radius described by
\begin{equation}
r_{\rm cri} = 0.07225a(q_2(1+q_2))^{1/4},
\end{equation}
where $q_2=M_2/M_1$  \citep{LubowShu1974,UlrichBurger1976, Hurley_2002},
we assume that the angular momentum of the transferred mass evaluated by using the critical radius {is added}  directly to the spin angular momentum of the accretor. 
Thus, the spin angular momentum transferred to the accretor is calculated as
\begin{equation}
\frac{dJ_{\rm sp,2}}{dt}=\dot{M}_2\sqrt{GM_2r_{\rm cri}} .
\end{equation}
Alternatively, if the transferred mass accretes through a disk, 
the spin angular momentum of the accretor increases assuming that the transferred mass falls onto the stellar surface of the accretor with the Keplerian velocity.
Then the spin angular momentum transferred via the accretion disk is estimated as
\begin{equation}
\frac{dJ_{\rm sp,2}}{dt}=\dot{M}_2\sqrt{GM_2R_2} \,.
\end{equation} 

\subsection{Common envelope}

If the companion star plunges into the primary star which has a core and envelope structure due to an eccentric orbit, or mass transfer becomes dynamically unstable ($\zeta_\mathrm{L}>\zeta_\mathrm{ad}$), 
the binary becomes a common envelope phase.
In this paper, we use the $\alpha\lambda$ formalism for calculating the common envelope phase
evolution \citep{Webbink1984}, and the orbital separation just after the common envelope phase $a_\mathrm{f}$ is
calculated by the following energy budget if the accretor star is not
a giant star, 
\begin{align}
   \alpha\left(\frac{GM_\mathrm{c,1}M_\mathrm{2}}{2a_\mathrm{f}}-\frac{GM_\mathrm{1}M_\mathrm{2}}{2a_\mathrm{i}}\right)=
   \frac{GM_\mathrm{1}M_\mathrm{env,1}}{\lambda R_\mathrm{1}}.
\end{align}
Here, $M_\mathrm{c,1}$ and $M_\mathrm{env,1}$
are the core and envelope mass of the donor star, 
$M_\mathrm{2}$ is the mass of the accretor star and 
$a_\mathrm{i}$ is the separation just before the common envelope phase. The common envelope parameters are $\alpha$ and $\lambda$,  where $\alpha$ is the parameter of the efficiency showing how much orbital energy is used to strip the stellar envelope, and 
$\lambda$ is the  parameter of the binding energy of the envelope.

When the accretor star is also a giant star, 
the orbital energy is used not only to strip the envelope of the donor star, but also to strip the envelope of the accretor star. In this case, 
the orbital separation just after the common envelope phase $a_\mathrm{f}$ is calculated as
\begin{equation}\label{eq: dce}
   \alpha\left(\frac{GM_\mathrm{1c}M_\mathrm{2c}}{2a_\mathrm{f}}-\frac{GM_\mathrm{1}M_\mathrm{2}}{2a_\mathrm{i}}\right)=
   \frac{GM_\mathrm{1}M_\mathrm{1env}}{\lambda R_\mathrm{1}}+
   \frac{GM_\mathrm{2}M_\mathrm{2env}}{\lambda R_\mathrm{2}},
\end{equation}
where $M_\mathrm{2c}$, $M_\mathrm{2env}$ and $R_\mathrm{2}$ are the core and envelope mass and radius of accretor star.

\subsection{Merged remnant and rotation effect}
When the common envelope phase occurs, we estimate the separation just after the common envelope phase $a_{\rm f}$, and check whether the binary has coalesced within the common envelope phase or not. When $a_{\rm f}$ is smaller than the sum of the remnant stellar radii, the binary has merged. Additionally, when the post-MS star does not reach the Hayashi track nor ignite helium burning, such a star, so-called a Hertzsprung gap star, might not have a clear core-envelope structure. In this case, we also assume the binary merges \citep{Taam_2000,Belczynski_2008}. If a binary merges before CCSNe, we treat the merged product as rapidly rotating with Kepler velocity.

Rapid rotation can enhance the material mixing inside the star. \cite{Horiuchi2021} showed the percentage increase of the carbon-oxygen (CO) core mass of pre-CCSN with respect to the non-rotating case, based on \cite{Takahashi2014, Limongi2017}. For massive stars with the zero-age main sequence mass $M_{\rm ZAMS}> 13\,M_\odot$, we consider the enhancing fraction of the CO core mass with respect to the non-rotating counterpart having the same total mass as
\begin{equation}
    f_L=53.4 \mZAMS^{-3/2}+0.847,
\end{equation}
from Limongi's models \citep{Limongi2017}.
For $\mZAMS<13\,M_\odot$, 
\begin{equation}
    f_T=0.123\mZAMS+0.392,
\end{equation}
from Takahashi model \citep{Horiuchi2021}.
Note that in all cases, if the CO core mass estimated by the above formulae exceeds the total stellar mass, we limit the CO core mass to the total stellar mass. 

Just after the merger, the merged remnant has a high angular momentum, from the orbital angular momentum. We assume the angular momentum of merged remnant $J_{\rm merge}$ as the Kepler angular momentum. 
On the other hand, since the merged remnant loses angular momentum by the stellar wind mass loss,
we consider the angular momentum mass loss described by
\begin{equation}
    \Delta J=\frac{2}{3}\Delta M R^2 \Omega_{\rm spin}=\frac{2}{3k}\frac{\Delta M}{M}J,
\end{equation}
where $\Delta M$ is the stellar wind mass loss, $M$, $R$, $\Omega_{\rm spin}$, $k$, and $J$ are the mass, radius, spin angular velocity, moment of inertia factor, and angular momentum of the merged remnant, respectively. We assume $k=0.15$ which is the value of the red giant branch.
The angular momentum of the merged remnant just before the SN is described as
\begin{equation}
    J_{\rm pre SN}=J_{\rm merge}\left(\frac{M_{\rm pre SN}}{M_{\rm merge}}\right)^{\frac{2}{3k}},
\end{equation}
where $M_{\rm preSN}$ and $M_{\rm merge}$ are the total stellar mass of the merged remnant just before SN and the mass of the merged remnant just after the merger.
We calculate the angular momentum of the CO core of the merged remnants just before SN as 
\begin{equation}
    J_{\rm CO}=\frac{\mCO R^2_{\rm CO}}{M_{\rm preSN}R_{\rm preSN}^2}J_{\rm preSN},
\end{equation}
where $\mCO$ and $R_{\rm CO}$ are the mass and the radius of the CO core, and $R_{\rm pre SN}$ is the radius of the star just before SN.
We estimate $R_{\rm CO}$ using the fitting formula,
\begin{equation}
    R_{\rm CO}=\frac{1.23\times10^{-3}+8.06\times10^{-2} \mCO-3.31\times10^{-3} \mCO^2}{1+0.467 \mCO-3.03\times10^{-2} \mCO^2}\rsun,
\end{equation}
based on the mass-radius relation from \cite{Nicola2018}.

\subsection{Determination of SN Type}

\begin{table}
\caption{SN types considered in this work.}
\label{tab:SN}
\begin{center}
\begin{tabular}{c|ccc}
\hline
 SN type&  CO core mass & $a/M$ & Envelope type\\
 \hline
  AIC & $\mCO<1.34\,\msun$ &  & WD \\ 
  II  & $1.34\,\msun<\mCO<5\,\msun$ &  & yes H \\
  Ibc & $1.34\,\msun<\mCO<5\,\msun$ & $<1$ & no H \\
  RSN & $1.34\,\msun<\mCO<5\,\msun$ & $>1$ & no H \\
  BH  & $5\,\msun<\mCO$& $<1$ & either \\
  GRB & $5\,\msun<\mCO$ & $>1$ & no H \\
 \hline
\end{tabular}
\end{center}
\end{table}

To study the SN explosions, we categorize the SN into six Types, based on the progenitor CO core mass ($\mCO$), angular momentum ($a_{\rm CO}/\mCO=cJ_{\rm CO}/G\mCO^2$), and the presence of a Hydrogen envelope (see Table~\ref{tab:SN}), extending the treatment in \cite{Yoon2006}. 

First, if the CO core mass is between 1.34\,$\msun$ and 5\,$\msun$, and the progenitor retains its hydrogen envelope, we classify them as Type II SNe, regardless of the angular momentum. 
Second and third involve the same CO core mass range but progenitors that have lost their hydrogen envelopes. If the angular momentum is less than 1 ($a_{\rm CO}/\mCO < 1$) we classify them as Type Ibc SNs, while if the angular momentum is greater than 1 ($a_{\rm CO}/\mCO > 1$) we classify them as Rapidly-rotating SNe (RSN). We expect the formation of rapidly rotating neutron stars, which may show different explosions from the normal SNe (MHD-driven explosion, see e.g., \citealt{Obergaulinger2014} or explosions driven by low-$T/W$ instability, see e.g., \citealt{Takiwaki2021}).
Fourth involves CO core masses less than 1.34 solar masses. If such progenitors undergo explosions due to accretion, they are designated as Accretion-Induced Collapse (AIC) SNe. 
Fifth and sixths concern progenitors with CO core masses greater than $5\,\msun$. If $a_{\rm CO}/\mCO < 1$, they are classified as Black Hole Formation events irrespective of the presence of a Hydrogen envelope or not. On the other hand, if $a_{\rm CO}/\mCO > 1$ and there are no Hydrogen envelope, they are identified as GRBs. 
We assume that if the star with high angular momentum has a Hydrogen envelope, it loses the angular momentum due to efficient mass loss, and it cannot explode as GRB or RSN.

We should keep in mind the limitation of such a simple classification.
\cite{Patton2022} employ more complex criteria \citep{Fryer2012,Ertl2016} and the results are shown in their Fig.~4.
Though the bifurcation of NS and BH is not the sole function of $\mCO$, we still see some general trend that BH appears where $\mCO>5\,\msun$.
\cite{Schneider2021} employ the criteria of \cite{Muller2016a}. In their Fig.~7, BH appears in $7\,\msun < \mCO < 8\,\msun$ and $14\,\msun < \mCO$.
\cite{Burrows2020} and other studies claim that black holes tend to appear in a mass range of 
$13\,\msun <M_{\rm ZAMS}<15\,\msun$, which correspond to $2\,\msun <\mCO<3\,\msun$.
Such new scenarios should be tested in the future.

It is hard to estimate the angular momentum of the final compact object, WD, NS, or BH from the angular momentum of the CO core.
This study treats the angular momentum in a qualitative way.
Here, we introduce a simple treatment in the previous studies for further improvement.
The angular momentum of CO core can be distributed to the central objects and accretion disks. See Section~8.3.4 in \cite{Fragos2023} for this issue. Numerical simulations would be also useful to map the angular momentum from the core to the compact object \citep{Sekiguchi2011,Fujibayashi2023}.


\section{Models \& Calculation Results} \label{sec:model}

\subsection{Parameter survey} \label{subsec:Qconst}

We first conduct a parameter survey of solar metal binary evolution, by performing binary evolution calculations with fixed mass ratios, $q=M_2/M_1$, and binary parameters in circular orbits. We explore the impacts of initial mass and initial separation on binary evolution and how it influences SN outcomes.
We calculate three fixed mass ratio models, $q=0.5$, $0.7$, and $0.9$.
In each model, we calculate the initial mass from 3\,$\msun$ to 100\,$\msun$ and the initial separation $a_{\rm ini}$ from 10\,$\rsun$ to $10^6\,\rsun$. The initial eccentricity is set to zero. We assume $\beta=1$, $\alpha\lambda=1$, and no pulsar kick. 

In this section, we focus on the $q=0.7$ model, and describe the
$q=0.5$ and $q=0.9$ models in Appendix~\ref{sec:Q}.
Figure~\ref{fig:Q07_channel} shows the progenitor of the SN. If SN progenitors do not merge before the core collapse, we split these into binary systems where both stars are core-collapse progenitors (double) and binary systems where only one of the stars are core-collapse progenitors (single).
If the binary stars merge before core collapse, we split these into binary systems where the progenitors become core-collapse progenitors as a result of rotational effects (merger rotation), and binary systems where the progenitor become core-collapse progenitors independently of rotational effects (merger) \citep[see][for similar classification]{Horiuchi2021}.

\begin{figure}
  \begin{center}
    \includegraphics[width=\hsize]{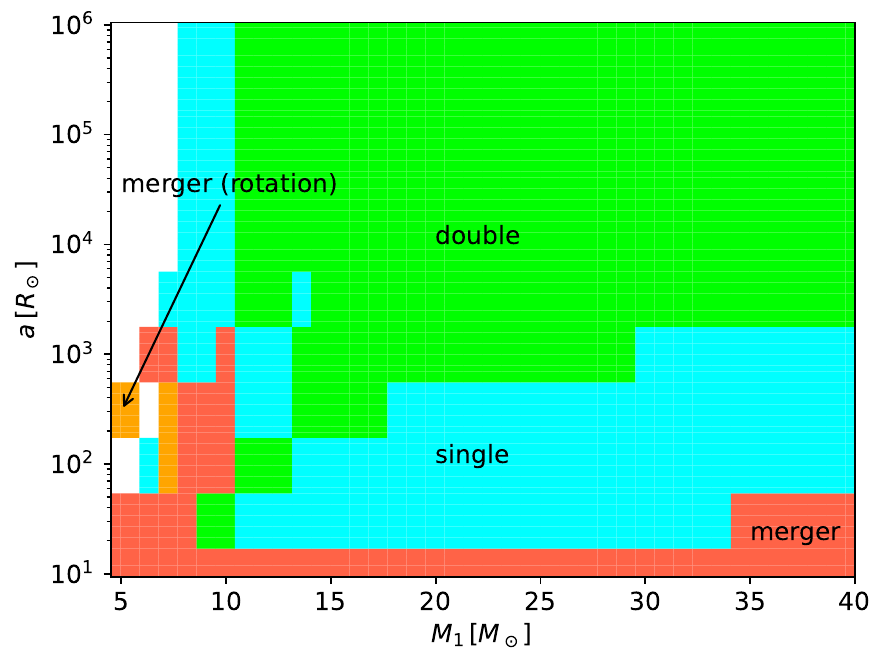}
  \end{center}
  \caption{The binary progenitors of SNe in the mass ratio $q=0.7$ case. Here, ``double'' and ``single'' refer to binary systems where both stars are core-collapse progenitors and where only one of the stars are core-collapse progenitors, respectively. Binary systems where the stars merge before core collapse are labeled ``merger''. Systems that merge and create a light mass star that only core collapse after considering rotation effects are labeled ``merger (rotation)''. See text for details. }
  \label{fig:Q07_channel}
\end{figure}

Binary calculations with $a=10^6\,\rsun$ are effectively single stellar evolutions. However, when the initial separation is less than $\sim 10^{4}\,\rsun$, the SN progenitors can interact with the companion star and qualitatively change the SN progenitors. Figure~\ref{fig:Q07_channel} shows that the influence of binary interactions depends strongly on the evolution separation. In the range $a\simeq10^{1.5}\,\rsun$ to $a\simeq10^{3}\,\rsun$, a massive binary where both stars are originally expected to undergo SNe becomes a single core-collapse system due to binary interactions. After the primary star become a SN or a core collapse, the secondary star reaches the Hertzsprung gap and makes a common envelope with the primary compact object. At that time, the secondary star disappears due to merging with the primary compact object during the common envelope phase. 
In the range $a\lesssim 10^{1.5}\,\rsun$, a massive binary will merge due to binary interactions before a SN.
In particular, even if the primary star starts with a mass lower than the criterion for SN explosion, the binary interaction can enable it to become a SN. We see this new channel appear for close binaries $ a\lesssim 10^3\,\rsun$. The effect of rotation amplifies this effect by increasing the core mass by mixing \citep{Horiuchi2021}.

\begin{figure}
    \centering
    \includegraphics[width=\hsize]{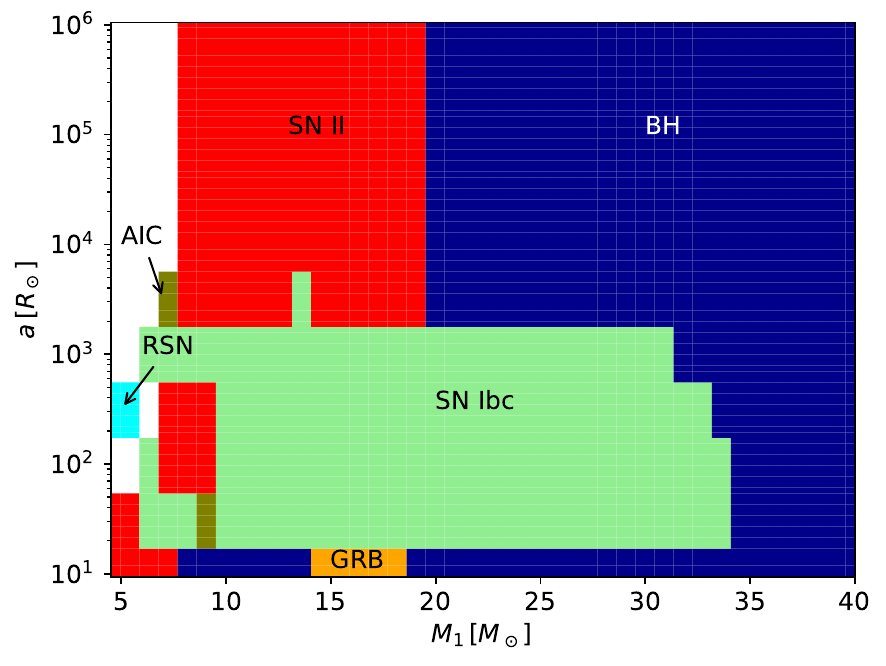}
    \caption{SN type of primary star in the $q=0.7$ case. See Table \ref{tab:SN} for SN type classifications.}
       \label{fig:Q07}
    \includegraphics[width=\hsize]{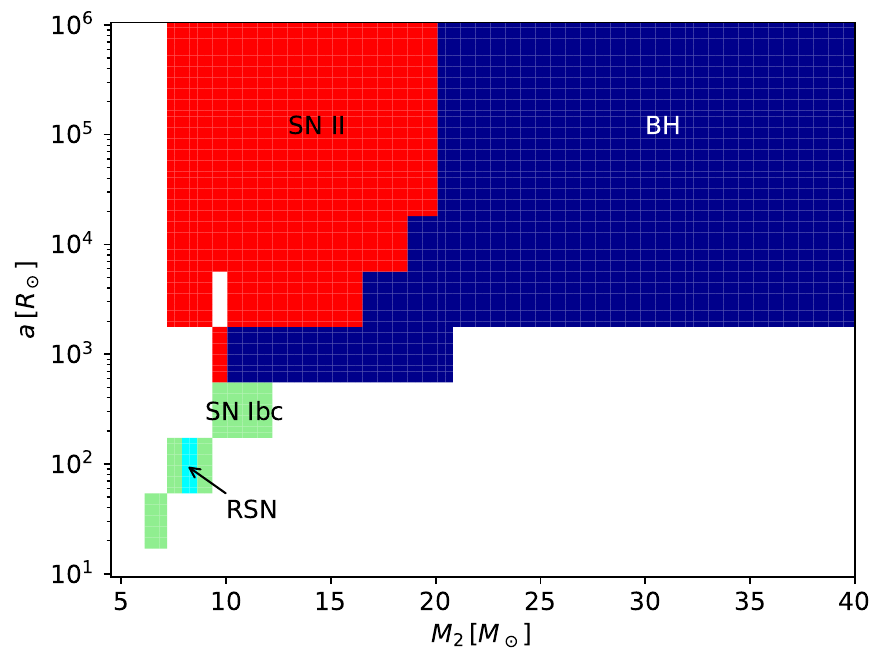}
    \caption{SN type of secondary star in the $q=0.7$ case. Note that the results of merged stars are omitted here and shown instead in Fig.~\ref{fig:Q07}. The parameter region shown is thus limited to the double system labeled in Fig.~\ref{fig:Q07_channel}. }
     \label{fig:Q07_2}
\end{figure}

\begin{figure}
        \centering
        \includegraphics[width=\hsize]{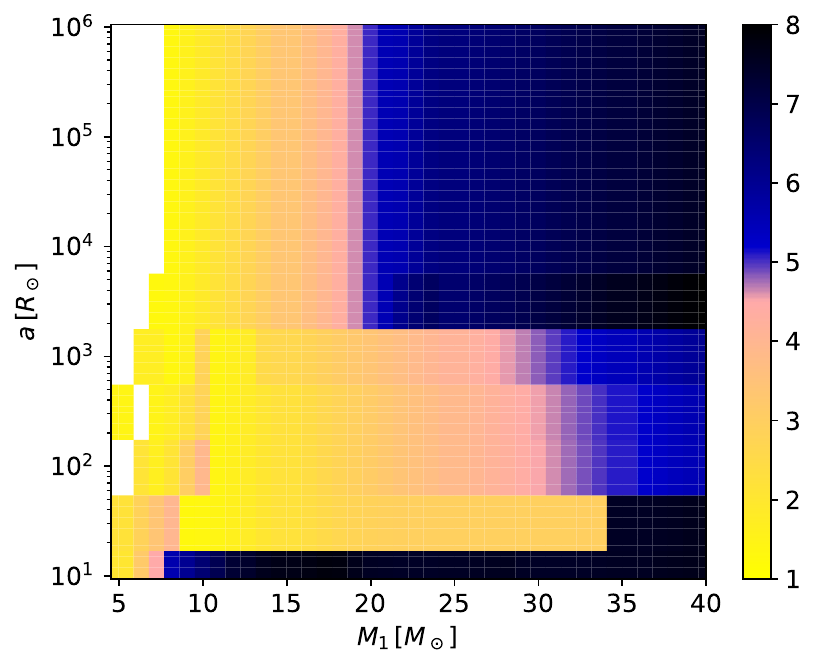}
    \caption{CO core mass of primary star in the $q=0.7$ case.}
       \label{fig:Q07_Mco}
        \includegraphics[width=\hsize]{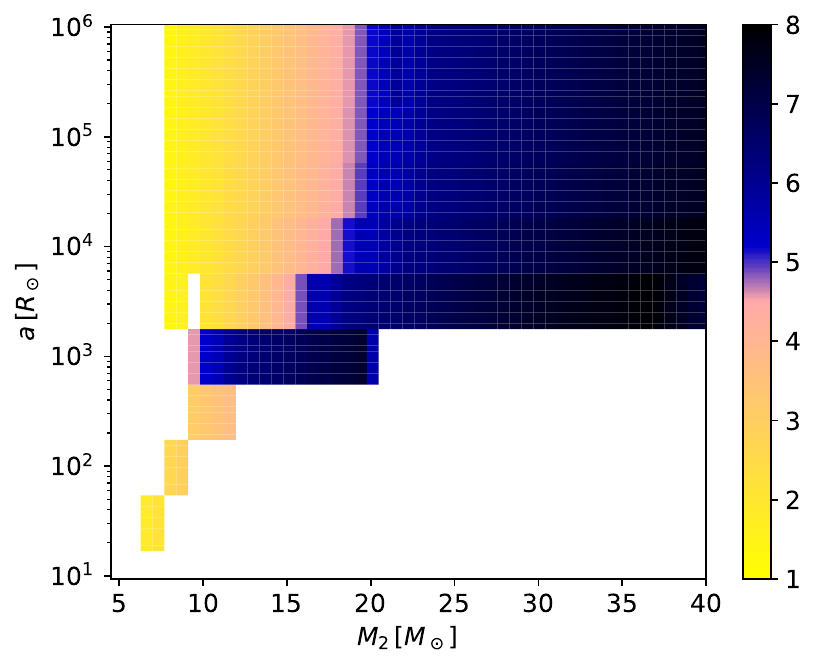}
    \caption{CO core mass of secondary star in the $q=0.7$ case.}
       \label{fig:Q07_2Mco}
\end{figure}

\begin{figure}
 \centering
          \includegraphics[width=\hsize]{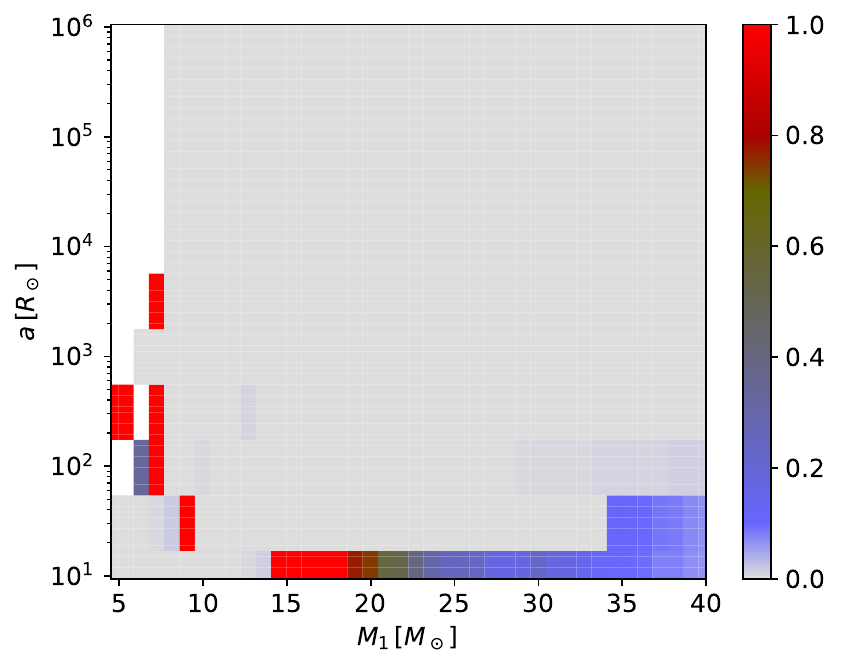}
    \caption{Angular momentum of primary star in the $q=0.7$ case.}
       \label{fig:Q07_am}
          \includegraphics[width=\hsize]{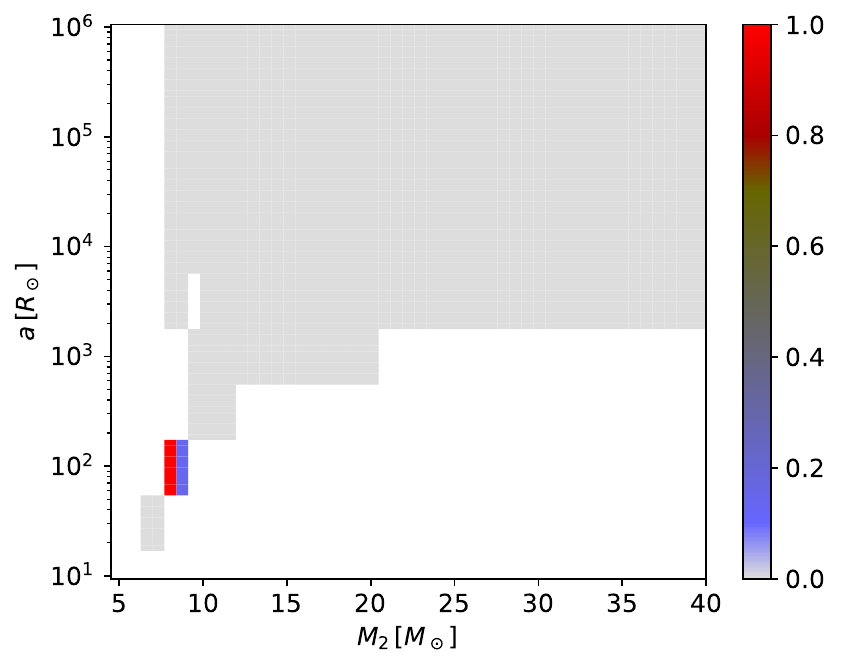}
    \caption{Angular momentum of secondary star in the $q=0.7$ case.}
       \label{fig:Q07_2am}
\end{figure}

In Figs.~\ref{fig:Q07} and \ref{fig:Q07_2}, we show the fates of the primary star and the secondary star, respectively, as functions of the ZAMS mass and binary separation. For the secondary, white regions with close separation represent cases where mass is lost in the common envelope phase and the secondary star does not experience core collapse.
If the SN progenitor is effectively a single star ($a_{\rm ini}\gtrsim 10^{4}\,\rsun$), they evolve only to Type~II SN or BH, depending on the CO core mass. 
Figures~\ref{fig:Q07_Mco} and \ref{fig:Q07_2Mco} show the CO core mass as functions of the ZAMS mass and binary separation, illustrating this dependence.  
In the high separation regime, the CO core mass is determined solely by the initial mass, with a larger initial mass leading to a larger CO core mass.
When $M_1$ or $M_2$ becomes larger than $\sim 20\,\msun$, $\mCO$ becomes $\ge 5\,\msun$ and BH is formed in our classification.
The angular momentum of the CO core is negligible in the case of effectively single stellar evolution, because the majority of the angular momentum is held by the hydrogen envelope, and there is angular momentum loss due to stellar wind mass loss.
We can confirm this in Figs.~\ref{fig:Q07_am} and \ref{fig:Q07_2am}, where we show the angular momentum as functions of the ZAMS mass and binary separation.

\cite{Fragos2023} shows similar results. In their Fig.~4, $\mCO$ is shown as a function of $M_{\rm ZAMS}$. In the range of $M_{\rm ZAMS}<40\,\msun$, $\mCO$ almost linearly increases and is consistent with our results.
In their Fig.~30, BH appear $20\,\msun<M_1$, which is consistent with our results. 
They use slightly different classifications of the fate of massive stars.
In their Fig.~29, they also classify electron capture SN with the criterion of $1.37\,\msun <  \mCO < 1.43\,\msun$. In this paper, we do not focus on these kinds of SNe (see Fig.~\ref{fig:Q07}).
They also consider pair-instability SN and pulsational pair-instability SN with certain criteria. Those SNe should appear in more massive stars, $M_{\rm ZAMS} > 50\,\msun$, and does not affect our results. Note that in a higher mass range, $M_{\rm ZAMS} > 30\,\msun$, the value of $\mCO$ significantly depends on the mass loss prescription of Wolf-Rayet stars (see also Fig.~1 of \citealt{Patton2022}).

\begin{figure}
  \begin{center}
    \includegraphics[width=8cm]{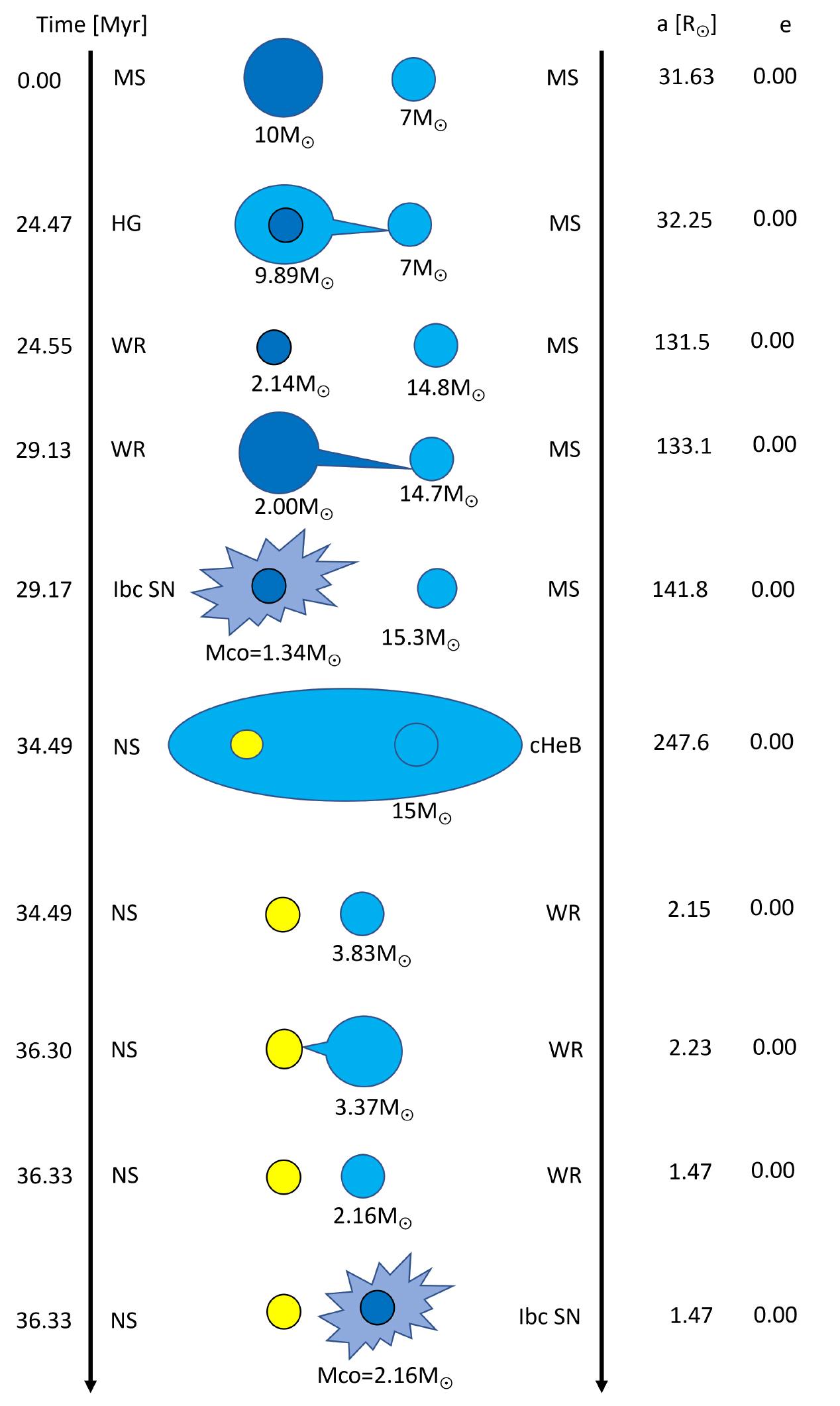}
  \end{center}
  \caption{Example of the Type Ibc SN progenitor evolutionary path. MS, HG, WR, and cHeB are main sequence, Hertzsprung gap, Wolf-Rayet star, and helium core burning phase, respectively.}
  \label{fig:SNIbc}
\end{figure}

If a binary is close enough so that the binary interaction is effective, more complex behavior is exhibited. In the parameter region of $10^{1.5}\,\rsun<a <10^{3.5}\,\rsun$, a SN progenitor becomes a Type Ibc SN due to mass loss by binary interactions. 
One typical evolutionary path of Type Ibc SN progenitors is shown in Fig.~\ref{fig:SNIbc}.
Since the primary star and the secondary star lose their envelopes via mass transfer and common envelope phase, respectively, they can become type Ibc SN due to mass loss. 
Another consequence is that binary effects reduce the $\mCO$ of the primary; see Fig.~\ref{fig:Q07_Mco} and compare $\mCO$ in $10^{1.5}\,\rsun<a <10^{3.5}\,\rsun$ to that in $10^{3.5}\,\rsun < a $. At $M_1 \sim 20\, \msun$, $\mCO\sim 2$--$3\,\msun$ for 
$10^{1.5}\,\rsun<a <10^{3.5}\,\rsun$ but $\mCO\sim 5\,\msun$ in  $10^{3.5}<a$. 

These results are consistent with \cite{Schneider2021}.
In their Fig.~3, Case A and Case B mass transfer makes the core mass lighter. The effect of mass transfer on the secondary star is complicated. See Fig.~\ref{fig:Q07_2Mco}, at $a\sim10^3\,\rsun$ and $ 10\,\msun < M_2 < 20 \,\msun$, the core mass is increased by the mass transfer from the primary star.
In the region with smaller $a$ and $M_2$,  core mass is also increased due to the same reason.
Note that the parameter regions of the figure are limited and correspond to the double in Fig.~\ref{fig:Q07_channel}.

If a binary is close enough to merge, there are generally two main scenarios, depending on the evolutionary stage of the primary star.
If the primary star has already become a compact object, the secondary star disappears through merging with the primary compact object.
In the range $a\simeq10^{2}\,\rsun$ to $a\simeq10^{3}\,\rsun$, a massive binary where both stars were originally expected to undergo SNe therefore only experience one SN due to the merger. 
On the other hand, if the primary star is not a compact object, the merger creates a rapidly rotating star. The CO core of the merged remnant tend to be increased by the rotation effect.
Subsequently, part or most of the angular momentum of the merged remnant can be lost by stellar wind mass loss.

\begin{figure}
  \begin{center}
    \includegraphics[width=8cm]{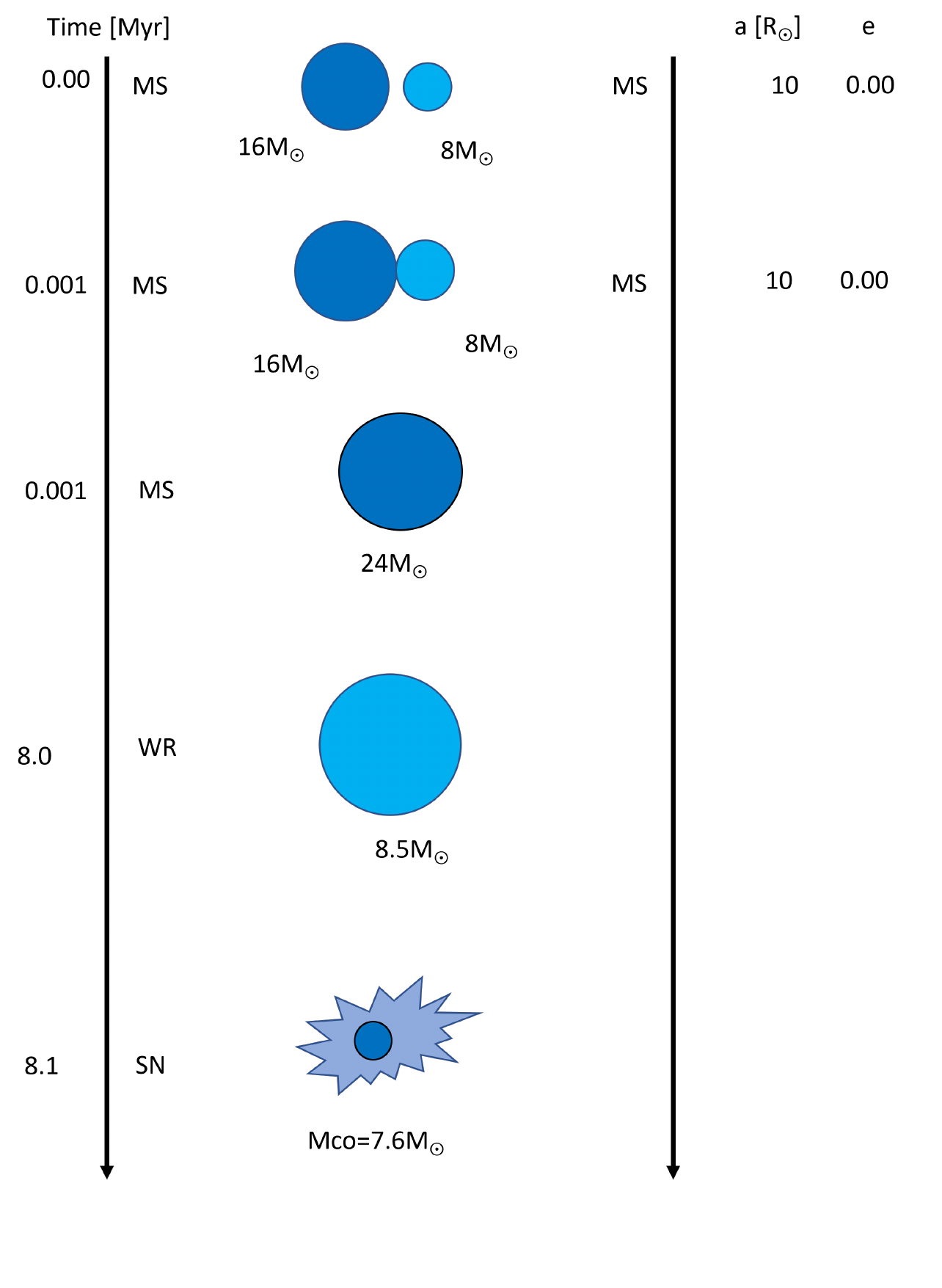}
  \end{center}
  \caption{Example of the GRB progenitor evolutionary path.}
  \label{fig:GRB}
\end{figure}

Merged remnants that lose their hydrogen envelope due to stellar winds while retaining sufficient angular momentum until the time of their explosion can lead to GRBs.
Note that \cite{Fragos2023} stop their simulation after the merger, and we cannot compare results.
GRBs appear in a limited parameter range, $a=10\,\rsun$ and  $M_1\sim15\,\msun$, as see in Fig.~\ref{fig:Q07}.
We show the main channel for the birth of GRB progenitors in Fig.~\ref{fig:GRB}.
Typically, GRB progenitors gain huge angular momenta from a binary merger.
If the mass of the merged progenitors is relatively massive, angular momentum loss due to strong stellar wind mass prevents the occurrence of a GRB. 
On the other hand, if the mass of merged progenitors is relatively low mass, they can evolve while retaining angular momentum until just before the gravitational collapse. Thus, systems whose mass $M_1\sim 15\,\msun$ can undergo gravitational collapse while retaining sufficient angular momentum of the CO core. 
If the merged progenitor mass is too low ($M_1\lesssim14\,\msun$), they can collapse with a hydrogen envelope. In this case, the star is a giant with a large radius, causing the majority of the angular momentum to be held by the envelope rather than the CO core. 
Hence, not only is it unable to produce a GRB, the angular momentum of the core also becomes significantly reduced (See Fig.~\ref{fig:Q07_am}). 

\begin{figure*}
  \begin{minipage}[b]{0.45\linewidth}
    \centering
    \includegraphics[width=\hsize]{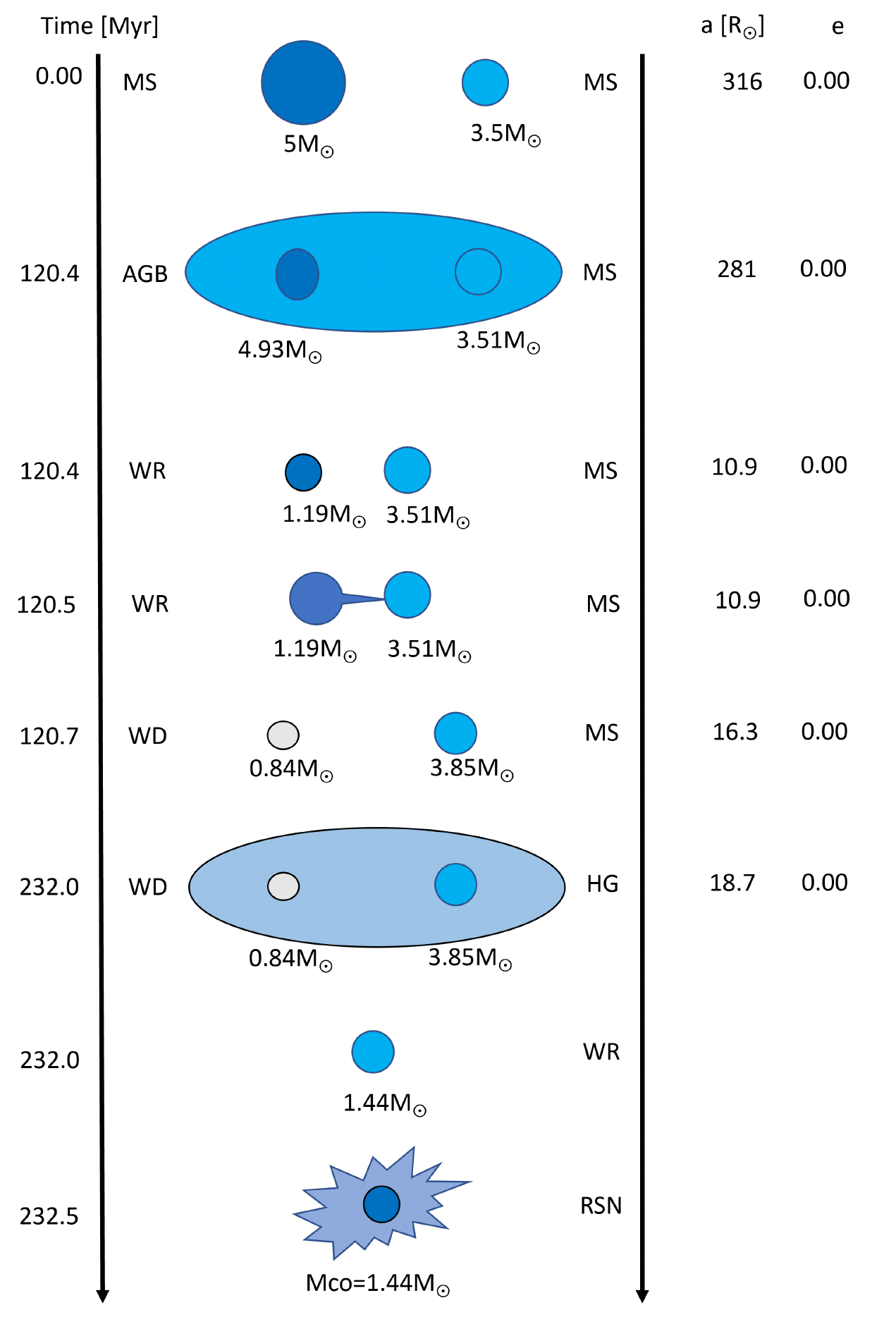}
       \label{fig:RSN1}
    \end{minipage}  
      \begin{minipage}[b]{0.45\linewidth}
    \centering
    \includegraphics[width=\hsize]{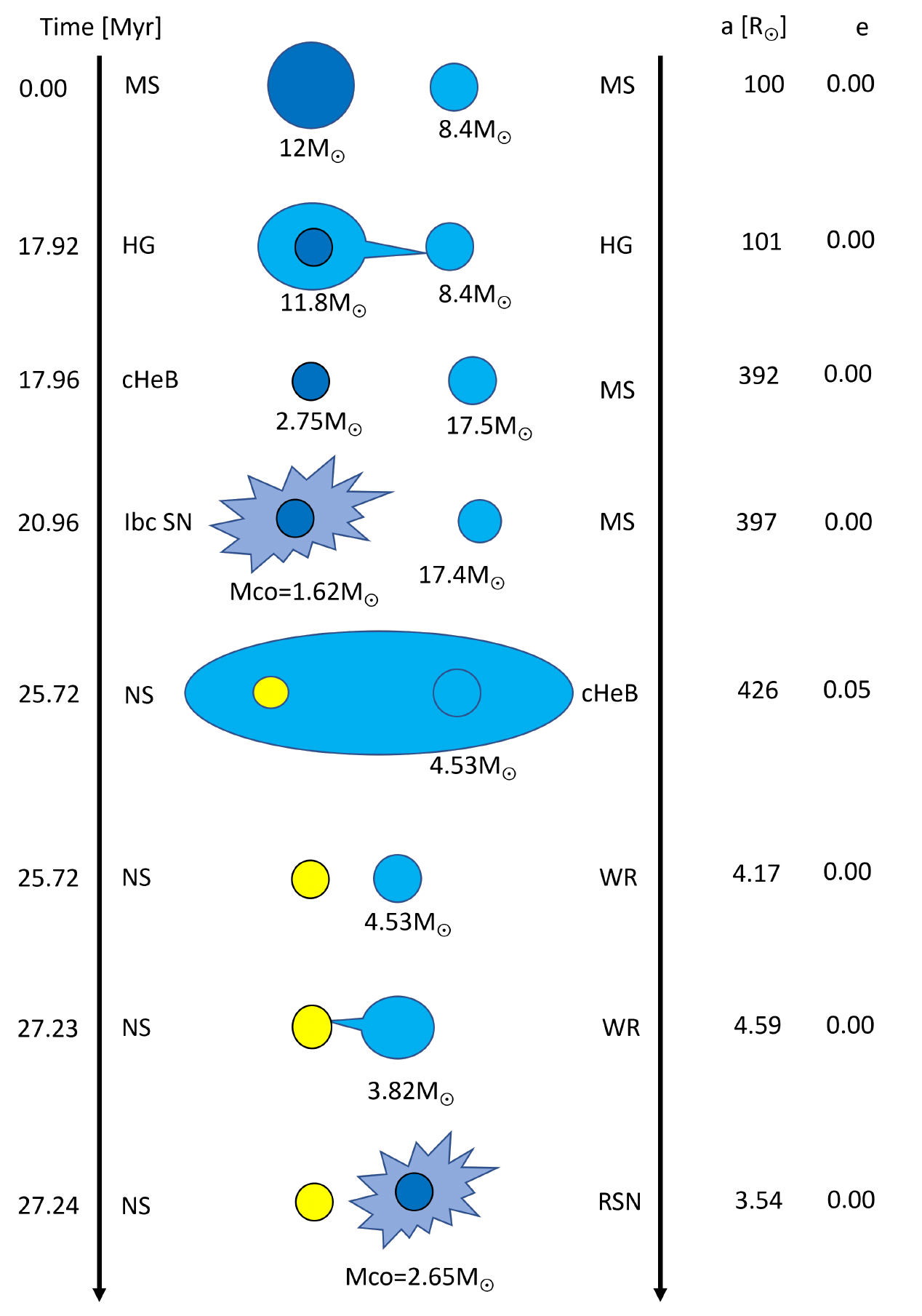}
       \label{fig:RSN2}
  \end{minipage}
        \caption{Examples of RSN progenitor evolutionary paths. The case for $q=0.7$ is shown. We have found another example as in Fig.~\ref{fig:RSN3} for $q=0.9$.}\label{fig:RSN}
\end{figure*}

We found two parameter regions for a Rapidly-rotating SN (RSN), i.e., $(M_1,~a)=(5\,\msun,~10^{3.5}\,\rsun)$ in Fig.~\ref{fig:Q07} and $(M_2, ~a)=(7\,\msun,~10^2\,\rsun)$  in Fig.~\ref{fig:Q07_2}. We show the two pathways in Fig.~\ref{fig:RSN}.
One of them is through a merger. For instance, in the left-hand side of Fig.~\ref{fig:RSN}, a binary system with an initial mass of 5 solar masses and an initial orbital semi-major axis of $a=10^{2.5}\,\rsun$  evolves into a binary system composed of a White Dwarf (WD) and a Red Giant (RG). After the binary merger, the remaining Wolf-Rayet (WR) star undergoes a SN explosion. Because the binary merger allows for the retention of a significant amount of angular momentum, it can lead to a RSN.
In the right-hand side of Fig.~\ref{fig:RSN}, the second pathway involves the binary evolution leading to a very close binary system ($a$ of a few $\rsun$), such as a NS-WR binary, in the late stages. In this scenario, the secondary star undergoes rapid rotation due to tidal effects, and it eventually explodes as a RSN.
In both cases, the core mass is not so large, i.e., $\mCO=1.44\,\msun$ or $2.65\,\msun$. It is interesting to note that heavier stars are often employed in simulations, e.g., $(\mZAMS, ~\mCO)=(27\,\msun,~\sim 7\,\msun)$ in \cite{Takiwaki2021}; $(\mZAMS, ~\mCO)=(35\,\msun,~20$--$30\,\msun)$ in \cite{Obergaulinger2020}. Although lighter stars are also considered
$(\mZAMS, \mCO)=(5\,\msun,4\,\msun)$ in \cite{Obergaulinger2022}, it is still heavier than what we observe due to binary effects.

\begin{figure*}
  \begin{minipage}[b]{0.45\linewidth}
    \centering
    \includegraphics[width=\hsize]{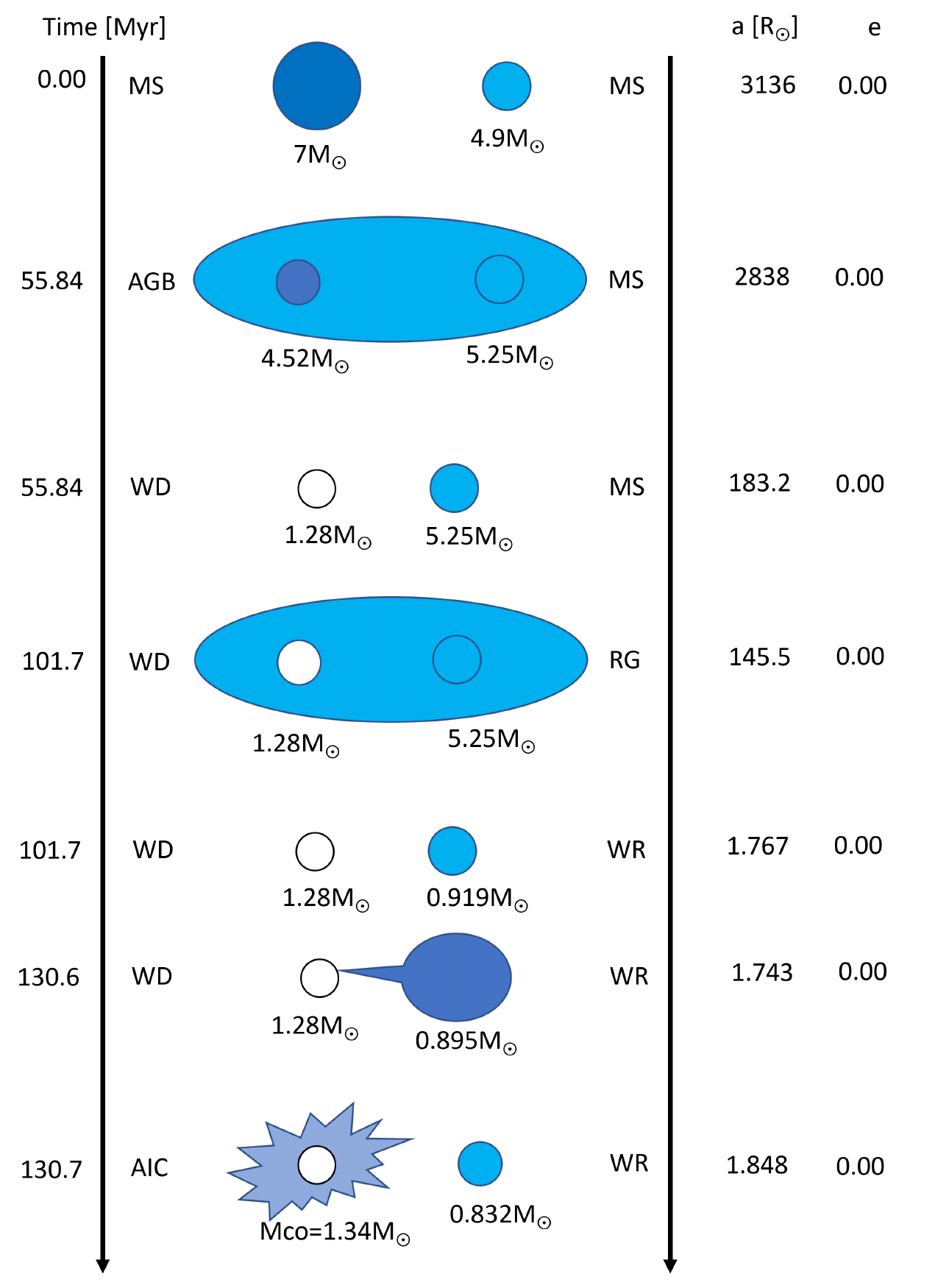}
       \label{fig:AIC1}
    \end{minipage}  
      \begin{minipage}[b]{0.45\linewidth}
    \centering
    \includegraphics[width=\hsize]{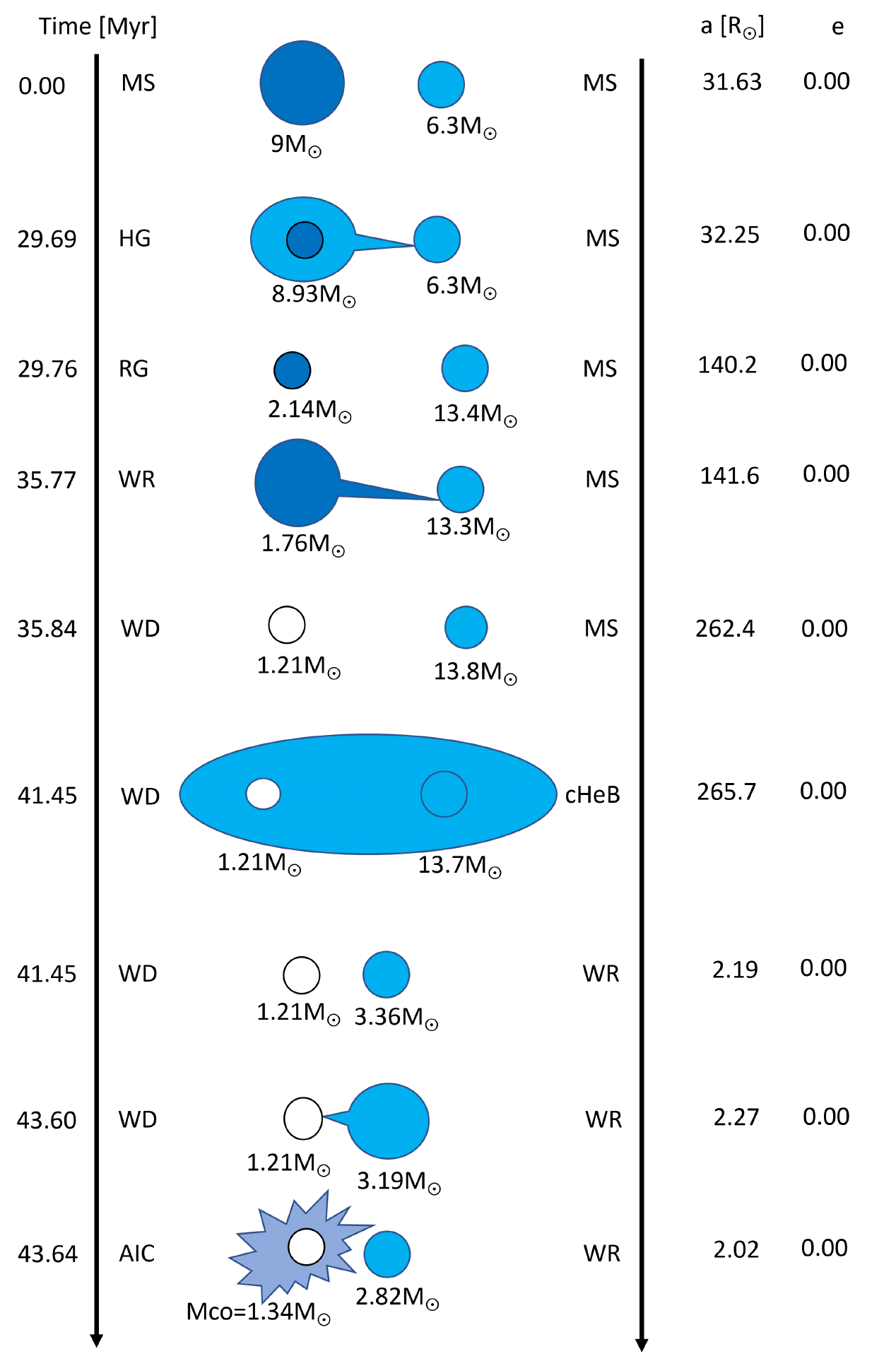}
       \label{fig:AIC2}
  \end{minipage}
        \caption{Examples of AIC progenitor evolutionary paths.}\label{fig:AIC}
\end{figure*}

There are two parameter regions corresponding to an accretion-induced collapse (AIC), see $(M_1,a)\sim(7\,\msun,3000\,\rsun)$ and $(9\,\msun,30\,\rsun)$ in Fig.~\ref{fig:Q07}. The two pathways are shown in Fig.~\ref{fig:AIC}.
One of them is that the primary star, whose initial mass is lower than the SN criterion (in single stellar evolution case), becomes an AIC due to mass accretion from the secondary star. 
The other is the case where the primary star's initial mass is more massive than the SN criterion mass (in single stellar evolution case), but it loses a lot of mass via mass transfers and it cannot become a SN. However, it can become an AIC due to mass accretion from the secondary star. Interestingly, the angular momentum of the core is high in both case, see Fig.~\ref{fig:Q07_am} and the points $(M_1,a)\sim(7\,\msun,3000\,\rsun), (9\,\msun,31.30\,\rsun)$. Recently, simulations of AIC were performed \cite[e.g.,][]{MoriM2023} and the effect of rotation is considered \citep{Abdikamalov2010,Longo2023}.

\subsection{Population synthesis calculation}

\subsubsection{Initial parameter distribution and binary parameters}

We conduct Binary Population Synthesis calculations for 12 models by varying the initial parameter distributions and binary evolution parameters with solar metallicity. The differences between the 12 models are combinations of variations in the initial condition distribution of binary systems, mass transfer parameter $\beta$, and common envelope parameters $\alpha\lambda$. Table \ref{tab:model} shows the parameters for our 12 models. For each model, we performed calculations for $10^5$ binary systems. 

Each initial condition model uses the same initial mass function (IMF) and the same initial mass ratio function (IMRF).
We adopt the Salpeter IMF, 
\begin{equation}
    f(M)=M^{-2.35},
\end{equation}
from $3\msun$ to $100\msun$ \cite{Salpeter1955}, and the flat IMRF from \cite{Kobulnicky2007},
\begin{equation}
    f(q)=\rm const,
\end{equation}
from 0 to 1, where $q=M_2/M_1$.
In the case of the initial orbit parameter distributions such as separation (period) and eccentricity, we use two initial conditions. One is the initial orbit condition distributions from \cite{Sana2012}.
In order to calculate the separation in this model, we use the initial period $P$ function, 
\begin{equation}
    f(\log P)=\left(\log \left[\frac{P}{\rm day}\right]\right)^{-0.55}, 
\end{equation}
from min[10$^{0.15}$day, $P_{\rm min}$] to $(P/\rm day)=10^{3.5}$, where $P_{\rm min}$ is the minimum period where the binary does not interact,
and use the initial eccentricity function,
\begin{equation}
    f(e)=e^{-0.5},
\end{equation}
from 0 to 1.
The other model adopts the orbital initial conditions by \cite{Abt_1983,Heggie_1975}.
In this model, we use the log flat separation distribution,
\begin{equation}
    f(a)=\frac{1}{a},
\end{equation}
from $a_{\rm min}$ to 10$^6\rsun$, where $a_{\rm min}$ is the minimum separation where the binary does not interact, and use the initial eccentricity function
\begin{equation}
    f(e)=e,
\end{equation}
from 0 to 1.

For the mass transfer parameter $\beta$, we use 0.5 and 1. Here, 
$\beta=1$ is a conservative mass transfer.
On the other hand, $\beta=0.5$ means that half of the transferred matter can be accreted to the companion star, and the rest is released out of the binary system.  

We use 0.1, 1, and 10 as the combination of the common envelope parameters $\alpha\lambda$.
When the common envelope parameter values are small, it means that there is less energy available to expel the envelope efficiently during the common envelope phase. As a result, the separation between the two stars tends to decrease because they remain closer together within the common envelope phase.
Conversely, when the common envelope parameter values are large, it indicates that more energy is available to expel the envelope. This can lead to a more effective expulsion of the envelope, allowing the two stars to separate further from each other after the common envelope phase.

\begin{table}
\caption{Parameters for our 12 binary synthesis models.}
\label{tab:model}
\begin{center}
\begin{tabular}{l|ccc}
\hline
 model &  Initial orbit population & $\beta$ & $\alpha\lambda$\\
 \hline
  Sana\_MT1\_CE01 & \cite{Sana2012} & 1 & 0.1 \\
  Sana\_MT1\_CE1 & \cite{Sana2012} & 1 & 1 \\
  Sana\_MT1\_CE10 & \cite{Sana2012} & 1 & 10 \\
  Sana\_MT05\_CE01 & \cite{Sana2012} & 0.5 & 0.1 \\
    Sana\_MT05\_CE1 & \cite{Sana2012} & 0.5 & 1 \\
  Sana\_MT05\_CE10 & \cite{Sana2012} & 0.5 & 10 \\
  Abt\_MT1\_CE01 & \cite{Abt_1983,Heggie_1975} & 1 & 0.1\\
  Abt\_MT1\_CE1 & \cite{Abt_1983,Heggie_1975} & 1 & 1  \\
  Abt\_MT1\_CE10 & \cite{Abt_1983,Heggie_1975} & 1 & 10 \\
  Abt\_MT05\_CE01 & \cite{Abt_1983,Heggie_1975} & 0.5 & 0.1 \\
  Abt\_MT05\_CE1 & \cite{Abt_1983,Heggie_1975} & 0.5 & 1  \\
  Abt\_MT05\_CE10 & \cite{Abt_1983,Heggie_1975}& 0.5 & 10 \\ 
 \hline
\end{tabular}
\end{center}
\end{table}

\subsubsection{Results of population synthesis calculation}
Table \ref{tab:SNtype number} shows the results of our 12 binary synthesis models, listing the numbers of each SN Type in 10$^5$ binaries. 
The main deference between the Sana initial orbit models and the Abt initail orbit models is the numbers of Type II and Type Ibc SNe.
The fraction of close binaries is much higher in the Sana initial orbit models. 
Close binaries are more likely to become Type Ibc SNe and less likely to become Type II SNe (See Fig.~\ref{fig:Q07}).
Thus, the number of Type Ibc SN is much higher in the Sana initial orbit models.

The mass transfer parameter $\beta$ impacts strongly the number of AIC.
Since AIC is caused by accretion, the small $\beta$ which reduces the mass transfer generally makes binaries harder to become AIC.

The common envelope parameter $\alpha\lambda$ influences the number of mergers, with smaller $\alpha\lambda$ increasing binary mergers.
Thus, there is a monotonic increases in the number of GRB with small $\alpha\lambda$. 
However, for AIC and RSN cases, the number of events peaks at $\alpha\lambda$ = 1, and decrease both as $\alpha\lambda$ is decreased or increased. Particularly, at $\alpha\lambda$ = 0.1, the number of events decreases significantly.
This is because both AIC and RSN scenarios require the binary separation to be reduced through a common envelope phase. When the primary star initiates the common envelope phase, and if $\alpha\lambda$ is too small, more binaries tend to merge at that stage. On the other hand, if $\alpha\lambda$ is too large, it does not shrinks the orbit much, leading to a slight decrease in the likelihood of forming close binary systems later on. 
The sharp increase in the number of Type II SNe and the sharp decrease in the number of Type Ibc SNe when $\alpha\lambda$ = 0.1 are attributed to an increase in binaries that fail to effectively shed the envelope during the common envelope phase and consequently merge.
After the merger, the presence of the remaining envelope makes it more likely for the star to become a Type II SN. 
\begin{table}
\caption{The numbers of each SN type for our 12 models. The number of binaries per model is 10$^5$.}
\label{tab:SNtype number}
\begin{center}
\begin{tabular}{l|cccccc}
\hline
 model &  AIC & Ibc & II & RSN & BH & GRB\\
 \hline
  Sana\_MT1\_CE01 & 31 & 6780 & 32646 & 75 & 14726 & 912\\
  Sana\_MT1\_CE1 & 794 & 26456 & 16460 & 1298 & 11412 & 729\\
  Sana\_MT1\_CE10 & 438 & 26322 & 12015 & 1261 & 10551 & 678 \\
  Sana\_MT05\_CE01 & 57 & 6769 & 30314 & 60 &11719 & 897\\
  Sana\_MT05\_CE1 & 189 & 20189 & 17619 & 1515 & 9677 & 701 \\
  Sana\_MT05\_CE10 & 138 & 24598 & 11432 & 1949 & 8740 & 663 \\
  Abt\_MT1\_CE01 & 210 & 2564 & 30560 & 71 & 12638 & 638\\
  Abt\_MT1\_CE1 & 451 & 11222 & 24659 & 546 & 10565 & 427  \\
  Abt\_MT1\_CE10 & 517 & 10515 & 22003 & 504 & 9976 & 390 \\
  Abt\_MT05\_CE01 & 265 & 2678 & 30258 & 62 & 11950 & 629 \\
  Abt\_MT05\_CE1 & 276 & 9453 & 24776 & 608 & 10107 & 416  \\
  Abt\_MT05\_CE10 & 356 & 9933 & 21831 & 644 & 9549 & 389 \\ 
 \hline
\end{tabular}
\end{center}
\end{table}

\begin{figure*}
    \centering
    \includegraphics[width=\hsize]{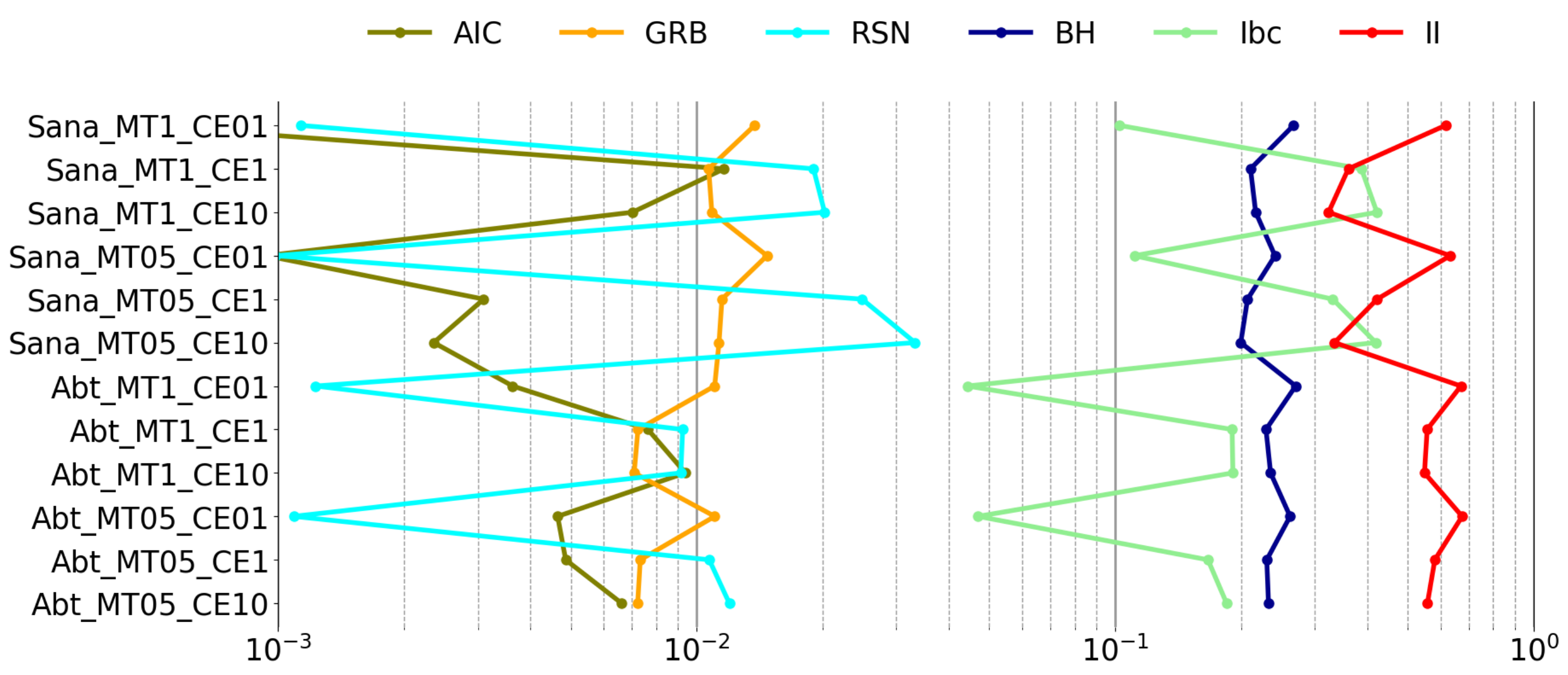}
    \caption{The fractions of each SN type for our 12 models, with binary fraction $f_b=$70\%.}
       \label{fig:fraction70}
\end{figure*}

\begin{figure*}
    \centering
    \includegraphics[width=\hsize]{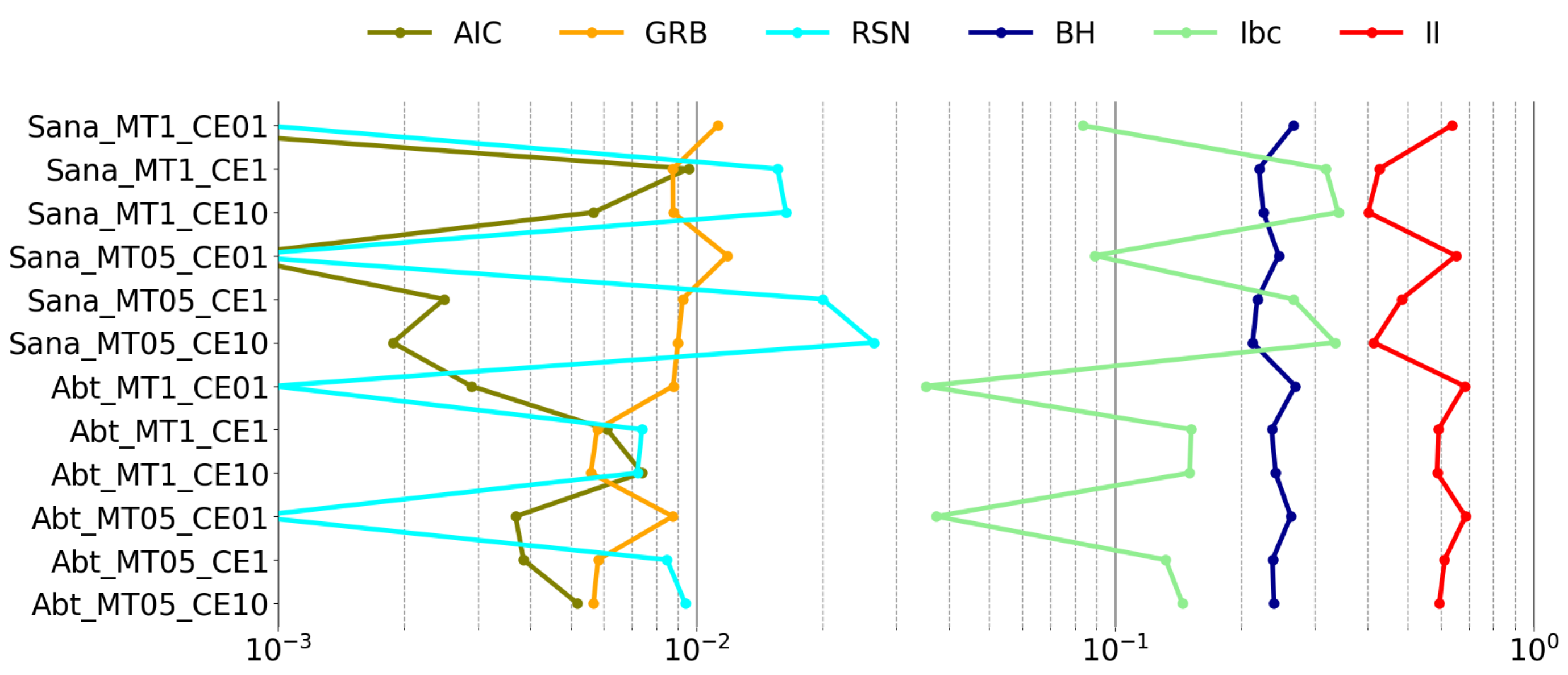}
    \caption{Same as Fig.~\ref{fig:fraction70} but with $f_b=$50\%.}
       \label{fig:fraction50}
\end{figure*}

In order to calculate the fractions of each SN type for our 12 models, we assume a binary fraction $f_b$ of 70\% \citep[e.g.,][]{Sana2012} and $f_b$=50\% \citep[e.g.,][]{Tian2018}. 
According to Fig.~\ref{fig:Q07}, for effectively single stars, relatively light stars ($8\,\msun< M<20\,\msun$) tend to become Type II SN, while more massive stars ($>20\,\msun$) tend to become BH.
If $f_b$=70\% (50\%), the number of Type II SN and the number of BH increase to 8163 (19049) and 2959 (6903), respectively.
Figures \ref{fig:fraction70} and \ref{fig:fraction50} show the fractions of each SN type for our 12 models with $f_b=$70\%, and $f_b=50\%$, respectively.
The numerical data are provided in Appendix \ref{fraction}.

\begin{figure}
    \centering
    \includegraphics[width=\hsize]{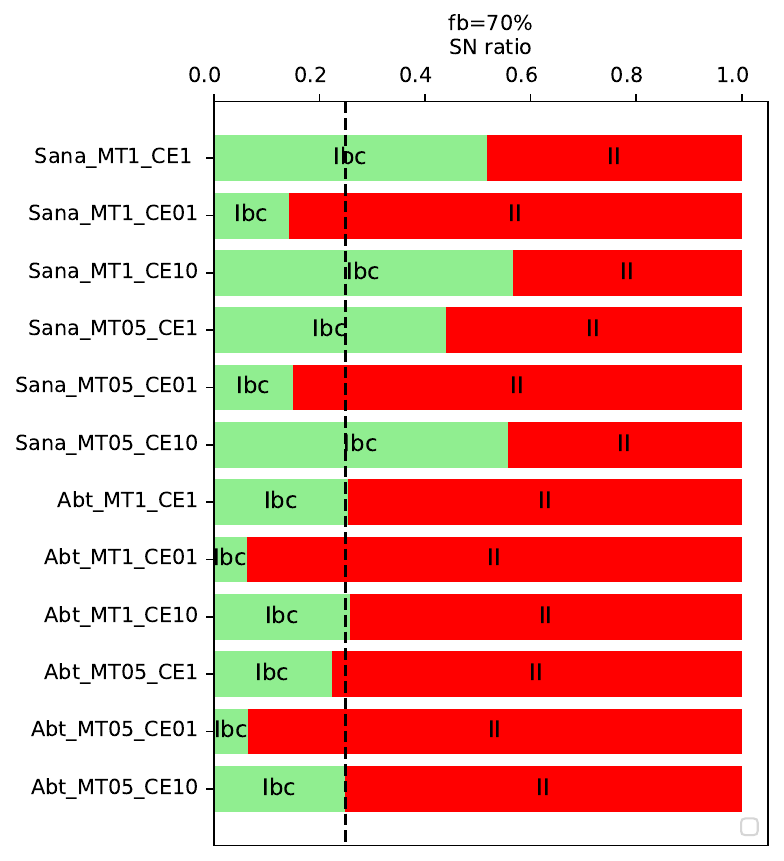}
    \caption{The ratio of Type II SNe to Type Ibc SNe for each binary synthesis model with binary fraction $f_b=70\%$. The black dashed line is the SN ratio from observations \citep{Li2011}.}
    \label{fig:SNration70}
\end{figure}

\begin{figure}
    \centering
    \includegraphics[width=\hsize]{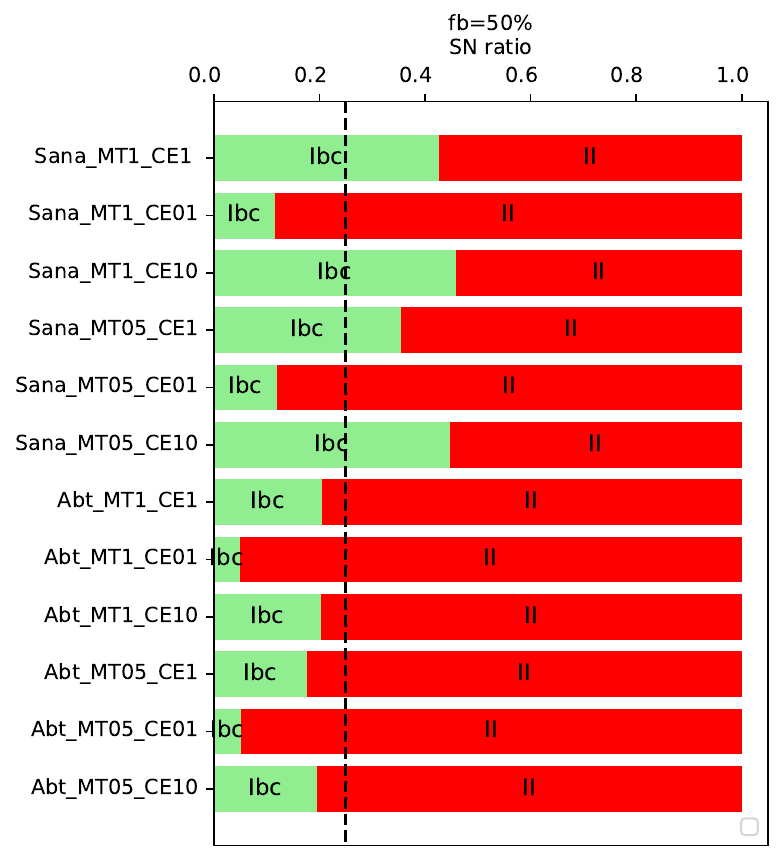}
    \caption{Same as Fig.~\ref{fig:SNration70} but for binary fraction $f_b=50\%$.}
    \label{fig:SNration50}
\end{figure}

Observations of SN at low redshifts reveal that the fraction of SN types is approximately II:Ibc = 75:25 \citep{Li2011}. We show in 
Figures \ref{fig:SNration70} and \ref{fig:SNration50} the II:Ibc ratio for each of our 12 model with $f_b=70\%$ and $f_b=50\%$, respectively. 
Among the models we calculated, the Abt\_MT05\_CE10 model with $f_b=70\%$ best matches observations. The next best-fitting model is the Abt\_MT1\_CE1 model with $f_b=70\%$. 
In the models using \cite{Sana2012} initial parameters, we find that due to a higher number of close binary systems, there is a tendency for more Type Ibc SNe compared to Type II SNe, in comparison to the models using \cite{Abt_1983,Heggie_1975} initial parameters.

The local SN rate in our galaxy $R_{\rm SN}$ can be calculated as,
\begin{equation}
    R_{\rm SN}=\frac{\int^{100}_{3}M^{-2.35}dM}{\int^{100}_{0.1}M^{-2.35}dM}\frac{N_{\rm Ibc}+N_{\rm II}+N_{\rm RSN}}{2N_{\rm binary}+N_{\rm single}}\frac{{\rm SFR}_{\rm gal}}{\langle M\rangle},
\end{equation}
where $N_{\rm Ibc}$, $N_{\rm II}$, and $N_{\rm RSN}$ are the numbers of Type Ibc SN, Type II SN, and RSN, respectively.
$N_{\rm binary}$ and $N_{\rm single}$ are the numbers of total binary systems and single star systems, respectively.
${\rm SFR}_{\rm gal}$ and $\langle M\rangle$ are the star formation late of our galaxy (assumed $2\,\rm\msun~yr^{-1}$), and the average mass of stars, respectively. 
In our simulation, we find $R_{\rm SN}$ falls from $1.14\times10^{-2}~\rm yr^{-1}$ to $1.57\times10^{-2}~\rm yr^{-1}$. 
These values are consistent with the core collapse SN rate in our galaxy from observation, $R_{\rm SN, obs}=3.2^{+7.3}_{-2.6}\times10^{-2}~\rm yr^{-1}$ \citep{Adams2013}.

The local GRB rate at $z\sim0$  can be calculated as,
\begin{equation}
    R_{\rm GRB}=f_B\frac{\int^{100}_{3}M^{-2.35}dM}{\int^{100}_{0.1}M^{-2.35}dM}\frac{N_{\rm GRB}}{2N_{\rm binary}+N_{\rm single}}\frac{{\rm SFR}}{\langle M\rangle},
\end{equation}
where $f_B$, $N_{\rm GRB}$, and ${\rm SFR}$ are the beaming factor of long GRB ($f_B=0.01$ estimated in \citealt{Liang2008, KinugawaAsano2017, KinugawaHA2019}), the number of GRB, and the star formation rate at $z\sim0$ (${\rm SFR}=10^{-1.82}$ estimated in \citealt{Wyder2005, Madau2014}), respectively. 
We find that the long GRB rate of our models is from 5.61\,$\rm yr^{-1}~Gpc^{-3}$ to 16.3\,$\rm yr^{-1}~Gpc^{-3}$.
On the one hand, the local long GRB rate from the GRB observation is $1.3^{+0.6}_{-0.7}\,\rm ~yr^{-1}~ Gpc^{-3}$ \citep{WP2010}, much lower than our estimates. 
On the other hand, observations suggest that there are many long GRBs with lower luminosities that are harder to detect \citep{Pian2006}, and the long GRB rate $1.3^{+0.6}_{-0.7}\,\rm ~yr^{-1}~ Gpc^{-1}$ \citep{WP2010} only includes the ``normal'' long GRBs with high luminosity $L>10^{49}\,\rm ergs~s^{-1}$. 
The low luminosity long GRB rate is estimated $R_{\rm LLGRB}\sim100$--$600~\rm yr^{-1}~ Gpc^{-1}$ \citep{Pian2006,Liang2007}.
If we assume that all GRB progenitors of our models become a low luminosity long GRB, then we estimated the low luminosity long GRB rate from 40.1\,$\rm yr^{-1}~ Gpc^{-1}$ to 116\,$\rm yr^{-1}~ Gpc^{-1}$ using a beaming factor for low luminosity long GRB $f_B=1/14$ 
\citep{Liang2007}.
According to \cite{Liang2007}, it is possible that $f_B$ of a low luminosity long GRB is greater than 1/14. In this case, this rate could become even larger.
Thus, the low luminosity long GRB rates of our models are roughly consistent with the low luminosity GRB rate from observations.    

Based on our population synthesis models, we are able to estimate the rates of RSN and AIC. To the best of our knowledge, this is the first time for the RSN rate to be estimated. \cite{Shibagaki2020} assume GRB rates as a proxy for the rate and we confirm that assumption is not wrong. In the best-fit models (Abt\_MT05\_CE10 model or Abt\_MT1\_CE1 model), the rate is close to the GRB rate.
Though it is a rare opportunity, we can expect neutrino and gravitational waves from nearby RSN, which would have time-variability of the frequency of proto-neutron star \citep[e.g.,][]{Takiwaki2021}.
Note that the rate strongly depends on the common envelope parameter. As discussed above, in the models with $\alpha\lambda = 0.1$, the rate significantly drops. The rate of AIC is similar to RSN.

{
We can also obtain the NS:BH ratios from our calculations. In all models, the NS:BH fraction remains relatively constant, ranging from 79:21 to 72:28. These values are approximately the same as a single star case (NS:BH=73:27) if we assume that progenitors with $8~\msun\le M \le 20~\msun$ become type II SNe (NSs) and progenitors with $20~\msun< M$ become BHs. These are broadly consistent with observations. In particular, there are a few candidate massive stars that have disappeared without obvious luminous SNe \citep{Gerke:2014ooa,Adams:2016ffj,Adams:2016hit,Basinger:2020iir,Neustadt:2021jjt}. If these are collapses to black holes, they can be combined with the number of disappearances coincident with SNe to yield the fraction of core collapses that fail to produce SNe \citep{Kochanek:2008mp}. Current estimates of such a black hole channel is $23.6^{+23.3}_{-15.7}$\% of massive stars undergoing collapse. Other indirect probes also indicate similar fractions \citep[e.g.,][]{Horiuchi:2014ska}. 
}




\section{Summary and Discussion}\label{sec:summary}
In this paper we aim to systematically investigate how binary interactions affect the progenitors of SNe. To this end, we first conducted binary evolution calculations with fixed mass ratios. With this setup, we explored the orbital separations and primary star masses where various types of SNe can be produced, and delineated the evolutionary pathways through which each SN explosion could occur. We found that binaries with orbital separation greater than the red giant radius evolve similarly to single stars. Relatively lighter stars in such binaries become Type II SNe, while heavier stars become BHs. 
On the other hand, we found that binaries with orbital separations roughly equal to or less than the red giant's radius develop into diverse types of SNe due to binary interactions, such as a common envelope phase and/or stable mass transfer. Rare explosion phenomenon such as rapidly rotating SNe and accretion induced collapse (AIC) are also observed in this separation range, $10^{1.5}\,\rsun<a <10^{3.5}\,\rsun$ with lighter mass progenitor $\mZAMS<9\,\msun$. Furthermore, we found that binaries with orbital separation $\lesssim10\,\rsun$ undergo stellar mergers before core collapse. This implies that such progenitor stars have relatively large angular momentum at core collapse, making them conducive as long GRB progenitors.

In a previous study, \cite{KinugawaAsano2017} calculated the long GRB rate of binary merger progenitor models \citep{Fryer2005}.
In their binary merger progenitor model, the stellar merger of naked He stars or the stellar mergers of naked He star and red giant as the long GRB progenitor were considered. 
\cite{KinugawaAsano2017} showed that the long GRB rate by the binary merger progenitor model is consistent with the long GRB rate from observation \citep{WP2010}, and the metallicity evolution effect to the binary merger progenitor model can explain the evolution of the redshift of the long GRB rate.
In our study, however, there is no long GRB progenitor of the binary merger progenitor model. 
The reason for this difference lies in the different treatments of the envelopes of the remnants that merged during the common envelope phase.
In \cite{KinugawaAsano2017}, it was assume that all the envelope will evaporate after the common envelope phase due to the huge angular momentum by the merger.
On the other hand, we assumed that the envelope is not completely blown away during the common envelope phase, and partially remains. 
The binary merger progenitor model considers the merging of He stars in the late stages of evolution, which implies that they might have more angular momentum compared to our model, where stars merge early in the main sequence and lose angular momentum due to stellar winds. 
Furthermore, the binary merger progenitor model calculation \citep{KinugawaAsano2017} is consistent with the normal long GRB rate, whereas our calculation models appear to be consistent with low luminosity long GRBs rather than normal long GRBs.
Therefore, it is plausible that the formation processes of normal long GRBs and low luminosity long GRBs are different.

In our calculations, close binaries tend to lead to Type Ibc SNe. Therefore, the models with \cite{Sana2012} initial parameters typically showed larger discrepancies in the ratio between Type Ibc and Type II SNe compared to observations, with an overabundance of Type Ibc SNe. While this can be offset with a lower common envelope parameter, which results in increased number of remnants merging within the common envelope and a significant increase in the number of Type II SNe, this outcome is highly dependent on the handling of the envelopes of the merging remnants in the process. If we assume the same hypothesis of \cite{KinugawaAsano2017}, the number of Type II SNe could decrease in this case. 
To increase the number of Type II SNe in the \cite{Sana2012} initial parameter models, several factors might be needed, e.g., an increase in the number of single stars, an increase in binaries with significantly wider separations where interactions don't play a significant role, or making the stripping of the hydrogen envelope by common envelope interactions less effective than it is now.

Binary population models like ours holds various possible applications. For example, there are growing numbers of observations of neutron stars \cite[see][and references therein]{Enoto2019}.
Previous studies have revealed valuable insights into the distributions of mass, rotation, magnetic fields, and kicks \citep[e.g.,][]{Noutsos2013,Igoshev2022}.
Though this paper does not discuss neutron star properties, it is important to compare our theoretical models with observations.
Recently, the correlation of  neutron star spins and spin-kick alignment is discussed in \cite{Janka2022}.

Also, since we calculate the number of explosive phenomena and CO core mass of the progenitor, our models allow for an updated prediction of DSNB flux such as previously demonstrated in \cite{Horiuchi2021}.
The DSNB is a promising method to investigate the properties of extragalactic core collapses \citep{Beacom2010,Lunardini2016,Ando:2023fcc}.
The SuperKamiokande (SK) water Cherenkov detector has already excluded some theoretical models and placed upper bounds on the DSNB flux \citep{Abe2021}. Recently, SK was upgraded with a gadolinium salt (SK-Gd), and expected to perform significantly more sensitive searches for the DSNB \citep{Harada2023}.
On the theoretical side, more realistic predictions of DNSB is ongoing. For example, 
\cite{Horiuchi2021} and \cite{Kresse2021} consider the binary effect. \cite{Ashida2023} uses a new modeling of galactic chemical evolution, where a variable stellar initial mass function depending on the galaxy type. \cite{Ekanger:2022neg} explore multiple schemes to estimate time-integrated spetra, while \cite{Ekanger2023} employs the neutrino spectrum calculated in 2D long-term simulations and updated star formation data. 

In this study, we simply connected the CO core mass to the fate of the core collapse. However other parameters could be important to determine the ultimate fate \citep{Pejcha2015,Ertl2016,Wang2022,Tsang2022,Takahashi2023}.
For improvements, systematic studies of core-collapse SNe could extend previous works in 1D simulations \citep{O'connor2011,Ugliano2012,Sukhbold2016,Ebinger2020,Couch2020,Boccioli2021}, 2D simulations \citep{NakamuraKo2015,Summa2016,Vartanyan2023}, and 3D simulations \citep{Burrows2020}.
Simultaneously, more sophisticated input physics should be required, e.g., neutrino oscillation \citep{Nagakura23, Ehring23}, neutrino reaction rate \citep{Kotake18, Sugiura2022}, and equation of state \citep{Fischer14, Fischer20}.
Using such data, we can develop a better phenomenological treatment of the explosions \citep[e.g.,][]{Belczynski2012}.

There is much room for improvement in the assumptions used in this study.
Even in the physics used in single star evolution we have uncertainties.
For example, we have several recipes of wind mass loss rate \citep{deJager1988,Vink2000,Vink2001,Vink2005,Nugis2000},
thereby impacting the core mass \citep[e.g.,][]{Woosley2020}.
Advanced stages of stellar evolution lack strict constraints on overshooting parameters \citep[e.g.,][]{Yoshida2019,Temaj2023} necessitating high-resolution simulations of convective shells for accurate determination \citep{Cristini2017}.
We use {\small REACLIB} for the reaction rate of nucleosynthesis \citep{Cyburt2010}, but uncertainties with some reaction channels potentially alter stellar structures \citep[e.g.,][]{Takahashi2016}.
The final stellar structure can also depend on numerical resolution \citep[e.g.,][]{Kato2020}.
Angular momentum transport inside stars requires much theoretical efforts. The Tayler--Spruit dynamo is usually assumed \citep[e.g.,][]{Heger2005} and wave-driven mechanism is also considered \citep{Fuller2015}.
On binary star evolution, there are uncertainties about the mass transfer rate and the angular momentum loss \citep{Hirai2023, Willcox2023}. 
Despite these uncertainties, we have been able to explore the progenitors behind major as well as rare SN types, and moreover their typical evolutionary paths. 
To make progress, it would be necessary to start from our study and systematically consider each uncertain factor, comparing and evaluating them in relation to observations one by one.

\section*{acknowledgments}
We thank K.~Takahashi, R.~Hirai, and C.~Lunardini for informative and stimulating discussions.
This work was supported by the Research Institute of Stellar Explosive Phenomena at Fukuoka University and also by JSPS KAKENHI
Grant Number (JP21K13915,  
JP21H01088, 
JP22H01223, 
JP22K03630,  
JP23H04899,  
JP23H01199, 
and JP23K03400). 
The work of S.H.~is also supported by the U.S.~Department of Energy Office of Science under award number DE-SC0020262, and U.S.~National Science Foundation Grant No.~AST1908960 and No.~PHY-2209420. 
This work was supported by World Premier International Research Center Initiative (WPI Initiative), MEXT, Japan.
This research was also supported by MEXT as “Program for Promoting researches on the Supercomputer Fugaku” (Structure and Evolution of the Universe Unraveled by Fusion of Simulation and AI; Grant Number JPMXP1020230406) and JICFuS.

\section*{Data Availability}
The data underlying this article will be shared on reasonable request to the corresponding author.

\bibliographystyle{mnras}
\bibliography{BinarySN}

\begin{thebibliography}{}
\makeatletter
\relax
\def\mn@urlcharsother{\let\do\@makeother \do\$\do\&\do\#\do\^\do\_\do\%\do\~}
\def\mn@doi{\begingroup\mn@urlcharsother \@ifnextchar [ {\mn@doi@} {\mn@doi@[]}}
\def\mn@doi@[#1]#2{\def\@tempa{#1}\ifx\@tempa\@empty \href {http://dx.doi.org/#2} {doi:#2}\else \href {http://dx.doi.org/#2} {#1}\fi \endgroup}
\def\mn@eprint#1#2{\mn@eprint@#1:#2::\@nil}
\def\mn@eprint@arXiv#1{\href {http://arxiv.org/abs/#1} {{\tt arXiv:#1}}}
\def\mn@eprint@dblp#1{\href {http://dblp.uni-trier.de/rec/bibtex/#1.xml} {dblp:#1}}
\def\mn@eprint@#1:#2:#3:#4\@nil{\def\@tempa {#1}\def\@tempb {#2}\def\@tempc {#3}\ifx \@tempc \@empty \let \@tempc \@tempb \let \@tempb \@tempa \fi \ifx \@tempb \@empty \def\@tempb {arXiv}\fi \@ifundefined {mn@eprint@\@tempb}{\@tempb:\@tempc}{\expandafter \expandafter \csname mn@eprint@\@tempb\endcsname \expandafter{\@tempc}}}

\bibitem[\protect\citeauthoryear{{Abdikamalov}, {Ott}, {Rezzolla}, {Dessart}, {Dimmelmeier}, {Marek}  \& {Janka}}{{Abdikamalov} et~al.}{2010}]{Abdikamalov2010}
{Abdikamalov} E.~B.,  {Ott} C.~D.,  {Rezzolla} L.,  {Dessart} L.,  {Dimmelmeier} H.,  {Marek} A.,   {Janka} H.~T.,  2010, \mn@doi [\prd] {10.1103/PhysRevD.81.044012}, \href {https://ui.adsabs.harvard.edu/abs/2010PhRvD..81d4012A} {81, 044012}

\bibitem[\protect\citeauthoryear{{Abe} et~al.,}{{Abe} et~al.}{2021}]{Abe2021}
{Abe} K.,  et~al., 2021, \mn@doi [\prd] {10.1103/PhysRevD.104.122002}, \href {https://ui.adsabs.harvard.edu/abs/2021PhRvD.104l2002A} {104, 122002}

\bibitem[\protect\citeauthoryear{{Abt}}{{Abt}}{1983}]{Abt_1983}
{Abt} H.~A.,  1983, \mn@doi [\araa] {10.1146/annurev.aa.21.090183.002015}, \href {http://adsabs.harvard.edu/abs/1983ARA%26A..21..343A} {21, 343}

\bibitem[\protect\citeauthoryear{{Adams}, {Kochanek}, {Beacom}, {Vagins}  \& {Stanek}}{{Adams} et~al.}{2013}]{Adams2013}
{Adams} S.~M.,  {Kochanek} C.~S.,  {Beacom} J.~F.,  {Vagins} M.~R.,   {Stanek} K.~Z.,  2013, \mn@doi [\apj] {10.1088/0004-637X/778/2/164}, \href {https://ui.adsabs.harvard.edu/abs/2013ApJ...778..164A} {778, 164}

\bibitem[\protect\citeauthoryear{Adams, Kochanek, Gerke, Stanek  \& Dai}{Adams et~al.}{2017a}]{Adams:2016ffj}
Adams S.~M.,  Kochanek C.~S.,  Gerke J.~R.,  Stanek K.~Z.,   Dai X.,  2017a, \mn@doi [Mon. Not. Roy. Astron. Soc.] {10.1093/mnras/stx816}, 468, 4968

\bibitem[\protect\citeauthoryear{Adams, Kochanek, Gerke  \& Stanek}{Adams et~al.}{2017b}]{Adams:2016hit}
Adams S.~M.,  Kochanek C.~S.,  Gerke J.~R.,   Stanek K.~Z.,  2017b, \mn@doi [Mon. Not. Roy. Astron. Soc.] {10.1093/mnras/stx898}, 469, 1445

\bibitem[\protect\citeauthoryear{{Aguilera-Dena}, {Langer}, {Moriya}  \& {Schootemeijer}}{{Aguilera-Dena} et~al.}{2018}]{Aguilera-Dena2018}
{Aguilera-Dena} D.~R.,  {Langer} N.,  {Moriya} T.~J.,   {Schootemeijer} A.,  2018, \mn@doi [\apj] {10.3847/1538-4357/aabfc1}, \href {https://ui.adsabs.harvard.edu/abs/2018ApJ...858..115A} {858, 115}

\bibitem[\protect\citeauthoryear{Ando, Ekanger, Horiuchi  \& Koshio}{Ando et~al.}{2023}]{Ando:2023fcc}
Ando S.,  Ekanger N.,  Horiuchi S.,   Koshio Y.,  2023, in Proceedings of the Japan Academy, Series B.  (\mn@eprint {arXiv} {2306.16076})

\bibitem[\protect\citeauthoryear{{Ashida}, {Nakazato}  \& {Tsujimoto}}{{Ashida} et~al.}{2023}]{Ashida2023}
{Ashida} Y.,  {Nakazato} K.,   {Tsujimoto} T.,  2023, \mn@doi [\apj] {10.3847/1538-4357/ace3ba}, \href {https://ui.adsabs.harvard.edu/abs/2023ApJ...953..151A} {953, 151}

\bibitem[\protect\citeauthoryear{Basinger, Kochanek, Adams, Dai  \& Stanek}{Basinger et~al.}{2021}]{Basinger:2020iir}
Basinger C.~M.,  Kochanek C.~S.,  Adams S.~M.,  Dai X.,   Stanek K.~Z.,  2021, \mn@doi [Mon. Not. Roy. Astron. Soc.] {10.1093/mnras/stab2620}, 508, 1156

\bibitem[\protect\citeauthoryear{{Beacom}}{{Beacom}}{2010}]{Beacom2010}
{Beacom} J.~F.,  2010, \mn@doi [Annual Review of Nuclear and Particle Science] {10.1146/annurev.nucl.010909.083331}, \href {https://ui.adsabs.harvard.edu/abs/2010ARNPS..60..439B} {60, 439}

\bibitem[\protect\citeauthoryear{{Belczynski}, {Kalogera}  \& {Bulik}}{{Belczynski} et~al.}{2002}]{Belczynski_2002}
{Belczynski} K.,  {Kalogera} V.,   {Bulik} T.,  2002, \mn@doi [\apj] {10.1086/340304}, \href {https://ui.adsabs.harvard.edu/abs/2002ApJ...572..407B} {572, 407}

\bibitem[\protect\citeauthoryear{{Belczynski}, {Kalogera}, {Rasio}, {Taam}, {Zezas}, {Bulik}, {Maccarone}  \& {Ivanova}}{{Belczynski} et~al.}{2008}]{Belczynski_2008}
{Belczynski} K.,  {Kalogera} V.,  {Rasio} F.~A.,  {Taam} R.~E.,  {Zezas} A.,  {Bulik} T.,  {Maccarone} T.~J.,   {Ivanova} N.,  2008, \mn@doi [\apjs] {10.1086/521026}, \href {https://ui.adsabs.harvard.edu/abs/2008ApJS..174..223B} {174, 223}

\bibitem[\protect\citeauthoryear{{Belczynski}, {Wiktorowicz}, {Fryer}, {Holz}  \& {Kalogera}}{{Belczynski} et~al.}{2012}]{Belczynski2012}
{Belczynski} K.,  {Wiktorowicz} G.,  {Fryer} C.~L.,  {Holz} D.~E.,   {Kalogera} V.,  2012, \mn@doi [\apj] {10.1088/0004-637X/757/1/91}, \href {https://ui.adsabs.harvard.edu/abs/2012ApJ...757...91B} {757, 91}

\bibitem[\protect\citeauthoryear{{Boccioli}, {Mathews}  \& {O'Connor}}{{Boccioli} et~al.}{2021}]{Boccioli2021}
{Boccioli} L.,  {Mathews} G.~J.,   {O'Connor} E.~P.,  2021, \mn@doi [\apj] {10.3847/1538-4357/abe767}, \href {https://ui.adsabs.harvard.edu/abs/2021ApJ...912...29B} {912, 29}

\bibitem[\protect\citeauthoryear{{Bollig}, {Yadav}, {Kresse}, {Janka}, {M{\"u}ller}  \& {Heger}}{{Bollig} et~al.}{2021}]{Bollig2021}
{Bollig} R.,  {Yadav} N.,  {Kresse} D.,  {Janka} H.-T.,  {M{\"u}ller} B.,   {Heger} A.,  2021, \mn@doi [\apj] {10.3847/1538-4357/abf82e}, \href {https://ui.adsabs.harvard.edu/abs/2021ApJ...915...28B} {915, 28}

\bibitem[\protect\citeauthoryear{{Breivik} et~al.,}{{Breivik} et~al.}{2020}]{Breivik2020}
{Breivik} K.,  et~al., 2020, \mn@doi [\apj] {10.3847/1538-4357/ab9d85}, \href {https://ui.adsabs.harvard.edu/abs/2020ApJ...898...71B} {898, 71}

\bibitem[\protect\citeauthoryear{{Bugli}, {Guilet}, {Foglizzo}  \& {Obergaulinger}}{{Bugli} et~al.}{2023}]{Bugli2023}
{Bugli} M.,  {Guilet} J.,  {Foglizzo} T.,   {Obergaulinger} M.,  2023, \mn@doi [\mnras] {10.1093/mnras/stad496}, \href {https://ui.adsabs.harvard.edu/abs/2023MNRAS.520.5622B} {520, 5622}

\bibitem[\protect\citeauthoryear{{Burrows}, {Radice}, {Vartanyan}, {Nagakura}, {Skinner}  \& {Dolence}}{{Burrows} et~al.}{2020}]{Burrows2020}
{Burrows} A.,  {Radice} D.,  {Vartanyan} D.,  {Nagakura} H.,  {Skinner} M.~A.,   {Dolence} J.~C.,  2020, \mn@doi [\mnras] {10.1093/mnras/stz3223}, \href {https://ui.adsabs.harvard.edu/abs/2020MNRAS.491.2715B} {491, 2715}

\bibitem[\protect\citeauthoryear{{Cantiello}, {Yoon}, {Langer}  \& {Livio}}{{Cantiello} et~al.}{2007}]{Cantiello2007}
{Cantiello} M.,  {Yoon} S.~C.,  {Langer} N.,   {Livio} M.,  2007, \mn@doi [\aap] {10.1051/0004-6361:20077115}, \href {https://ui.adsabs.harvard.edu/abs/2007A&A...465L..29C} {465, L29}

\bibitem[\protect\citeauthoryear{{Chatzopoulos}, {Frank}, {Marcello}  \& {Clayton}}{{Chatzopoulos} et~al.}{2020}]{Chatzopoulos2020}
{Chatzopoulos} E.,  {Frank} J.,  {Marcello} D.~C.,   {Clayton} G.~C.,  2020, \mn@doi [\apj] {10.3847/1538-4357/ab91bb}, \href {https://ui.adsabs.harvard.edu/abs/2020ApJ...896...50C} {896, 50}

\bibitem[\protect\citeauthoryear{{Chieffi} \& {Limongi}}{{Chieffi} \& {Limongi}}{2020}]{Chieffi2020}
{Chieffi} A.,  {Limongi} M.,  2020, \mn@doi [\apj] {10.3847/1538-4357/ab6739}, \href {https://ui.adsabs.harvard.edu/abs/2020ApJ...890...43C} {890, 43}

\bibitem[\protect\citeauthoryear{{Chini}, {Hoffmeister}, {Nasseri}, {Stahl}  \& {Zinnecker}}{{Chini} et~al.}{2012}]{Chiini2012}
{Chini} R.,  {Hoffmeister} V.~H.,  {Nasseri} A.,  {Stahl} O.,   {Zinnecker} H.,  2012, \mn@doi [\mnras] {10.1111/j.1365-2966.2012.21317.x}, \href {https://ui.adsabs.harvard.edu/abs/2012MNRAS.424.1925C} {424, 1925}

\bibitem[\protect\citeauthoryear{{Cigan} et~al.,}{{Cigan} et~al.}{2019}]{Cigan2019}
{Cigan} P.,  et~al., 2019, \mn@doi [\apj] {10.3847/1538-4357/ab4b46}, \href {https://ui.adsabs.harvard.edu/abs/2019ApJ...886...51C} {886, 51}

\bibitem[\protect\citeauthoryear{{Conroy}}{{Conroy}}{2013}]{Conroy2013}
{Conroy} C.,  2013, \mn@doi [\araa] {10.1146/annurev-astro-082812-141017}, \href {https://ui.adsabs.harvard.edu/abs/2013ARA&A..51..393C} {51, 393}

\bibitem[\protect\citeauthoryear{{Couch}, {Chatzopoulos}, {Arnett}  \& {Timmes}}{{Couch} et~al.}{2015}]{Couch2015}
{Couch} S.~M.,  {Chatzopoulos} E.,  {Arnett} W.~D.,   {Timmes} F.~X.,  2015, \mn@doi [\apjl] {10.1088/2041-8205/808/1/L21}, \href {https://ui.adsabs.harvard.edu/abs/2015ApJ...808L..21C} {808, L21}

\bibitem[\protect\citeauthoryear{{Couch}, {Warren}  \& {O'Connor}}{{Couch} et~al.}{2020}]{Couch2020}
{Couch} S.~M.,  {Warren} M.~L.,   {O'Connor} E.~P.,  2020, \mn@doi [\apj] {10.3847/1538-4357/ab609e}, \href {https://ui.adsabs.harvard.edu/abs/2020ApJ...890..127C} {890, 127}

\bibitem[\protect\citeauthoryear{{Cristini}, {Meakin}, {Hirschi}, {Arnett}, {Georgy}, {Viallet}  \& {Walkington}}{{Cristini} et~al.}{2017}]{Cristini2017}
{Cristini} A.,  {Meakin} C.,  {Hirschi} R.,  {Arnett} D.,  {Georgy} C.,  {Viallet} M.,   {Walkington} I.,  2017, \mn@doi [\mnras] {10.1093/mnras/stx1535}, \href {https://ui.adsabs.harvard.edu/abs/2017MNRAS.471..279C} {471, 279}

\bibitem[\protect\citeauthoryear{{Cyburt} et~al.,}{{Cyburt} et~al.}{2010}]{Cyburt2010}
{Cyburt} R.~H.,  et~al., 2010, \mn@doi [\apjs] {10.1088/0067-0049/189/1/240}, \href {https://ui.adsabs.harvard.edu/abs/2010ApJS..189..240C} {189, 240}

\bibitem[\protect\citeauthoryear{{De Marco} \& {Izzard}}{{De Marco} \& {Izzard}}{2017}]{DeMarco2017}
{De Marco} O.,  {Izzard} R.~G.,  2017, \mn@doi [\pasa] {10.1017/pasa.2016.52}, \href {https://ui.adsabs.harvard.edu/abs/2017PASA...34....1D} {34, e001}

\bibitem[\protect\citeauthoryear{{Dominik}, {Belczynski}, {Fryer}, {Holz}, {Berti}, {Bulik}, {Mandel}  \& {O'Shaughnessy}}{{Dominik} et~al.}{2013}]{Dominik_2013}
{Dominik} M.,  {Belczynski} K.,  {Fryer} C.,  {Holz} D.~E.,  {Berti} E.,  {Bulik} T.,  {Mandel} I.,   {O'Shaughnessy} R.,  2013, \mn@doi [\apj] {10.1088/0004-637X/779/1/72}, \href {http://adsabs.harvard.edu/abs/2013ApJ...779...72D} {779, 72}

\bibitem[\protect\citeauthoryear{{Ebinger}, {Curtis}, {Ghosh}, {Fr{\"o}hlich}, {Hempel}, {Perego}, {Liebend{\"o}rfer}  \& {Thielemann}}{{Ebinger} et~al.}{2020}]{Ebinger2020}
{Ebinger} K.,  {Curtis} S.,  {Ghosh} S.,  {Fr{\"o}hlich} C.,  {Hempel} M.,  {Perego} A.,  {Liebend{\"o}rfer} M.,   {Thielemann} F.-K.,  2020, \mn@doi [\apj] {10.3847/1538-4357/ab5dcb}, \href {https://ui.adsabs.harvard.edu/abs/2020ApJ...888...91E} {888, 91}

\bibitem[\protect\citeauthoryear{{Eggleton}}{{Eggleton}}{1983}]{Eggleton_1983}
{Eggleton} P.~P.,  1983, \mn@doi [\apj] {10.1086/160960}, \href {http://adsabs.harvard.edu/abs/1983ApJ...268..368E} {268, 368}

\bibitem[\protect\citeauthoryear{{Eggleton}}{{Eggleton}}{2011}]{Eggleton2011}
{Eggleton} P.,  2011, {Evolutionary Processes in Binary and Multiple Stars}.
{Cambridge University Press}

\bibitem[\protect\citeauthoryear{{Ehring}, {Abbar}, {Janka}, {Raffelt}  \& {Tamborra}}{{Ehring} et~al.}{2023}]{Ehring23}
{Ehring} J.,  {Abbar} S.,  {Janka} H.-T.,  {Raffelt} G.,   {Tamborra} I.,  2023, \mn@doi [\prd] {10.1103/PhysRevD.107.103034}, \href {https://ui.adsabs.harvard.edu/abs/2023PhRvD.107j3034E} {107, 103034}

\bibitem[\protect\citeauthoryear{Ekanger, Horiuchi, Kotake  \& Sumiyoshi}{Ekanger et~al.}{2022}]{Ekanger:2022neg}
Ekanger N.,  Horiuchi S.,  Kotake K.,   Sumiyoshi K.,  2022, \mn@doi [Phys. Rev. D] {10.1103/PhysRevD.106.043026}, 106, 043026

\bibitem[\protect\citeauthoryear{{Ekanger}, {Horiuchi}, {Nagakura}  \& {Reitz}}{{Ekanger} et~al.}{2023}]{Ekanger2023}
{Ekanger} N.,  {Horiuchi} S.,  {Nagakura} H.,   {Reitz} S.,  2023, \mn@doi [arXiv e-prints] {10.48550/arXiv.2310.15254}, \href {https://ui.adsabs.harvard.edu/abs/2023arXiv231015254E} {p. arXiv:2310.15254}

\bibitem[\protect\citeauthoryear{{Ekstr{\"o}m} et~al.,}{{Ekstr{\"o}m} et~al.}{2012}]{Ekstrom2012}
{Ekstr{\"o}m} S.,  et~al., 2012, \mn@doi [\aap] {10.1051/0004-6361/201117751}, \href {https://ui.adsabs.harvard.edu/abs/2012A&A...537A.146E} {537, A146}

\bibitem[\protect\citeauthoryear{{Eldridge} \& {Stanway}}{{Eldridge} \& {Stanway}}{2022}]{Eldridge2022}
{Eldridge} J.~J.,  {Stanway} E.~R.,  2022, \mn@doi [\araa] {10.1146/annurev-astro-052920-100646}, \href {https://ui.adsabs.harvard.edu/abs/2022ARA&A..60..455E} {60, 455}

\bibitem[\protect\citeauthoryear{{Enoto}, {Kisaka}  \& {Shibata}}{{Enoto} et~al.}{2019}]{Enoto2019}
{Enoto} T.,  {Kisaka} S.,   {Shibata} S.,  2019, \mn@doi [Reports on Progress in Physics] {10.1088/1361-6633/ab3def}, \href {https://ui.adsabs.harvard.edu/abs/2019RPPh...82j6901E} {82, 106901}

\bibitem[\protect\citeauthoryear{{Ertl}, {Janka}, {Woosley}, {Sukhbold}  \& {Ugliano}}{{Ertl} et~al.}{2016}]{Ertl2016}
{Ertl} T.,  {Janka} H.~T.,  {Woosley} S.~E.,  {Sukhbold} T.,   {Ugliano} M.,  2016, \mn@doi [\apj] {10.3847/0004-637X/818/2/124}, \href {https://ui.adsabs.harvard.edu/abs/2016ApJ...818..124E} {818, 124}

\bibitem[\protect\citeauthoryear{{Fields}}{{Fields}}{2022}]{Fields2022}
{Fields} C.~E.,  2022, \mn@doi [\apjl] {10.3847/2041-8213/ac460c}, \href {https://ui.adsabs.harvard.edu/abs/2022ApJ...924L..15F} {924, L15}

\bibitem[\protect\citeauthoryear{{Fischer}, {Hempel}, {Sagert}, {Suwa}  \& {Schaffner-Bielich}}{{Fischer} et~al.}{2014}]{Fischer14}
{Fischer} T.,  {Hempel} M.,  {Sagert} I.,  {Suwa} Y.,   {Schaffner-Bielich} J.,  2014, \mn@doi [European Physical Journal A] {10.1140/epja/i2014-14046-5}, \href {https://ui.adsabs.harvard.edu/abs/2014EPJA...50...46F} {50, 46}

\bibitem[\protect\citeauthoryear{{Fischer}, {Wu}, {Wehmeyer}, {Bastian}, {Mart{\'\i}nez-Pinedo}  \& {Thielemann}}{{Fischer} et~al.}{2020}]{Fischer20}
{Fischer} T.,  {Wu} M.-R.,  {Wehmeyer} B.,  {Bastian} N.-U.~F.,  {Mart{\'\i}nez-Pinedo} G.,   {Thielemann} F.-K.,  2020, \mn@doi [\apj] {10.3847/1538-4357/ab86b0}, \href {https://ui.adsabs.harvard.edu/abs/2020ApJ...894....9F} {894, 9}

\bibitem[\protect\citeauthoryear{{Fragos} et~al.,}{{Fragos} et~al.}{2023}]{Fragos2023}
{Fragos} T.,  et~al., 2023, \mn@doi [\apjs] {10.3847/1538-4365/ac90c1}, \href {https://ui.adsabs.harvard.edu/abs/2023ApJS..264...45F} {264, 45}

\bibitem[\protect\citeauthoryear{{Fryer} \& {Heger}}{{Fryer} \& {Heger}}{2005}]{Fryer2005}
{Fryer} C.~L.,  {Heger} A.,  2005, \mn@doi [\apj] {10.1086/428379}, \href {https://ui.adsabs.harvard.edu/abs/2005ApJ...623..302F} {623, 302}

\bibitem[\protect\citeauthoryear{{Fryer}, {Belczynski}, {Wiktorowicz}, {Dominik}, {Kalogera}  \& {Holz}}{{Fryer} et~al.}{2012}]{Fryer2012}
{Fryer} C.~L.,  {Belczynski} K.,  {Wiktorowicz} G.,  {Dominik} M.,  {Kalogera} V.,   {Holz} D.~E.,  2012, \mn@doi [\apj] {10.1088/0004-637X/749/1/91}, \href {https://ui.adsabs.harvard.edu/abs/2012ApJ...749...91F} {749, 91}

\bibitem[\protect\citeauthoryear{{Fujibayashi}, {Sekiguchi}, {Shibata}  \& {Wanajo}}{{Fujibayashi} et~al.}{2023}]{Fujibayashi2023}
{Fujibayashi} S.,  {Sekiguchi} Y.,  {Shibata} M.,   {Wanajo} S.,  2023, \mn@doi [\apj] {10.3847/1538-4357/acf5e5}, \href {https://ui.adsabs.harvard.edu/abs/2023ApJ...956..100F} {956, 100}

\bibitem[\protect\citeauthoryear{{Fuller}, {Cantiello}, {Lecoanet}  \& {Quataert}}{{Fuller} et~al.}{2015}]{Fuller2015}
{Fuller} J.,  {Cantiello} M.,  {Lecoanet} D.,   {Quataert} E.,  2015, \mn@doi [\apj] {10.1088/0004-637X/810/2/101}, \href {https://ui.adsabs.harvard.edu/abs/2015ApJ...810..101F} {810, 101}

\bibitem[\protect\citeauthoryear{{Gehrels}, {Ramirez-Ruiz}  \& {Fox}}{{Gehrels} et~al.}{2009}]{Gehrels2009}
{Gehrels} N.,  {Ramirez-Ruiz} E.,   {Fox} D.~B.,  2009, \mn@doi [\araa] {10.1146/annurev.astro.46.060407.145147}, \href {https://ui.adsabs.harvard.edu/abs/2009ARA&A..47..567G} {47, 567}

\bibitem[\protect\citeauthoryear{{Georgy}, {Ekstr{\"o}m}, {Meynet}, {Massey}, {Levesque}, {Hirschi}, {Eggenberger}  \& {Maeder}}{{Georgy} et~al.}{2012}]{Georgy2012}
{Georgy} C.,  {Ekstr{\"o}m} S.,  {Meynet} G.,  {Massey} P.,  {Levesque} E.~M.,  {Hirschi} R.,  {Eggenberger} P.,   {Maeder} A.,  2012, \mn@doi [\aap] {10.1051/0004-6361/201118340}, \href {https://ui.adsabs.harvard.edu/abs/2012A&A...542A..29G} {542, A29}

\bibitem[\protect\citeauthoryear{Gerke, Kochanek  \& Stanek}{Gerke et~al.}{2015}]{Gerke:2014ooa}
Gerke J.~R.,  Kochanek C.~S.,   Stanek K.~Z.,  2015, \mn@doi [Mon. Not. Roy. Astron. Soc.] {10.1093/mnras/stv776}, 450, 3289

\bibitem[\protect\citeauthoryear{Giacobbo, Mapelli  \& Spera}{Giacobbo et~al.}{2017}]{Nicola2018}
Giacobbo N.,  Mapelli M.,   Spera M.,  2017, \mn@doi [Monthly Notices of the Royal Astronomical Society] {10.1093/mnras/stx2933}, 474, 2959

\bibitem[\protect\citeauthoryear{{Harada} et~al.,}{{Harada} et~al.}{2023}]{Harada2023}
{Harada} M.,  et~al., 2023, \mn@doi [\apjl] {10.3847/2041-8213/acdc9e}, \href {https://ui.adsabs.harvard.edu/abs/2023ApJ...951L..27H} {951, L27}

\bibitem[\protect\citeauthoryear{{Heger}, {Langer}  \& {Woosley}}{{Heger} et~al.}{2000}]{Heger2000}
{Heger} A.,  {Langer} N.,   {Woosley} S.~E.,  2000, \mn@doi [\apj] {10.1086/308158}, \href {https://ui.adsabs.harvard.edu/abs/2000ApJ...528..368H} {528, 368}

\bibitem[\protect\citeauthoryear{{Heger}, {Fryer}, {Woosley}, {Langer}  \& {Hartmann}}{{Heger} et~al.}{2003}]{Heger2003}
{Heger} A.,  {Fryer} C.~L.,  {Woosley} S.~E.,  {Langer} N.,   {Hartmann} D.~H.,  2003, \mn@doi [\apj] {10.1086/375341}, \href {https://ui.adsabs.harvard.edu/abs/2003ApJ...591..288H} {591, 288}

\bibitem[\protect\citeauthoryear{{Heger}, {Woosley}  \& {Spruit}}{{Heger} et~al.}{2005}]{Heger2005}
{Heger} A.,  {Woosley} S.~E.,   {Spruit} H.~C.,  2005, \mn@doi [\apj] {10.1086/429868}, \href {https://ui.adsabs.harvard.edu/abs/2005ApJ...626..350H} {626, 350}

\bibitem[\protect\citeauthoryear{{Heggie}}{{Heggie}}{1975}]{Heggie_1975}
{Heggie} D.~C.,  1975, \mn@doi [\mnras] {10.1093/mnras/173.3.729}, \href {http://adsabs.harvard.edu/abs/1975MNRAS.173..729H} {173, 729}

\bibitem[\protect\citeauthoryear{{Hijikawa}, {Kinugawa}, {Yoshida}  \& {Umeda}}{{Hijikawa} et~al.}{2019}]{Hijikawa2019}
{Hijikawa} K.,  {Kinugawa} T.,  {Yoshida} T.,   {Umeda} H.,  2019, \mn@doi [\apj] {10.3847/1538-4357/ab321c}, \href {https://ui.adsabs.harvard.edu/abs/2019ApJ...882...93H} {882, 93}

\bibitem[\protect\citeauthoryear{{Hirai}}{{Hirai}}{2023}]{Hirai2023}
{Hirai} R.,  2023, \mn@doi [\mnras] {10.1093/mnras/stad1856}, \href {https://ui.adsabs.harvard.edu/abs/2023MNRAS.523.6011H} {523, 6011}

\bibitem[\protect\citeauthoryear{{Hirai}, {Sato}, {Podsiadlowski}, {Vigna-G{\'o}mez}  \& {Mandel}}{{Hirai} et~al.}{2020}]{Hirai2020}
{Hirai} R.,  {Sato} T.,  {Podsiadlowski} P.,  {Vigna-G{\'o}mez} A.,   {Mandel} I.,  2020, \mn@doi [\mnras] {10.1093/mnras/staa2898}, \href {https://ui.adsabs.harvard.edu/abs/2020MNRAS.499.1154H} {499, 1154}

\bibitem[\protect\citeauthoryear{{Hjellming}}{{Hjellming}}{1989}]{Hjellming_1989}
{Hjellming} M.~S.,  1989, PhD thesis, Illinois Univ.~at Urbana-Champaign, Savoy.

\bibitem[\protect\citeauthoryear{Horiuchi, Nakamura, Takiwaki, Kotake  \& Tanaka}{Horiuchi et~al.}{2014}]{Horiuchi:2014ska}
Horiuchi S.,  Nakamura K.,  Takiwaki T.,  Kotake K.,   Tanaka M.,  2014, \mn@doi [Mon. Not. Roy. Astron. Soc.] {10.1093/mnrasl/slu146}, 445, L99

\bibitem[\protect\citeauthoryear{{Horiuchi}, {Kinugawa}, {Takiwaki}, {Takahashi}  \& {Kotake}}{{Horiuchi} et~al.}{2021}]{Horiuchi2021}
{Horiuchi} S.,  {Kinugawa} T.,  {Takiwaki} T.,  {Takahashi} K.,   {Kotake} K.,  2021, \mn@doi [\prd] {10.1103/PhysRevD.103.043003}, \href {https://ui.adsabs.harvard.edu/abs/2021PhRvD.103d3003H} {103, 043003}

\bibitem[\protect\citeauthoryear{{Hsieh}, {Cabez{\'o}n}, {Ma}  \& {Pan}}{{Hsieh} et~al.}{2023}]{Hsieh2023}
{Hsieh} H.-F.,  {Cabez{\'o}n} R.,  {Ma} L.-T.,   {Pan} K.-C.,  2023, \mn@doi [arXiv e-prints] {10.48550/arXiv.2310.20411}, \href {https://ui.adsabs.harvard.edu/abs/2023arXiv231020411H} {p. arXiv:2310.20411}

\bibitem[\protect\citeauthoryear{{Huang}, {Gies}  \& {McSwain}}{{Huang} et~al.}{2010}]{Huang2010}
{Huang} W.,  {Gies} D.~R.,   {McSwain} M.~V.,  2010, \mn@doi [\apj] {10.1088/0004-637X/722/1/605}, \href {https://ui.adsabs.harvard.edu/abs/2010ApJ...722..605H} {722, 605}

\bibitem[\protect\citeauthoryear{{Hurley}, {Pols}  \& {Tout}}{{Hurley} et~al.}{2000}]{Hurley2000}
{Hurley} J.~R.,  {Pols} O.~R.,   {Tout} C.~A.,  2000, \mn@doi [\mnras] {10.1046/j.1365-8711.2000.03426.x}, \href {https://ui.adsabs.harvard.edu/abs/2000MNRAS.315..543H} {315, 543}

\bibitem[\protect\citeauthoryear{{Hurley}, {Tout}  \& {Pols}}{{Hurley} et~al.}{2002}]{Hurley_2002}
{Hurley} J.~R.,  {Tout} C.~A.,   {Pols} O.~R.,  2002, \mn@doi [\mnras] {10.1046/j.1365-8711.2002.05038.x}, \href {http://adsabs.harvard.edu/abs/2002MNRAS.329..897H} {329, 897}

\bibitem[\protect\citeauthoryear{{Hut}}{{Hut}}{1981}]{Hut1981}
{Hut} P.,  1981, \aap, \href {https://ui.adsabs.harvard.edu/abs/1981A&A....99..126H} {99, 126}

\bibitem[\protect\citeauthoryear{{Igoshev}, {Frantsuzova}, {Gourgouliatos}, {Tsichli}, {Konstantinou}  \& {Popov}}{{Igoshev} et~al.}{2022}]{Igoshev2022}
{Igoshev} A.~P.,  {Frantsuzova} A.,  {Gourgouliatos} K.~N.,  {Tsichli} S.,  {Konstantinou} L.,   {Popov} S.~B.,  2022, \mn@doi [\mnras] {10.1093/mnras/stac1648}, \href {https://ui.adsabs.harvard.edu/abs/2022MNRAS.514.4606I} {514, 4606}

\bibitem[\protect\citeauthoryear{{Ivanova}, {Podsiadlowski}  \& {Spruit}}{{Ivanova} et~al.}{2002}]{Ivanova_2002}
{Ivanova} N.,  {Podsiadlowski} P.,   {Spruit} H.,  2002, \mn@doi [\mnras] {10.1046/j.1365-8711.2002.05543.x}, \href {https://ui.adsabs.harvard.edu/abs/2002MNRAS.334..819I} {334, 819}

\bibitem[\protect\citeauthoryear{{Iwakami}, {Nagakura}  \& {Yamada}}{{Iwakami} et~al.}{2014}]{Iwakami2014}
{Iwakami} W.,  {Nagakura} H.,   {Yamada} S.,  2014, \mn@doi [\apj] {10.1088/0004-637X/793/1/5}, \href {https://ui.adsabs.harvard.edu/abs/2014ApJ...793....5I} {793, 5}

\bibitem[\protect\citeauthoryear{{Janka}, {Wongwathanarat}  \& {Kramer}}{{Janka} et~al.}{2022}]{Janka2022}
{Janka} H.-T.,  {Wongwathanarat} A.,   {Kramer} M.,  2022, \mn@doi [\apj] {10.3847/1538-4357/ac403c}, \href {https://ui.adsabs.harvard.edu/abs/2022ApJ...926....9J} {926, 9}

\bibitem[\protect\citeauthoryear{{Kato}, {Hirai}  \& {Nagakura}}{{Kato} et~al.}{2020}]{Kato2020}
{Kato} C.,  {Hirai} R.,   {Nagakura} H.,  2020, \mn@doi [\mnras] {10.1093/mnras/staa1738}, \href {https://ui.adsabs.harvard.edu/abs/2020MNRAS.496.3961K} {496, 3961}

\bibitem[\protect\citeauthoryear{{Keszthelyi}}{{Keszthelyi}}{2023}]{Keszthelyi2023}
{Keszthelyi} Z.,  2023, \mn@doi [Galaxies] {10.3390/galaxies11020040}, \href {https://ui.adsabs.harvard.edu/abs/2023Galax..11...40K} {11, 40}

\bibitem[\protect\citeauthoryear{{Kinugawa} \& {Asano}}{{Kinugawa} \& {Asano}}{2017}]{KinugawaAsano2017}
{Kinugawa} T.,  {Asano} K.,  2017, \mn@doi [\apjl] {10.3847/2041-8213/aa95bb}, \href {https://ui.adsabs.harvard.edu/abs/2017ApJ...849L..29K} {849, L29}

\bibitem[\protect\citeauthoryear{{Kinugawa}, {Inayoshi}, {Hotokezaka}, {Nakauchi}  \& {Nakamura}}{{Kinugawa} et~al.}{2014}]{Kinugawa2014}
{Kinugawa} T.,  {Inayoshi} K.,  {Hotokezaka} K.,  {Nakauchi} D.,   {Nakamura} T.,  2014, \mn@doi [\mnras] {10.1093/mnras/stu1022}, \href {http://adsabs.harvard.edu/abs/2014MNRAS.442.2963K} {442, 2963}

\bibitem[\protect\citeauthoryear{{Kinugawa}, {Miyamoto}, {Kanda}  \& {Nakamura}}{{Kinugawa} et~al.}{2016}]{Kinugawa2016}
{Kinugawa} T.,  {Miyamoto} A.,  {Kanda} N.,   {Nakamura} T.,  2016, \mn@doi [\mnras] {10.1093/mnras/stv2624}, \href {http://adsabs.harvard.edu/abs/2016MNRAS.456.1093K} {456, 1093}

\bibitem[\protect\citeauthoryear{{Kinugawa}, {Harikane}  \& {Asano}}{{Kinugawa} et~al.}{2019}]{KinugawaHA2019}
{Kinugawa} T.,  {Harikane} Y.,   {Asano} K.,  2019, \mn@doi [\apj] {10.3847/1538-4357/ab2188}, \href {https://ui.adsabs.harvard.edu/abs/2019ApJ...878..128K} {878, 128}

\bibitem[\protect\citeauthoryear{{Kinugawa}, {Nakamura}  \& {Nakano}}{{Kinugawa} et~al.}{2020}]{Kinugawa2020}
{Kinugawa} T.,  {Nakamura} T.,   {Nakano} H.,  2020, \mn@doi [\mnras] {10.1093/mnras/staa2511}, \href {https://ui.adsabs.harvard.edu/abs/2020MNRAS.498.3946K} {498, 3946}

\bibitem[\protect\citeauthoryear{{Kobulnicky} \& {Fryer}}{{Kobulnicky} \& {Fryer}}{2007}]{Kobulnicky2007}
{Kobulnicky} H.~A.,  {Fryer} C.~L.,  2007, \mn@doi [\apj] {10.1086/522073}, \href {https://ui.adsabs.harvard.edu/abs/2007ApJ...670..747K} {670, 747}

\bibitem[\protect\citeauthoryear{{Kobulnicky} et~al.,}{{Kobulnicky} et~al.}{2014}]{Kobulnicky2014}
{Kobulnicky} H.~A.,  et~al., 2014, \mn@doi [\apjs] {10.1088/0067-0049/213/2/34}, \href {https://ui.adsabs.harvard.edu/abs/2014ApJS..213...34K} {213, 34}

\bibitem[\protect\citeauthoryear{Kochanek, Beacom, Kistler, Prieto, Stanek, Thompson  \& Yuksel}{Kochanek et~al.}{2008}]{Kochanek:2008mp}
Kochanek C.~S.,  Beacom J.~F.,  Kistler M.~D.,  Prieto J.~L.,  Stanek K.~Z.,  Thompson T.~A.,   Yuksel H.,  2008, \mn@doi [Astrophys. J.] {10.1086/590053}, 684, 1336

\bibitem[\protect\citeauthoryear{{Kotake}, {Takiwaki}, {Fischer}, {Nakamura}  \& {Mart{\'\i}nez-Pinedo}}{{Kotake} et~al.}{2018}]{Kotake18}
{Kotake} K.,  {Takiwaki} T.,  {Fischer} T.,  {Nakamura} K.,   {Mart{\'\i}nez-Pinedo} G.,  2018, \mn@doi [\apj] {10.3847/1538-4357/aaa716}, \href {https://ui.adsabs.harvard.edu/abs/2018ApJ...853..170K} {853, 170}

\bibitem[\protect\citeauthoryear{{Kresse}, {Ertl}  \& {Janka}}{{Kresse} et~al.}{2021}]{Kresse2021}
{Kresse} D.,  {Ertl} T.,   {Janka} H.-T.,  2021, \mn@doi [\apj] {10.3847/1538-4357/abd54e}, \href {https://ui.adsabs.harvard.edu/abs/2021ApJ...909..169K} {909, 169}

\bibitem[\protect\citeauthoryear{{Kuroda}, {Arcones}, {Takiwaki}  \& {Kotake}}{{Kuroda} et~al.}{2020}]{Kuroda2020}
{Kuroda} T.,  {Arcones} A.,  {Takiwaki} T.,   {Kotake} K.,  2020, \mn@doi [\apj] {10.3847/1538-4357/ab9308}, \href {https://ui.adsabs.harvard.edu/abs/2020ApJ...896..102K} {896, 102}

\bibitem[\protect\citeauthoryear{{Langer}}{{Langer}}{2012}]{Langer2012}
{Langer} N.,  2012, \mn@doi [\araa] {10.1146/annurev-astro-081811-125534}, \href {https://ui.adsabs.harvard.edu/abs/2012ARA&A..50..107L} {50, 107}

\bibitem[\protect\citeauthoryear{{Laplace}, {Justham}, {Renzo}, {G{\"o}tberg}, {Farmer}, {Vartanyan}  \& {de Mink}}{{Laplace} et~al.}{2021}]{Laplace2021}
{Laplace} E.,  {Justham} S.,  {Renzo} M.,  {G{\"o}tberg} Y.,  {Farmer} R.,  {Vartanyan} D.,   {de Mink} S.~E.,  2021, \mn@doi [\aap] {10.1051/0004-6361/202140506}, \href {https://ui.adsabs.harvard.edu/abs/2021A&A...656A..58L} {656, A58}

\bibitem[\protect\citeauthoryear{{Larsson} et~al.,}{{Larsson} et~al.}{2023}]{Lasson2023}
{Larsson} J.,  et~al., 2023, \mn@doi [\apjl] {10.3847/2041-8213/acd555}, \href {https://ui.adsabs.harvard.edu/abs/2023ApJ...949L..27L} {949, L27}

\bibitem[\protect\citeauthoryear{{Li} et~al.,}{{Li} et~al.}{2011}]{Li2011}
{Li} W.,  et~al., 2011, \mn@doi [Monthly Notices of the Royal Astronomical Society] {10.1111/j.1365-2966.2011.18160.x}, \href {https://ui.adsabs.harvard.edu/abs/2011MNRAS.412.1441L} {412, 1441}

\bibitem[\protect\citeauthoryear{{Liang}, {Zhang}, {Virgili}  \& {Dai}}{{Liang} et~al.}{2007}]{Liang2007}
{Liang} E.,  {Zhang} B.,  {Virgili} F.,   {Dai} Z.~G.,  2007, \mn@doi [\apj] {10.1086/517959}, \href {https://ui.adsabs.harvard.edu/abs/2007ApJ...662.1111L} {662, 1111}

\bibitem[\protect\citeauthoryear{{Liang}, {Racusin}, {Zhang}, {Zhang}  \& {Burrows}}{{Liang} et~al.}{2008}]{Liang2008}
{Liang} E.-W.,  {Racusin} J.~L.,  {Zhang} B.,  {Zhang} B.-B.,   {Burrows} D.~N.,  2008, \mn@doi [\apj] {10.1086/524701}, \href {https://ui.adsabs.harvard.edu/abs/2008ApJ...675..528L} {675, 528}

\bibitem[\protect\citeauthoryear{{Limongi}}{{Limongi}}{2017}]{Limongi2017}
{Limongi} M.,  2017, {Supernovae from Massive Stars}.
pringer International Publishing, p.~513, \mn@doi{10.1007/978-3-319-21846-5_119}

\bibitem[\protect\citeauthoryear{{Longo Micchi}, {Radice}  \& {Chirenti}}{{Longo Micchi} et~al.}{2023}]{Longo2023}
{Longo Micchi} L.~F.,  {Radice} D.,   {Chirenti} C.,  2023, \mn@doi [\mnras] {10.1093/mnras/stad2420}, \href {https://ui.adsabs.harvard.edu/abs/2023MNRAS.525.6359L} {525, 6359}

\bibitem[\protect\citeauthoryear{{Lubow} \& {Shu}}{{Lubow} \& {Shu}}{1975}]{LubowShu1974}
{Lubow} S.~H.,  {Shu} F.~H.,  1975, \mn@doi [\apj] {10.1086/153614}, \href {https://ui.adsabs.harvard.edu/abs/1975ApJ...198..383L} {198, 383}

\bibitem[\protect\citeauthoryear{{Lunardini}}{{Lunardini}}{2016}]{Lunardini2016}
{Lunardini} C.,  2016, \mn@doi [Astroparticle Physics] {10.1016/j.astropartphys.2016.02.005}, \href {https://ui.adsabs.harvard.edu/abs/2016APh....79...49L} {79, 49}

\bibitem[\protect\citeauthoryear{{Madau} \& {Dickinson}}{{Madau} \& {Dickinson}}{2014}]{Madau2014}
{Madau} P.,  {Dickinson} M.,  2014, \mn@doi [\araa] {10.1146/annurev-astro-081811-125615}, \href {https://ui.adsabs.harvard.edu/abs/2014ARA&A..52..415M} {52, 415}

\bibitem[\protect\citeauthoryear{{Maeder}}{{Maeder}}{2009}]{Maeder2009}
{Maeder} A.,  2009, {Physics, Formation and Evolution of Rotating Stars}.
Springer, \mn@doi{10.1007/978-3-540-76949-1}

\bibitem[\protect\citeauthoryear{{Mason}, {Hartkopf}, {Gies}, {Henry}  \& {Helsel}}{{Mason} et~al.}{2009}]{Mason2009}
{Mason} B.~D.,  {Hartkopf} W.~I.,  {Gies} D.~R.,  {Henry} T.~J.,   {Helsel} J.~W.,  2009, \mn@doi [\aj] {10.1088/0004-6256/137/2/3358}, \href {https://ui.adsabs.harvard.edu/abs/2009AJ....137.3358M} {137, 3358}

\bibitem[\protect\citeauthoryear{{Matsumoto}, {Asahina}, {Takiwaki}, {Kotake}  \& {Takahashi}}{{Matsumoto} et~al.}{2022}]{Matsumoto2022}
{Matsumoto} J.,  {Asahina} Y.,  {Takiwaki} T.,  {Kotake} K.,   {Takahashi} H.~R.,  2022, \mn@doi [\mnras] {10.1093/mnras/stac2335}, \href {https://ui.adsabs.harvard.edu/abs/2022MNRAS.516.1752M} {516, 1752}

\bibitem[\protect\citeauthoryear{{McNeill} \& {M{\"u}ller}}{{McNeill} \& {M{\"u}ller}}{2022}]{Mcneill2022}
{McNeill} L.~O.,  {M{\"u}ller} B.,  2022, \mn@doi [\mnras] {10.1093/mnras/stab3076}, \href {https://ui.adsabs.harvard.edu/abs/2022MNRAS.509..818M} {509, 818}

\bibitem[\protect\citeauthoryear{{Menon} \& {Heger}}{{Menon} \& {Heger}}{2017}]{Menon2017}
{Menon} A.,  {Heger} A.,  2017, \mn@doi [\mnras] {10.1093/mnras/stx818}, \href {https://ui.adsabs.harvard.edu/abs/2017MNRAS.469.4649M} {469, 4649}

\bibitem[\protect\citeauthoryear{{Moc{\'a}k}, {Meakin}, {Campbell}  \& {Arnett}}{{Moc{\'a}k} et~al.}{2018}]{MOcak2018}
{Moc{\'a}k} M.,  {Meakin} C.,  {Campbell} S.~W.,   {Arnett} W.~D.,  2018, \mn@doi [\mnras] {10.1093/mnras/sty2392}, \href {https://ui.adsabs.harvard.edu/abs/2018MNRAS.481.2918M} {481, 2918}

\bibitem[\protect\citeauthoryear{{Moe} \& {Di Stefano}}{{Moe} \& {Di Stefano}}{2017}]{Moe2017}
{Moe} M.,  {Di Stefano} R.,  2017, \mn@doi [\apjs] {10.3847/1538-4365/aa6fb6}, \href {https://ui.adsabs.harvard.edu/abs/2017ApJS..230...15M} {230, 15}

\bibitem[\protect\citeauthoryear{{Mori}, {Sawada}, {Suwa}, {Tanikawa}, {Kashiyama}  \& {Murase}}{{Mori} et~al.}{2023}]{MoriM2023}
{Mori} M.,  {Sawada} R.,  {Suwa} Y.,  {Tanikawa} A.,  {Kashiyama} K.,   {Murase} K.,  2023, \mn@doi [arXiv e-prints] {10.48550/arXiv.2306.17381}, \href {https://ui.adsabs.harvard.edu/abs/2023arXiv230617381M} {p. arXiv:2306.17381}

\bibitem[\protect\citeauthoryear{{M{\"u}ller}, {Heger}, {Liptai}  \& {Cameron}}{{M{\"u}ller} et~al.}{2016a}]{Muller2016a}
{M{\"u}ller} B.,  {Heger} A.,  {Liptai} D.,   {Cameron} J.~B.,  2016a, \mn@doi [\mnras] {10.1093/mnras/stw1083}, \href {https://ui.adsabs.harvard.edu/abs/2016MNRAS.460..742M} {460, 742}

\bibitem[\protect\citeauthoryear{{M{\"u}ller}, {Viallet}, {Heger}  \& {Janka}}{{M{\"u}ller} et~al.}{2016b}]{Muller2016b}
{M{\"u}ller} B.,  {Viallet} M.,  {Heger} A.,   {Janka} H.-T.,  2016b, \mn@doi [\apj] {10.3847/1538-4357/833/1/124}, \href {https://ui.adsabs.harvard.edu/abs/2016ApJ...833..124M} {833, 124}

\bibitem[\protect\citeauthoryear{{M{\"u}ller}, {Gay}, {Heger}, {Tauris}  \& {Sim}}{{M{\"u}ller} et~al.}{2018}]{Muller2018}
{M{\"u}ller} B.,  {Gay} D.~W.,  {Heger} A.,  {Tauris} T.~M.,   {Sim} S.~A.,  2018, \mn@doi [\mnras] {10.1093/mnras/sty1683}, \href {https://ui.adsabs.harvard.edu/abs/2018MNRAS.479.3675M} {479, 3675}

\bibitem[\protect\citeauthoryear{{Nagakura}}{{Nagakura}}{2023}]{Nagakura23}
{Nagakura} H.,  2023, \mn@doi [\prl] {10.1103/PhysRevLett.130.211401}, \href {https://ui.adsabs.harvard.edu/abs/2023PhRvL.130u1401N} {130, 211401}

\bibitem[\protect\citeauthoryear{{Nakamura}, {Takiwaki}, {Kuroda}  \& {Kotake}}{{Nakamura} et~al.}{2015}]{NakamuraKo2015}
{Nakamura} K.,  {Takiwaki} T.,  {Kuroda} T.,   {Kotake} K.,  2015, \mn@doi [\pasj] {10.1093/pasj/psv073}, \href {https://ui.adsabs.harvard.edu/abs/2015PASJ...67..107N} {67, 107}

\bibitem[\protect\citeauthoryear{{Nakamura}, {Takiwaki}  \& {Kotake}}{{Nakamura} et~al.}{2022}]{Nakamura2022}
{Nakamura} K.,  {Takiwaki} T.,   {Kotake} K.,  2022, \mn@doi [\mnras] {10.1093/mnras/stac1586}, \href {https://ui.adsabs.harvard.edu/abs/2022MNRAS.514.3941N} {514, 3941}

\bibitem[\protect\citeauthoryear{Neustadt, Kochanek, Stanek, Basinger, Jayasinghe, Garling, Adams  \& Gerke}{Neustadt et~al.}{2021}]{Neustadt:2021jjt}
Neustadt J. M.~M.,  Kochanek C.~S.,  Stanek K.~Z.,  Basinger C.~M.,  Jayasinghe T.,  Garling C.~T.,  Adams S.~M.,   Gerke J.,  2021, \mn@doi [Mon. Not. Roy. Astron. Soc.] {10.1093/mnras/stab2605}, 508, 516

\bibitem[\protect\citeauthoryear{{Nomoto}, {Kobayashi}  \& {Tominaga}}{{Nomoto} et~al.}{2013}]{Nomoto2013}
{Nomoto} K.,  {Kobayashi} C.,   {Tominaga} N.,  2013, \mn@doi [\araa] {10.1146/annurev-astro-082812-140956}, \href {https://ui.adsabs.harvard.edu/abs/2013ARA&A..51..457N} {51, 457}

\bibitem[\protect\citeauthoryear{{Noutsos}, {Schnitzeler}, {Keane}, {Kramer}  \& {Johnston}}{{Noutsos} et~al.}{2013}]{Noutsos2013}
{Noutsos} A.,  {Schnitzeler} D.~H.~F.~M.,  {Keane} E.~F.,  {Kramer} M.,   {Johnston} S.,  2013, \mn@doi [\mnras] {10.1093/mnras/stt047}, \href {https://ui.adsabs.harvard.edu/abs/2013MNRAS.430.2281N} {430, 2281}

\bibitem[\protect\citeauthoryear{{Nugis} \& {Lamers}}{{Nugis} \& {Lamers}}{2000}]{Nugis2000}
{Nugis} T.,  {Lamers} H.~J.~G.~L.~M.,  2000, \aap, \href {https://ui.adsabs.harvard.edu/abs/2000A&A...360..227N} {360, 227}

\bibitem[\protect\citeauthoryear{{O'Connor} \& {Ott}}{{O'Connor} \& {Ott}}{2011}]{O'connor2011}
{O'Connor} E.,  {Ott} C.~D.,  2011, \mn@doi [\apj] {10.1088/0004-637X/730/2/70}, \href {https://ui.adsabs.harvard.edu/abs/2011ApJ...730...70O} {730, 70}

\bibitem[\protect\citeauthoryear{{Obergaulinger} \& {Aloy}}{{Obergaulinger} \& {Aloy}}{2020}]{Obergaulinger2020}
{Obergaulinger} M.,  {Aloy} M.~{\'A}.,  2020, \mn@doi [\mnras] {10.1093/mnras/staa096}, \href {https://ui.adsabs.harvard.edu/abs/2020MNRAS.492.4613O} {492, 4613}

\bibitem[\protect\citeauthoryear{{Obergaulinger} \& {Aloy}}{{Obergaulinger} \& {Aloy}}{2022}]{Obergaulinger2022}
{Obergaulinger} M.,  {Aloy} M.~{\'A}.,  2022, \mn@doi [\mnras] {10.1093/mnras/stac613}, \href {https://ui.adsabs.harvard.edu/abs/2022MNRAS.512.2489O} {512, 2489}

\bibitem[\protect\citeauthoryear{{Obergaulinger}, {Janka}  \& {Aloy}}{{Obergaulinger} et~al.}{2014}]{Obergaulinger2014}
{Obergaulinger} M.,  {Janka} H.~T.,   {Aloy} M.~A.,  2014, \mn@doi [\mnras] {10.1093/mnras/stu1969}, \href {https://ui.adsabs.harvard.edu/abs/2014MNRAS.445.3169O} {445, 3169}

\bibitem[\protect\citeauthoryear{{Ogata}, {Okawa}, {Fujisawa}, {Yasutake}, {Yamamoto}  \& {Yamada}}{{Ogata} et~al.}{2023}]{Ogata2023}
{Ogata} M.,  {Okawa} H.,  {Fujisawa} K.,  {Yasutake} N.,  {Yamamoto} Y.,   {Yamada} S.,  2023, \mn@doi [\mnras] {10.1093/mnras/stad647}, \href {https://ui.adsabs.harvard.edu/abs/2023MNRAS.521.2561O} {521, 2561}

\bibitem[\protect\citeauthoryear{{Ono}, {Nagataki}, {Ferrand}, {Takahashi}, {Umeda}, {Yoshida}, {Orlando}  \& {Miceli}}{{Ono} et~al.}{2020}]{Ono2020}
{Ono} M.,  {Nagataki} S.,  {Ferrand} G.,  {Takahashi} K.,  {Umeda} H.,  {Yoshida} T.,  {Orlando} S.,   {Miceli} M.,  2020, \mn@doi [\apj] {10.3847/1538-4357/ab5dba}, \href {https://ui.adsabs.harvard.edu/abs/2020ApJ...888..111O} {888, 111}

\bibitem[\protect\citeauthoryear{{Patton} \& {Sukhbold}}{{Patton} \& {Sukhbold}}{2020}]{Patton2020}
{Patton} R.~A.,  {Sukhbold} T.,  2020, \mn@doi [\mnras] {10.1093/mnras/staa3029}, \href {https://ui.adsabs.harvard.edu/abs/2020MNRAS.499.2803P} {499, 2803}

\bibitem[\protect\citeauthoryear{{Patton}, {Sukhbold}  \& {Eldridge}}{{Patton} et~al.}{2022}]{Patton2022}
{Patton} R.~A.,  {Sukhbold} T.,   {Eldridge} J.~J.,  2022, \mn@doi [\mnras] {10.1093/mnras/stab3797}, \href {https://ui.adsabs.harvard.edu/abs/2022MNRAS.511..903P} {511, 903}

\bibitem[\protect\citeauthoryear{{Pejcha} \& {Thompson}}{{Pejcha} \& {Thompson}}{2015}]{Pejcha2015}
{Pejcha} O.,  {Thompson} T.~A.,  2015, \mn@doi [\apj] {10.1088/0004-637X/801/2/90}, \href {https://ui.adsabs.harvard.edu/abs/2015ApJ...801...90P} {801, 90}

\bibitem[\protect\citeauthoryear{{Peters}}{{Peters}}{1964}]{Peters1964}
{Peters} P.~C.,  1964, \mn@doi [Physical Review] {10.1103/PhysRev.136.B1224}, \href {https://ui.adsabs.harvard.edu/abs/1964PhRv..136.1224P} {136, 1224}

\bibitem[\protect\citeauthoryear{{Peters} \& {Mathews}}{{Peters} \& {Mathews}}{1963}]{Peter_Mathews_1963}
{Peters} P.~C.,  {Mathews} J.,  1963, \mn@doi [Physical Review] {10.1103/PhysRev.131.435}, \href {http://adsabs.harvard.edu/abs/1963PhRv..131..435P} {131, 435}

\bibitem[\protect\citeauthoryear{{Pian} et~al.,}{{Pian} et~al.}{2006}]{Pian2006}
{Pian} E.,  et~al., 2006, \mn@doi [\nat] {10.1038/nature05082}, \href {https://ui.adsabs.harvard.edu/abs/2006Natur.442.1011P} {442, 1011}

\bibitem[\protect\citeauthoryear{{Rasio}, {Tout}, {Lubow}  \& {Livio}}{{Rasio} et~al.}{1996}]{Rasio1996}
{Rasio} F.~A.,  {Tout} C.~A.,  {Lubow} S.~H.,   {Livio} M.,  1996, \mn@doi [\apj] {10.1086/177941}, \href {https://ui.adsabs.harvard.edu/abs/1996ApJ...470.1187R} {470, 1187}

\bibitem[\protect\citeauthoryear{{Riley} et~al.,}{{Riley} et~al.}{2022}]{Riley2022}
{Riley} J.,  et~al., 2022, \mn@doi [\apjs] {10.3847/1538-4365/ac416c}, \href {https://ui.adsabs.harvard.edu/abs/2022ApJS..258...34R} {258, 34}

\bibitem[\protect\citeauthoryear{{Salpeter}}{{Salpeter}}{1955}]{Salpeter1955}
{Salpeter} E.~E.,  1955, \mn@doi [\apj] {10.1086/145971}, \href {https://ui.adsabs.harvard.edu/abs/1955ApJ...121..161S} {121, 161}

\bibitem[\protect\citeauthoryear{{Sana} et~al.,}{{Sana} et~al.}{2012}]{Sana2012}
{Sana} H.,  et~al., 2012, \mn@doi [Science] {10.1126/science.1223344}, \href {https://ui.adsabs.harvard.edu/abs/2012Sci...337..444S} {337, 444}

\bibitem[\protect\citeauthoryear{{Sana} et~al.,}{{Sana} et~al.}{2013}]{Sana2013}
{Sana} H.,  et~al., 2013, \mn@doi [\aap] {10.1051/0004-6361/201219621}, \href {https://ui.adsabs.harvard.edu/abs/2013A&A...550A.107S} {550, A107}

\bibitem[\protect\citeauthoryear{{Schneider}, {Podsiadlowski}  \& {M{\"u}ller}}{{Schneider} et~al.}{2021}]{Schneider2021}
{Schneider} F.~R.~N.,  {Podsiadlowski} P.,   {M{\"u}ller} B.,  2021, \mn@doi [\aap] {10.1051/0004-6361/202039219}, \href {https://ui.adsabs.harvard.edu/abs/2021A&A...645A...5S} {645, A5}

\bibitem[\protect\citeauthoryear{{Sekiguchi} \& {Shibata}}{{Sekiguchi} \& {Shibata}}{2011}]{Sekiguchi2011}
{Sekiguchi} Y.,  {Shibata} M.,  2011, \mn@doi [\apj] {10.1088/0004-637X/737/1/6}, \href {https://ui.adsabs.harvard.edu/abs/2011ApJ...737....6S} {737, 6}

\bibitem[\protect\citeauthoryear{{Shibagaki}, {Kuroda}, {Kotake}  \& {Takiwaki}}{{Shibagaki} et~al.}{2020}]{Shibagaki2020}
{Shibagaki} S.,  {Kuroda} T.,  {Kotake} K.,   {Takiwaki} T.,  2020, \mn@doi [\mnras] {10.1093/mnrasl/slaa021}, \href {https://ui.adsabs.harvard.edu/abs/2020MNRAS.493L.138S} {493, L138}

\bibitem[\protect\citeauthoryear{{Shibagaki}, {Kuroda}, {Kotake}, {Takiwaki}  \& {Fischer}}{{Shibagaki} et~al.}{2023}]{Shibagaki2023}
{Shibagaki} S.,  {Kuroda} T.,  {Kotake} K.,  {Takiwaki} T.,   {Fischer} T.,  2023, \mn@doi [arXiv e-prints] {10.48550/arXiv.2309.05161}, \href {https://ui.adsabs.harvard.edu/abs/2023arXiv230905161S} {p. arXiv:2309.05161}

\bibitem[\protect\citeauthoryear{{Smartt}}{{Smartt}}{2009}]{Smartt2009}
{Smartt} S.~J.,  2009, \mn@doi [\araa] {10.1146/annurev-astro-082708-101737}, \href {https://ui.adsabs.harvard.edu/abs/2009ARA&A..47...63S} {47, 63}

\bibitem[\protect\citeauthoryear{{Smith}}{{Smith}}{2014}]{Smith2014}
{Smith} N.,  2014, \mn@doi [\araa] {10.1146/annurev-astro-081913-040025}, \href {https://ui.adsabs.harvard.edu/abs/2014ARA&A..52..487S} {52, 487}

\bibitem[\protect\citeauthoryear{{Spera}, {Mapelli}, {Giacobbo}, {Trani}, {Bressan}  \& {Costa}}{{Spera} et~al.}{2019}]{Spera2019}
{Spera} M.,  {Mapelli} M.,  {Giacobbo} N.,  {Trani} A.~A.,  {Bressan} A.,   {Costa} G.,  2019, \mn@doi [\mnras] {10.1093/mnras/stz359}, \href {https://ui.adsabs.harvard.edu/abs/2019MNRAS.485..889S} {485, 889}

\bibitem[\protect\citeauthoryear{{Stanway} \& {Eldridge}}{{Stanway} \& {Eldridge}}{2018}]{Stanway2018}
{Stanway} E.~R.,  {Eldridge} J.~J.,  2018, \mn@doi [\mnras] {10.1093/mnras/sty1353}, \href {https://ui.adsabs.harvard.edu/abs/2018MNRAS.479...75S} {479, 75}

\bibitem[\protect\citeauthoryear{{Sugiura}, {Furusawa}, {Sumiyoshi}  \& {Yamada}}{{Sugiura} et~al.}{2022}]{Sugiura2022}
{Sugiura} K.,  {Furusawa} S.,  {Sumiyoshi} K.,   {Yamada} S.,  2022, \mn@doi [Progress of Theoretical and Experimental Physics] {10.1093/ptep/ptac118}, \href {https://ui.adsabs.harvard.edu/abs/2022PTEP.2022k3E01S} {2022, 113E01}

\bibitem[\protect\citeauthoryear{{Sukhbold}, {Ertl}, {Woosley}, {Brown}  \& {Janka}}{{Sukhbold} et~al.}{2016}]{Sukhbold2016}
{Sukhbold} T.,  {Ertl} T.,  {Woosley} S.~E.,  {Brown} J.~M.,   {Janka} H.~T.,  2016, \mn@doi [\apj] {10.3847/0004-637X/821/1/38}, \href {https://ui.adsabs.harvard.edu/abs/2016ApJ...821...38S} {821, 38}

\bibitem[\protect\citeauthoryear{{Sukhbold}, {Woosley}  \& {Heger}}{{Sukhbold} et~al.}{2018}]{Sukhbold2018}
{Sukhbold} T.,  {Woosley} S.~E.,   {Heger} A.,  2018, \mn@doi [\apj] {10.3847/1538-4357/aac2da}, \href {https://ui.adsabs.harvard.edu/abs/2018ApJ...860...93S} {860, 93}

\bibitem[\protect\citeauthoryear{{Summa}, {Hanke}, {Janka}, {Melson}, {Marek}  \& {M{\"u}ller}}{{Summa} et~al.}{2016}]{Summa2016}
{Summa} A.,  {Hanke} F.,  {Janka} H.-T.,  {Melson} T.,  {Marek} A.,   {M{\"u}ller} B.,  2016, \mn@doi [\apj] {10.3847/0004-637X/825/1/6}, \href {https://ui.adsabs.harvard.edu/abs/2016ApJ...825....6S} {825, 6}

\bibitem[\protect\citeauthoryear{{Summa}, {Janka}, {Melson}  \& {Marek}}{{Summa} et~al.}{2018}]{Summa2018}
{Summa} A.,  {Janka} H.-T.,  {Melson} T.,   {Marek} A.,  2018, \mn@doi [\apj] {10.3847/1538-4357/aa9ce8}, \href {https://ui.adsabs.harvard.edu/abs/2018ApJ...852...28S} {852, 28}

\bibitem[\protect\citeauthoryear{{Taam} \& {Sandquist}}{{Taam} \& {Sandquist}}{2000}]{Taam_2000}
{Taam} R.~E.,  {Sandquist} E.~L.,  2000, \mn@doi [\araa] {10.1146/annurev.astro.38.1.113}, \href {http://adsabs.harvard.edu/abs/2000ARA%26A..38..113T} {38, 113}

\bibitem[\protect\citeauthoryear{{Takahashi} \& {Langer}}{{Takahashi} \& {Langer}}{2021}]{Takahashi2021}
{Takahashi} K.,  {Langer} N.,  2021, \mn@doi [\aap] {10.1051/0004-6361/202039253}, \href {https://ui.adsabs.harvard.edu/abs/2021A&A...646A..19T} {646, A19}

\bibitem[\protect\citeauthoryear{{Takahashi}, {Umeda}  \& {Yoshida}}{{Takahashi} et~al.}{2014}]{Takahashi2014}
{Takahashi} K.,  {Umeda} H.,   {Yoshida} T.,  2014, \mn@doi [\apj] {10.1088/0004-637X/794/1/40}, \href {https://ui.adsabs.harvard.edu/abs/2014ApJ...794...40T} {794, 40}

\bibitem[\protect\citeauthoryear{{Takahashi}, {Yoshida}, {Umeda}, {Sumiyoshi}  \& {Yamada}}{{Takahashi} et~al.}{2016}]{Takahashi2016}
{Takahashi} K.,  {Yoshida} T.,  {Umeda} H.,  {Sumiyoshi} K.,   {Yamada} S.,  2016, \mn@doi [\mnras] {10.1093/mnras/stv2649}, \href {https://ui.adsabs.harvard.edu/abs/2016MNRAS.456.1320T} {456, 1320}

\bibitem[\protect\citeauthoryear{{Takahashi}, {Takiwaki}  \& {Yoshida}}{{Takahashi} et~al.}{2023}]{Takahashi2023}
{Takahashi} K.,  {Takiwaki} T.,   {Yoshida} T.,  2023, \mn@doi [\apj] {10.3847/1538-4357/acb8b3}, \href {https://ui.adsabs.harvard.edu/abs/2023ApJ...945...19T} {945, 19}

\bibitem[\protect\citeauthoryear{{Takiwaki}, {Kotake}  \& {Foglizzo}}{{Takiwaki} et~al.}{2021}]{Takiwaki2021}
{Takiwaki} T.,  {Kotake} K.,   {Foglizzo} T.,  2021, \mn@doi [\mnras] {10.1093/mnras/stab2607}, \href {https://ui.adsabs.harvard.edu/abs/2021MNRAS.508..966T} {508, 966}

\bibitem[\protect\citeauthoryear{{Tanikawa}, {Yoshida}, {Kinugawa}, {Takahashi}  \& {Umeda}}{{Tanikawa} et~al.}{2020}]{Tanikawa2020}
{Tanikawa} A.,  {Yoshida} T.,  {Kinugawa} T.,  {Takahashi} K.,   {Umeda} H.,  2020, \mn@doi [\mnras] {10.1093/mnras/staa1417}, \href {https://ui.adsabs.harvard.edu/abs/2020MNRAS.495.4170T} {495, 4170}

\bibitem[\protect\citeauthoryear{Tauris et~al.,}{Tauris et~al.}{2017}]{Tauris2017}
Tauris T.~M.,  et~al., 2017, \mn@doi [The Astrophysical Journal] {10.3847/1538-4357/aa7e89}, 846, 170

\bibitem[\protect\citeauthoryear{{Temaj}, {Schneider}, {Laplace}, {Wei}  \& {Podsiadlowski}}{{Temaj} et~al.}{2023}]{Temaj2023}
{Temaj} D.,  {Schneider} F.~R.~N.,  {Laplace} E.,  {Wei} D.,   {Podsiadlowski} P.,  2023, \mn@doi [arXiv e-prints] {10.48550/arXiv.2311.05701}, \href {https://ui.adsabs.harvard.edu/abs/2023arXiv231105701T} {p. arXiv:2311.05701}

\bibitem[\protect\citeauthoryear{{Tian} et~al.,}{{Tian} et~al.}{2018}]{Tian2018}
{Tian} Z.-J.,  et~al., 2018, \mn@doi [Research in Astronomy and Astrophysics] {10.1088/1674-4527/18/5/52}, \href {https://ui.adsabs.harvard.edu/abs/2018RAA....18...52T} {18, 052}

\bibitem[\protect\citeauthoryear{{Tsang}, {Vartanyan}  \& {Burrows}}{{Tsang} et~al.}{2022}]{Tsang2022}
{Tsang} B. T.~H.,  {Vartanyan} D.,   {Burrows} A.,  2022, \mn@doi [\apjl] {10.3847/2041-8213/ac8f4b}, \href {https://ui.adsabs.harvard.edu/abs/2022ApJ...937L..15T} {937, L15}

\bibitem[\protect\citeauthoryear{{Ugliano}, {Janka}, {Marek}  \& {Arcones}}{{Ugliano} et~al.}{2012}]{Ugliano2012}
{Ugliano} M.,  {Janka} H.-T.,  {Marek} A.,   {Arcones} A.,  2012, \mn@doi [\apj] {10.1088/0004-637X/757/1/69}, \href {https://ui.adsabs.harvard.edu/abs/2012ApJ...757...69U} {757, 69}

\bibitem[\protect\citeauthoryear{{Ulrich} \& {Burger}}{{Ulrich} \& {Burger}}{1976}]{UlrichBurger1976}
{Ulrich} R.~K.,  {Burger} H.~L.,  1976, \mn@doi [\apj] {10.1086/154406}, \href {https://ui.adsabs.harvard.edu/abs/1976ApJ...206..509U} {206, 509}

\bibitem[\protect\citeauthoryear{{Umeda} \& {Nomoto}}{{Umeda} \& {Nomoto}}{2008}]{Umeda2008}
{Umeda} H.,  {Nomoto} K.,  2008, \mn@doi [\apj] {10.1086/524767}, \href {https://ui.adsabs.harvard.edu/abs/2008ApJ...673.1014U} {673, 1014}

\bibitem[\protect\citeauthoryear{{Umeda}, {Yoshida}  \& {Takahashi}}{{Umeda} et~al.}{2012}]{Umeda2012}
{Umeda} H.,  {Yoshida} T.,   {Takahashi} K.,  2012, \mn@doi [Progress of Theoretical and Experimental Physics] {10.1093/ptep/pts017}, \href {https://ui.adsabs.harvard.edu/abs/2012PTEP.2012aA302U} {2012, 01A302}

\bibitem[\protect\citeauthoryear{{Urushibata}, {Takahashi}, {Umeda}  \& {Yoshida}}{{Urushibata} et~al.}{2018}]{Urushibata2018}
{Urushibata} T.,  {Takahashi} K.,  {Umeda} H.,   {Yoshida} T.,  2018, \mn@doi [\mnras] {10.1093/mnrasl/slx166}, \href {https://ui.adsabs.harvard.edu/abs/2018MNRAS.473L.101U} {473, L101}

\bibitem[\protect\citeauthoryear{{Utrobin}, {Wongwathanarat}, {Janka}, {M{\"u}ller}, {Ertl}, {Menon}  \& {Heger}}{{Utrobin} et~al.}{2021}]{Utrobin2021}
{Utrobin} V.~P.,  {Wongwathanarat} A.,  {Janka} H.~T.,  {M{\"u}ller} E.,  {Ertl} T.,  {Menon} A.,   {Heger} A.,  2021, \mn@doi [\apj] {10.3847/1538-4357/abf4c5}, \href {https://ui.adsabs.harvard.edu/abs/2021ApJ...914....4U} {914, 4}

\bibitem[\protect\citeauthoryear{{Varma} \& {M{\"u}ller}}{{Varma} \& {M{\"u}ller}}{2021}]{Varma2021}
{Varma} V.,  {M{\"u}ller} B.,  2021, \mn@doi [\mnras] {10.1093/mnras/stab883}, \href {https://ui.adsabs.harvard.edu/abs/2021MNRAS.504..636V} {504, 636}

\bibitem[\protect\citeauthoryear{{Varma}, {M{\"u}ller}  \& {Schneider}}{{Varma} et~al.}{2023}]{Varma2023}
{Varma} V.,  {M{\"u}ller} B.,   {Schneider} F. R.~N.,  2023, \mn@doi [\mnras] {10.1093/mnras/stac3247}, \href {https://ui.adsabs.harvard.edu/abs/2023MNRAS.518.3622V} {518, 3622}

\bibitem[\protect\citeauthoryear{{Vartanyan} \& {Burrows}}{{Vartanyan} \& {Burrows}}{2023}]{Vartanyan2023}
{Vartanyan} D.,  {Burrows} A.,  2023, \mn@doi [arXiv e-prints] {10.48550/arXiv.2307.08735}, \href {https://ui.adsabs.harvard.edu/abs/2023arXiv230708735V} {p. arXiv:2307.08735}

\bibitem[\protect\citeauthoryear{{Vartanyan}, {Laplace}, {Renzo}, {G{\"o}tberg}, {Burrows}  \& {de Mink}}{{Vartanyan} et~al.}{2021}]{Vartanian2021}
{Vartanyan} D.,  {Laplace} E.,  {Renzo} M.,  {G{\"o}tberg} Y.,  {Burrows} A.,   {de Mink} S.~E.,  2021, \mn@doi [\apjl] {10.3847/2041-8213/ac0b42}, \href {https://ui.adsabs.harvard.edu/abs/2021ApJ...916L...5V} {916, L5}

\bibitem[\protect\citeauthoryear{{Vink} \& {de Koter}}{{Vink} \& {de Koter}}{2005}]{Vink2005}
{Vink} J.~S.,  {de Koter} A.,  2005, \mn@doi [\aap] {10.1051/0004-6361:20052862}, \href {https://ui.adsabs.harvard.edu/abs/2005A&A...442..587V} {442, 587}

\bibitem[\protect\citeauthoryear{{Vink}, {de Koter}  \& {Lamers}}{{Vink} et~al.}{2000}]{Vink2000}
{Vink} J.~S.,  {de Koter} A.,   {Lamers} H.~J.~G.~L.~M.,  2000, \mn@doi [\aap] {10.48550/arXiv.astro-ph/0008183}, \href {https://ui.adsabs.harvard.edu/abs/2000A&A...362..295V} {362, 295}

\bibitem[\protect\citeauthoryear{{Vink}, {de Koter}  \& {Lamers}}{{Vink} et~al.}{2001}]{Vink2001}
{Vink} J.~S.,  {de Koter} A.,   {Lamers} H.~J.~G.~L.~M.,  2001, \mn@doi [\aap] {10.1051/0004-6361:20010127}, \href {https://ui.adsabs.harvard.edu/abs/2001A&A...369..574V} {369, 574}

\bibitem[\protect\citeauthoryear{{Wanderman} \& {Piran}}{{Wanderman} \& {Piran}}{2010}]{WP2010}
{Wanderman} D.,  {Piran} T.,  2010, \mn@doi [\mnras] {10.1111/j.1365-2966.2010.16787.x}, \href {https://ui.adsabs.harvard.edu/abs/2010MNRAS.406.1944W} {406, 1944}

\bibitem[\protect\citeauthoryear{{Wang}, {Vartanyan}, {Burrows}  \& {Coleman}}{{Wang} et~al.}{2022}]{Wang2022}
{Wang} T.,  {Vartanyan} D.,  {Burrows} A.,   {Coleman} M. S.~B.,  2022, \mn@doi [\mnras] {10.1093/mnras/stac2691}, \href {https://ui.adsabs.harvard.edu/abs/2022MNRAS.517..543W} {517, 543}

\bibitem[\protect\citeauthoryear{{Webbink}}{{Webbink}}{1984}]{Webbink1984}
{Webbink} R.~F.,  1984, \mn@doi [\apj] {10.1086/161701}, \href {https://ui.adsabs.harvard.edu/abs/1984ApJ...277..355W} {277, 355}

\bibitem[\protect\citeauthoryear{{Willcox}, {MacLeod}, {Mandel}  \& {Hirai}}{{Willcox} et~al.}{2023}]{Willcox2023}
{Willcox} R.,  {MacLeod} M.,  {Mandel} I.,   {Hirai} R.,  2023, \mn@doi [arXiv e-prints] {10.48550/arXiv.2308.06666}, \href {https://ui.adsabs.harvard.edu/abs/2023arXiv230806666W} {p. arXiv:2308.06666}

\bibitem[\protect\citeauthoryear{{Wolff}, {Strom}, {Dror}, {Lanz}  \& {Venn}}{{Wolff} et~al.}{2006}]{Wolff2006}
{Wolff} S.~C.,  {Strom} S.~E.,  {Dror} D.,  {Lanz} L.,   {Venn} K.,  2006, \mn@doi [\aj] {10.1086/505534}, \href {https://ui.adsabs.harvard.edu/abs/2006AJ....132..749W} {132, 749}

\bibitem[\protect\citeauthoryear{{Woosley} \& {Bloom}}{{Woosley} \& {Bloom}}{2006}]{Woosley2006ARA}
{Woosley} S.~E.,  {Bloom} J.~S.,  2006, \mn@doi [\araa] {10.1146/annurev.astro.43.072103.150558}, \href {https://ui.adsabs.harvard.edu/abs/2006ARA&A..44..507W} {44, 507}

\bibitem[\protect\citeauthoryear{{Woosley} \& {Heger}}{{Woosley} \& {Heger}}{2006}]{Woosley_Heger_2006}
{Woosley} S.~E.,  {Heger} A.,  2006, \mn@doi [\apj] {10.1086/498500}, \href {http://adsabs.harvard.edu/abs/2006ApJ...637..914W} {637, 914}

\bibitem[\protect\citeauthoryear{{Woosley} \& {Heger}}{{Woosley} \& {Heger}}{2007}]{Woosley2007}
{Woosley} S.~E.,  {Heger} A.,  2007, \mn@doi [\physrep] {10.1016/j.physrep.2007.02.009}, \href {https://ui.adsabs.harvard.edu/abs/2007PhR...442..269W} {442, 269}

\bibitem[\protect\citeauthoryear{{Woosley}, {Langer}  \& {Weaver}}{{Woosley} et~al.}{1995}]{Woosley1995}
{Woosley} S.~E.,  {Langer} N.,   {Weaver} T.~A.,  1995, \mn@doi [\apj] {10.1086/175963}, \href {http://adsabs.harvard.edu/abs/1995ApJ...448..315W} {448, 315}

\bibitem[\protect\citeauthoryear{{Woosley}, {Heger}  \& {Weaver}}{{Woosley} et~al.}{2002}]{Woosley2002}
{Woosley} S.~E.,  {Heger} A.,   {Weaver} T.~A.,  2002, \mn@doi [Reviews of Modern Physics] {10.1103/RevModPhys.74.1015}, \href {https://ui.adsabs.harvard.edu/abs/2002RvMP...74.1015W} {74, 1015}

\bibitem[\protect\citeauthoryear{{Woosley}, {Sukhbold}  \& {Janka}}{{Woosley} et~al.}{2020}]{Woosley2020}
{Woosley} S.~E.,  {Sukhbold} T.,   {Janka} H.~T.,  2020, \mn@doi [\apj] {10.3847/1538-4357/ab8cc1}, \href {https://ui.adsabs.harvard.edu/abs/2020ApJ...896...56W} {896, 56}

\bibitem[\protect\citeauthoryear{{Wyder} et~al.,}{{Wyder} et~al.}{2005}]{Wyder2005}
{Wyder} T.~K.,  et~al., 2005, \mn@doi [\apjl] {10.1086/424735}, \href {https://ui.adsabs.harvard.edu/abs/2005ApJ...619L..15W} {619, L15}

\bibitem[\protect\citeauthoryear{{Yadav}, {M{\"u}ller}, {Janka}, {Melson}  \& {Heger}}{{Yadav} et~al.}{2020}]{Yadav2020}
{Yadav} N.,  {M{\"u}ller} B.,  {Janka} H.~T.,  {Melson} T.,   {Heger} A.,  2020, \mn@doi [\apj] {10.3847/1538-4357/ab66bb}, \href {https://ui.adsabs.harvard.edu/abs/2020ApJ...890...94Y} {890, 94}

\bibitem[\protect\citeauthoryear{{Yoon}, {Langer}  \& {Norman}}{{Yoon} et~al.}{2006}]{Yoon2006}
{Yoon} S.~C.,  {Langer} N.,   {Norman} C.,  2006, \mn@doi [\aap] {10.1051/0004-6361:20065912}, \href {https://ui.adsabs.harvard.edu/abs/2006A&A...460..199Y} {460, 199}

\bibitem[\protect\citeauthoryear{{Yoshida}, {Suwa}, {Umeda}, {Shibata}  \& {Takahashi}}{{Yoshida} et~al.}{2017}]{Yoshida2017}
{Yoshida} T.,  {Suwa} Y.,  {Umeda} H.,  {Shibata} M.,   {Takahashi} K.,  2017, \mn@doi [\mnras] {10.1093/mnras/stx1738}, \href {https://ui.adsabs.harvard.edu/abs/2017MNRAS.471.4275Y} {471, 4275}

\bibitem[\protect\citeauthoryear{{Yoshida}, {Takiwaki}, {Kotake}, {Takahashi}, {Nakamura}  \& {Umeda}}{{Yoshida} et~al.}{2019}]{Yoshida2019}
{Yoshida} T.,  {Takiwaki} T.,  {Kotake} K.,  {Takahashi} K.,  {Nakamura} K.,   {Umeda} H.,  2019, \mn@doi [\apj] {10.3847/1538-4357/ab2b9d}, \href {https://ui.adsabs.harvard.edu/abs/2019ApJ...881...16Y} {881, 16}

\bibitem[\protect\citeauthoryear{{Yoshida}, {Takiwaki}, {Aguilera-Dena}, {Kotake}, {Takahashi}, {Nakamura}, {Umeda}  \& {Langer}}{{Yoshida} et~al.}{2021a}]{Yoshida2021b}
{Yoshida} T.,  {Takiwaki} T.,  {Aguilera-Dena} D.~R.,  {Kotake} K.,  {Takahashi} K.,  {Nakamura} K.,  {Umeda} H.,   {Langer} N.,  2021a, \mn@doi [\mnras] {10.1093/mnrasl/slab067}, \href {https://ui.adsabs.harvard.edu/abs/2021MNRAS.506L..20Y} {506, L20}

\bibitem[\protect\citeauthoryear{{Yoshida}, {Takiwaki}, {Kotake}, {Takahashi}, {Nakamura}  \& {Umeda}}{{Yoshida} et~al.}{2021b}]{Yoshida2021a}
{Yoshida} T.,  {Takiwaki} T.,  {Kotake} K.,  {Takahashi} K.,  {Nakamura} K.,   {Umeda} H.,  2021b, \mn@doi [\apj] {10.3847/1538-4357/abd3a3}, \href {https://ui.adsabs.harvard.edu/abs/2021ApJ...908...44Y} {908, 44}

\bibitem[\protect\citeauthoryear{{Zahn}}{{Zahn}}{1975}]{Zahn1975}
{Zahn} J.~P.,  1975, \aap, \href {https://ui.adsabs.harvard.edu/abs/1975A&A....41..329Z} {41, 329}

\bibitem[\protect\citeauthoryear{{Zahn}}{{Zahn}}{1977}]{Zahn1977}
{Zahn} J.~P.,  1977, \aap, \href {https://ui.adsabs.harvard.edu/abs/1977A&A....57..383Z} {500, 121}

\bibitem[\protect\citeauthoryear{{Zapartas} et~al.,}{{Zapartas} et~al.}{2017}]{Zapartas2017}
{Zapartas} E.,  et~al., 2017, \mn@doi [\aap] {10.1051/0004-6361/201629685}, \href {https://ui.adsabs.harvard.edu/abs/2017A&A...601A..29Z} {601, A29}

\bibitem[\protect\citeauthoryear{{de Jager}, {Nieuwenhuijzen}  \& {van der Hucht}}{{de Jager} et~al.}{1988}]{deJager1988}
{de Jager} C.,  {Nieuwenhuijzen} H.,   {van der Hucht} K.~A.,  1988, \aaps, \href {https://ui.adsabs.harvard.edu/abs/1988A&AS...72..259D} {72, 259}

\makeatother
\end{thebibliography}

\appendix

\section{Binary interactions for spin and orbit}\label{sec:tidaletc}

\subsection{Tidal interaction}
Tidal interaction changes the orbit and the spins of the binary system. There are two mechanisms for the dissipation of the tidal kinetic energy. One mechanism is the convective damping on the equilibrium tide for the stars with an outer convection envelope such as red giants. The other mechanism is the radiative damping on the dynamical tide for the stars with an outer radiative zone \citep{Zahn1977}. The time evolution of the separation, the eccentricity, and the spin are calculated as
\begin{align}
\frac{da}{dt}=&-6\frac{k}{T}q(1+q)\left(\frac{R_i}{a}\right)^8 \frac{a}{(1-e^2)^{\frac{15}{2}}}\notag\\
&\times\left[f_1(e^2)-(1-e^2)^{\frac{3}{2}}f_2(e^2)\frac{\Omega_{{\rm spin},i}}{\Omega_{\rm orb}}\right]\label{eq:dadt} \,, \\
\frac{de}{dt}=&-27\frac{k}{T}q(1+q)\left(\frac{R_i}{a}\right)^8\frac{e}{(1-e^2)^{\frac{13}{2}}}\notag\\
&\times\left[f_3(e^2)-\frac{11}{18}(1-e^2)^{\frac{3}{2}}f_4(e^2)\frac{\Omega_{{\rm spin},i}}{\Omega_{\rm orb}}\right] \,, \label{eq:dedt}\\
\frac{d\Omega_{{\rm spin},i}}{dt}=&3\frac{k}{T}\frac{q^2}{r_g^2}\left(\frac{R_i}{a}\right)^6\frac{\Omega_{\rm orb}}{(1-e^2)^6}\notag\\
&\times\left[f_2(e^2)-(1-e^2)^{\frac{3}{2}}f_5(e^2)\frac{\Omega_{{\rm spin},i}}{\Omega_{\rm orb}}\right] \,, \label{eq:dodt}
\end{align}

where
\begin{align}
f_1(e^2)&=1+\frac{31}{2}e^2+\frac{255}{8}e^4+\frac{185}{16}e^6+\frac{25}{64}e^8 \,, \label{eq:f1} \\
f_2(e^2)&=1+\frac{15}{2}e^2+\frac{45}{8}e^4+\frac{5}{16}e^6 \,, \label{eq:f2}\\
f_3(e^2)&=1+\frac{15}{4}e^2+\frac{15}{8}e^4+\frac{5}{64}e^6 \,, \label{eq:f3}\\
f_4(e^2)&=1+\frac{3}{2}e^2+\frac{1}{8}e^4 \,, \label{eq:f4}\\
f_5(e^2)&=1+3e^2+\frac{3}{8}e^4 \,. \label{eq:f5}
\end{align}
where $k/T,~q,~R_i,~\Omega_{{\rm spin},i},~\Omega_{\rm orb},$ and $r_g$ are the coupling parameter depending on the tidal interaction mechanism, the mass ratio, the radius of the star, the spin angular velocity of the star, the angular velocity of the orbit, and the dimensionless gyration radius of the star, respectively \citep{Zahn1977, Hut1981}.
Notation $i$ means 1 (primary star) or 2 (secondary star).

If the star is a red giant, the convective damping of the equilibrium tide is effective.
In this case, $k/T$ is calculated as
\begin{equation}
\frac{k}{T}=\frac{2}{21}\frac{
f_{\rm{con}}}{\tau_{\rm{con}}}\frac{M_{{\rm env},i}}{M_{\rm{i}}} \,,
\end{equation}
where $M_{{\rm env},i}$, $f_{\rm{con}}$, and $\tau_{\rm{con}}$ are
the stellar envelope mass, the correction factor of the tidal torque, and 
 the eddy turnover timescale \cite[e.g.,][]{Rasio1996,Hurley_2002}, respectively.
They are calculated as
\begin{align}
\tau_{\rm{con}}&=\left[\frac{M_{{\rm env},i}R_{{\rm env},i}
\left(R_i-\frac{1}{2}R_{{\rm env},i}\right)}{3L_i}\right]^{1/3} \,,
\\
f_{\rm{con}}&={\rm{min}}\left[1,\left(\frac{\pi|\Omega_{\rm{orb}}-\Omega_{{\rm spin},i}|^{-1}}{\tau_{\rm{con}}}\right)^2\right] \,.
\end{align}
where $L_i$, and $R_{{\rm env},i}$ are
the stellar luminosity and the radius of the stellar envelope, respectively.

On the other hand, if the star have the radiative envelope i.e. the evolution stage is not a red giant, the tidal mechanism is the radiative damping of the dynamical tide \citep{Zahn1975}.
$k/T$  is 
\begin{align}
\frac{k}{T}=&4.3118\times10^{-8}\left(\frac{M_i}{\msun}\right)\left(\frac{R_i}{{\rm R}_{\odot}}\right)^2\notag\\
&\times\left(\frac{a}{1\,{\rm{AU}}}\right)^{-5}(1+q)^{5/6}E_2~\rm{yr^{-1}} \,,
\end{align}
where $E_2$ is the tidal coefficient factor \citep{Zahn1977, Hurley_2002}.

\begin{figure}
  \begin{center}
    \includegraphics[width=\hsize]{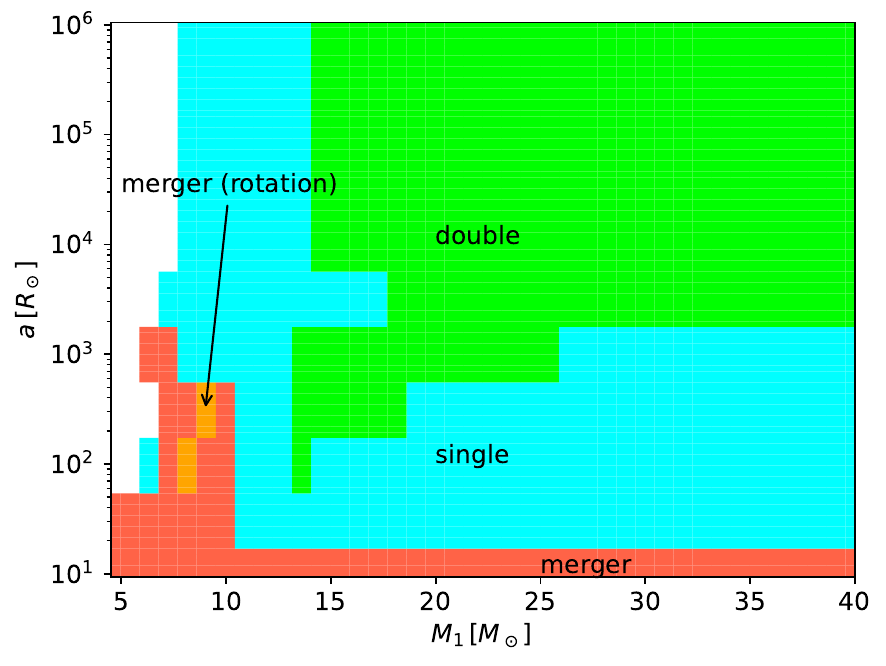}
  \end{center}
  \caption{The binary progenitors of SNe in the $q=0.5$ case. }
  \label{fig:Q05_channel}
\end{figure}

\subsection{Magnetic braking}
When a rotating star loses its mass, magnetic braking removes angular momentum from the rotating star via a magnetic field.
The spin angular momentum loss by the magnetic braking is
\begin{equation}
\dot{J}_{\mathrm{spin},i}=-5.83 \times 10^{-16} \frac{M_{\mathrm{env},i}}{M_i}\left(R_i \Omega_{\mathrm{spin},i}\right)^3 \mathrm{M}_{\odot} \mathrm{R}_{\odot}^2 \mathrm{yr}^{-2}
\end{equation}
where masses and the radius are in solar units and $\Omega_{\text {spin }}$ in units of years \citep{Hurley2000, Hurley_2002}. 
\subsection{Gravitational radiation}
After the stars in a binary explode or collapse at the end of their lifetime, the compact star binary is formed. The compact binary loses the angular momentum and the orbital energy by the gravitational radiation. We use the weak-field approximation formalism given by \cite{Peter_Mathews_1963, Peters1964}. The loss of angular momentum, the orbital separation and the eccentricity are described as
\begin{equation}
    \frac{j}{J}=-\frac{32 G^3 M_1 M_2 M_{\text {total }}}{5 c^5 a^4} \frac{1+\frac{7}{8} e^2}{\left(1-e^2\right)^{5 / 2}},
\end{equation}

\begin{figure}
    \centering
    \includegraphics[width=\hsize]{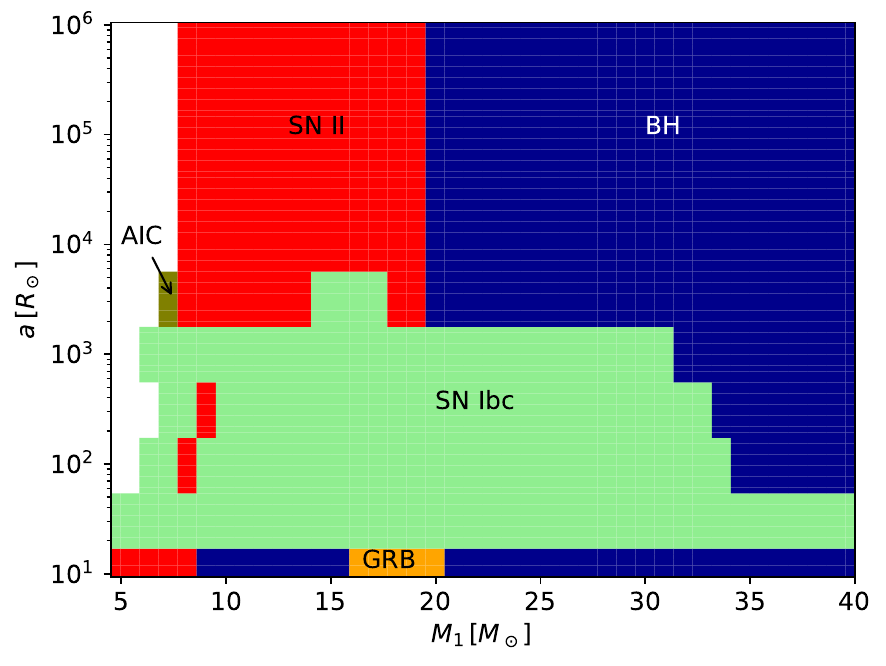}
    \caption{SN type of primary star in the $q=0.5$ case.}
       \label{fig:Q05}
    \end{figure}

\begin{figure}
    \centering
    \includegraphics[width=\hsize]{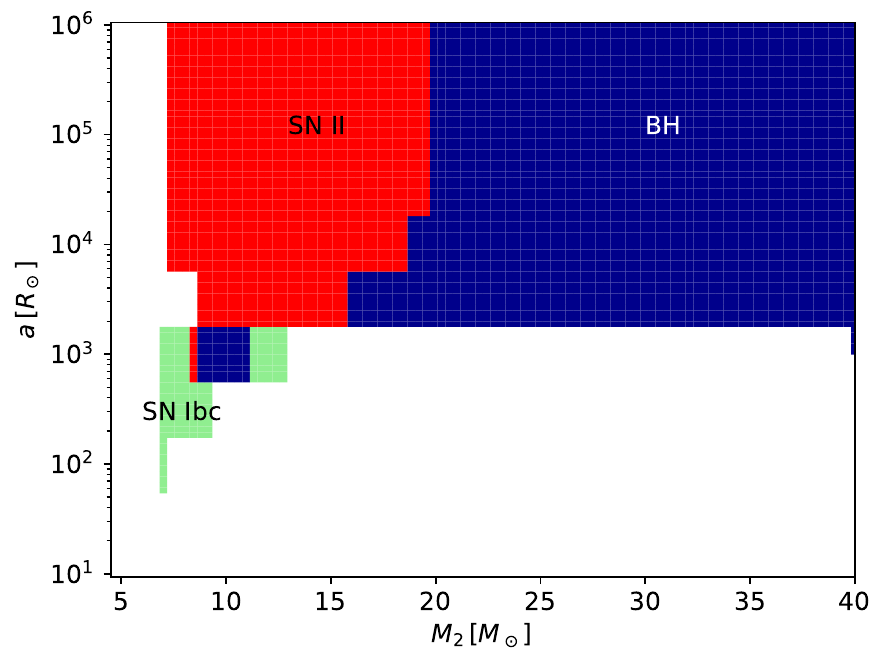}
    \caption{SN type of secondary star in the $q=0.5$ case.}
       \label{fig:Q05_2}
    \end{figure}

\begin{equation}
    \frac{\dot{a}}{a}=-\frac{64 G^3 M_1 M_2 M_{\text {total }}}{5 c^5 a^4} \frac{1+\frac{73}{24} e^2+\frac{37}{96} e^4}{\left(1-e^2\right)^{7 / 2}}
\end{equation}
and
\begin{equation}
\frac{\dot{e}}{e}=-\frac{304 G^3 M_1 M_2 M_{\text {total }}}{15 c^5 a^4} \frac{1+\frac{121}{304} e^2}{\left(1-e^2\right)^{5 / 2}}.
\end{equation}

\section{Parameter survey of q=0.5, and  0.9 cases}\label{sec:Q}

We show the parameter dependence of SN types, CO core mass, and angular momentum in the mass ratio of $q=0.5$ and  $q=0.9$ while the case for $q=0.7$ is discussed in Section~\ref{subsec:Qconst}.

\subsection{q=0.5 case}

    \begin{figure}
        \centering
        \includegraphics[width=\hsize]{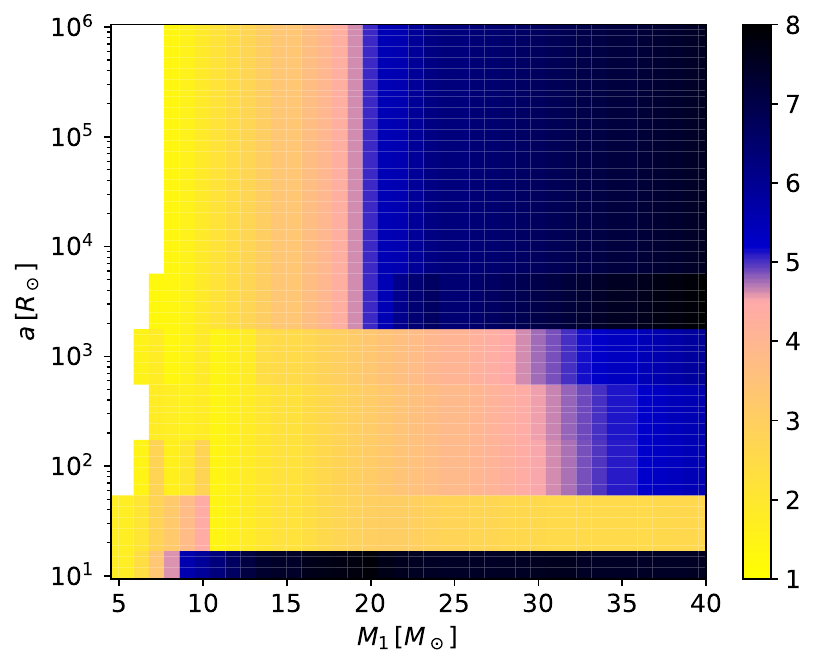}
    \caption{CO core mass of primary star in the $q=0.5$ case.}
       \label{fig:Q05_Mco}
  \end{figure}
    \begin{figure}
        \centering
        \includegraphics[width=\hsize]{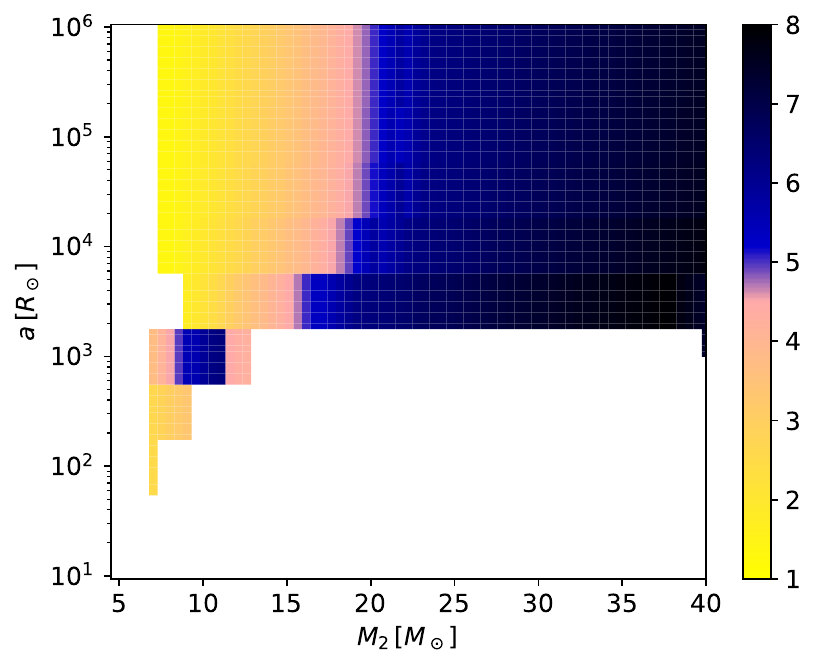}
    \caption{CO core mass of secondary star in the $q=0.5$ case.}
       \label{fig:Q05_2Mco}
  \end{figure}
    
\begin{figure}
 \centering
          \includegraphics[width=\hsize]{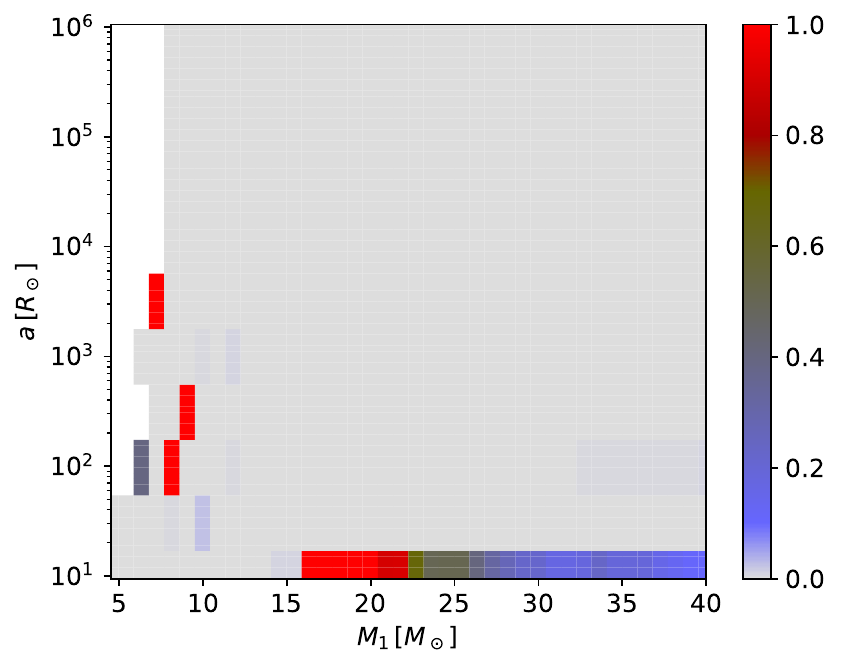}
    \caption{Angular momentum of primary star in the $q=0.5$ case.}
       \label{fig:Q05_am}
  \end{figure}

\begin{figure}
 \centering
          \includegraphics[width=\hsize]{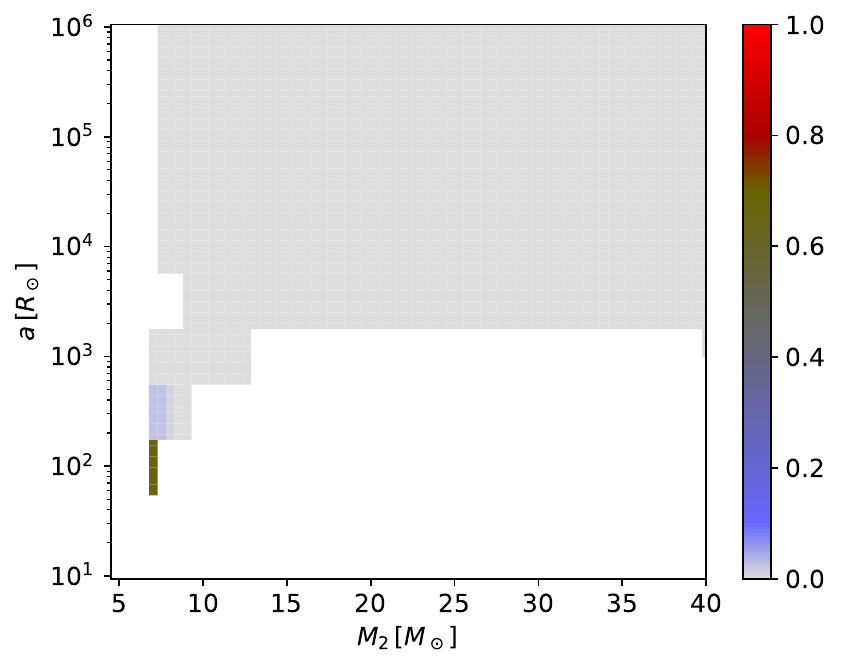}
    \caption{Angular momentum of secondary star in the $q=0.5$ case. }
       \label{fig:Q05_2am}
  \end{figure}

Figure~\ref{fig:Q05_channel} shows the progenitor of the SN in the mass ratio $q=0.5$.
Figures~\ref{fig:Q05} and \ref{fig:Q05_2} show the fate of the primary star and the secondary star, respectively, as a function of their ZAMS mass and binary separation.
Figures~\ref{fig:Q05_Mco} and \ref{fig:Q05_2Mco} shows CO core mass as a function of their ZAMS mass and binary separation.
Figures~\ref{fig:Q05_am} and \ref{fig:Q05_2am} show the angular momentum as a function of their ZAMS mass and binary separation.

The one of the main differences between this case and the $q=0.7$ case is the absence of a pathway leading to RSN.
See points $(M_1,a)=(8\,\msun, 100\,\rsun)$, and $(9\,\msun, 10^{2.5}\,\rsun)$ in the Fig.~\ref{fig:Q05}, they collapse via merger (rotation).
These progenitors evolve via a similar evolution path of RSN progenitor in the $q=0.7$ case (left-hand side of Fig.~\ref{fig:RSN}).
However, in these cases, the hydrogen envelopes remain and they collapse via type II SNe, although their CO cores have large angular momentum.

The other difference is that the type Ibc SNe occurs by secondary stars whose masses are $11\,\msun\le M_2\le 12.5\,\msun$ with $a=10^3\,\rsun$ (see Fig.~\ref{fig:Q05_2}).
The SNe do not occur with the parameter range in $q=0.7$ (see Fig.~\ref{fig:Q07_2}).
BHs are formed from secondary stars whose masses are $9\,\msun\le M_2\le 11\,\msun$ with $a=10^3\,\rsun$ and that is similar to case of $q=0.7$.
When secondary masses are $9\,\msun\le M_2\le 11\,\msun$ with $a=10^3\,\rsun$, the secondary stars can become BHs due to the mass gain from the primary stars.
However, when secondary masses are $11\,\msun\le M_2\le 12.5\,\msun$ with $a=10^3\,\rsun$, they get such a lot of masses from the primary stars that they become a common envelope and loses hydrogen envelopes.
After the common envelope phase, the Wolf-Rayet stars lose a lot of masses due to the strong stellar wind mass loss and they cannot collapse to BHs.

\subsection{q=0.9 case.}

\begin{figure}[h]
  \begin{center}
    \includegraphics[width=\hsize]{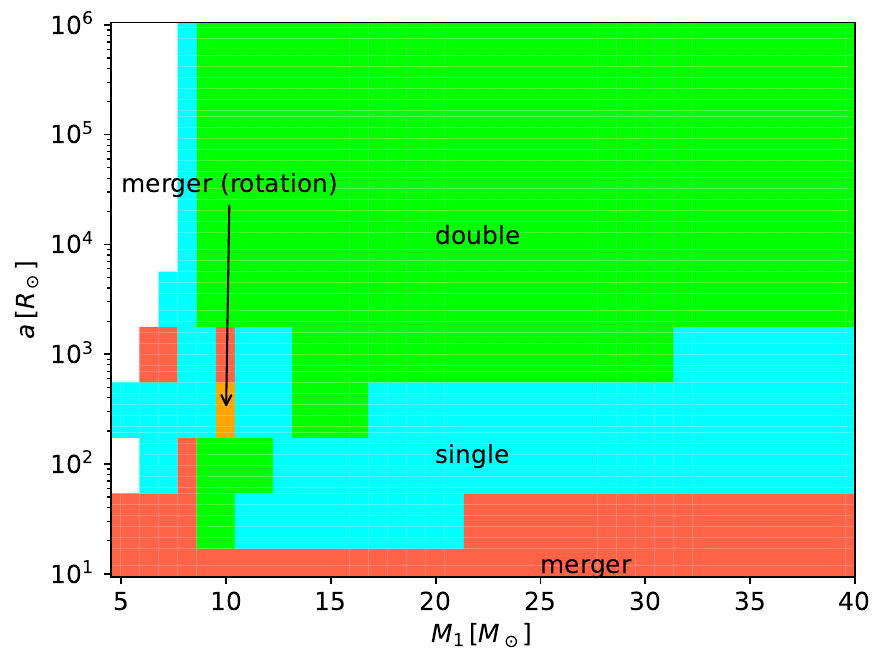}
  \end{center}
  \caption{The binary progenitors of SNe in the $q=0.9$ case. }
  \label{fig:Q09_channel}
\end{figure}

\begin{figure}[h]
    \centering
    \includegraphics[width=\hsize]{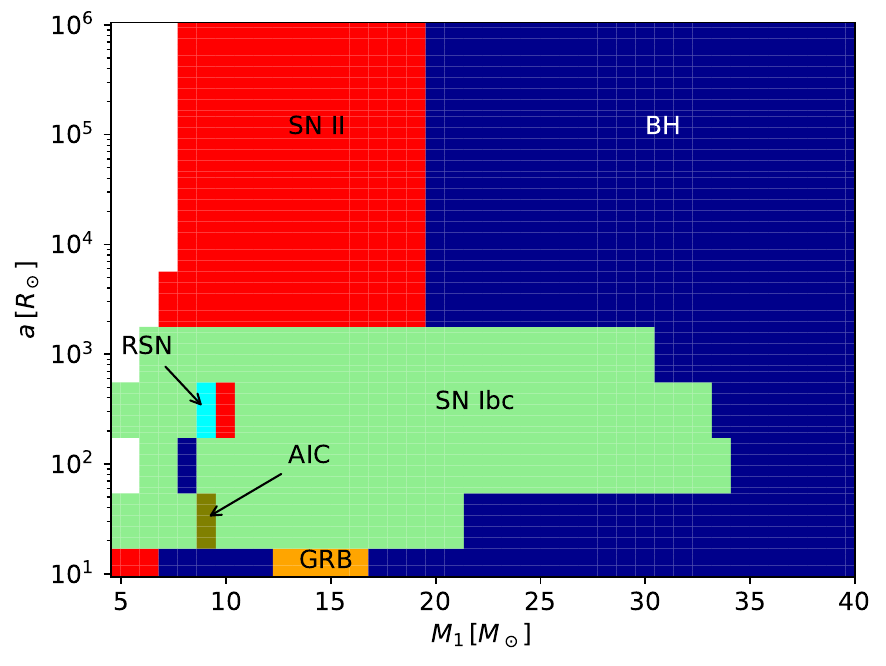}
    \caption{SN type of primary star in the $q=0.9$ case.}
       \label{fig:Q09}
    \centering
    \includegraphics[width=\hsize]{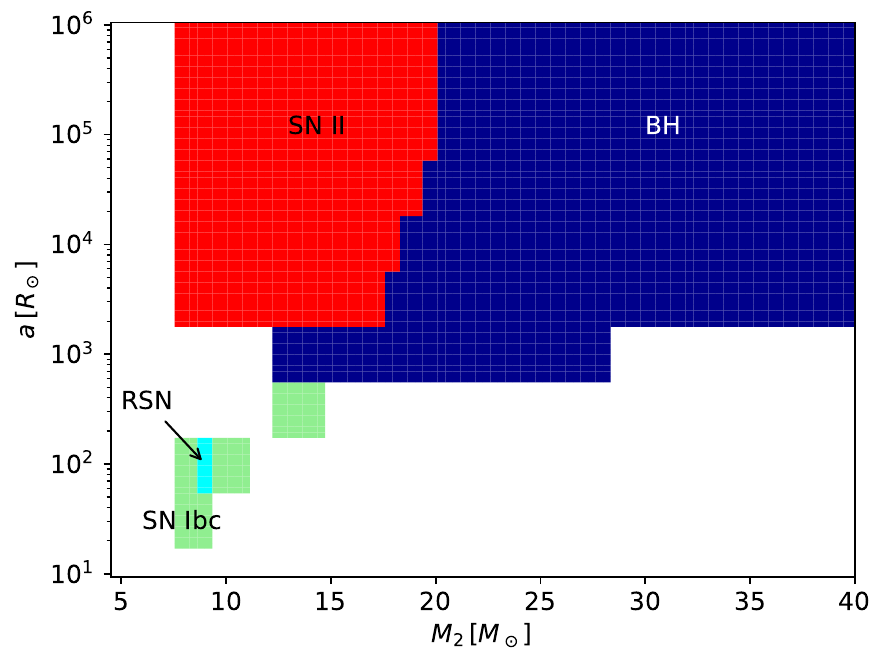}
    \caption{SN type of secondary star in the $q=0.9$ case.}
       \label{fig:Q09_2}
\end{figure}    

\begin{figure}[h]
        \centering
        \includegraphics[width=\hsize]{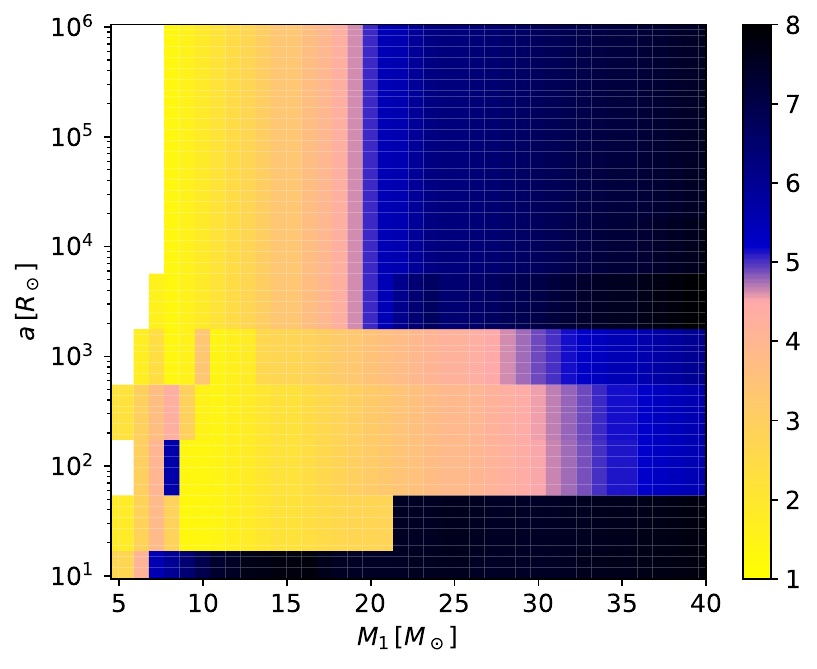}
    \caption{CO core mass of primary star in the $q=0.9$ case.}
       \label{fig:Q09_Mco}

        \centering
        \includegraphics[width=\hsize]{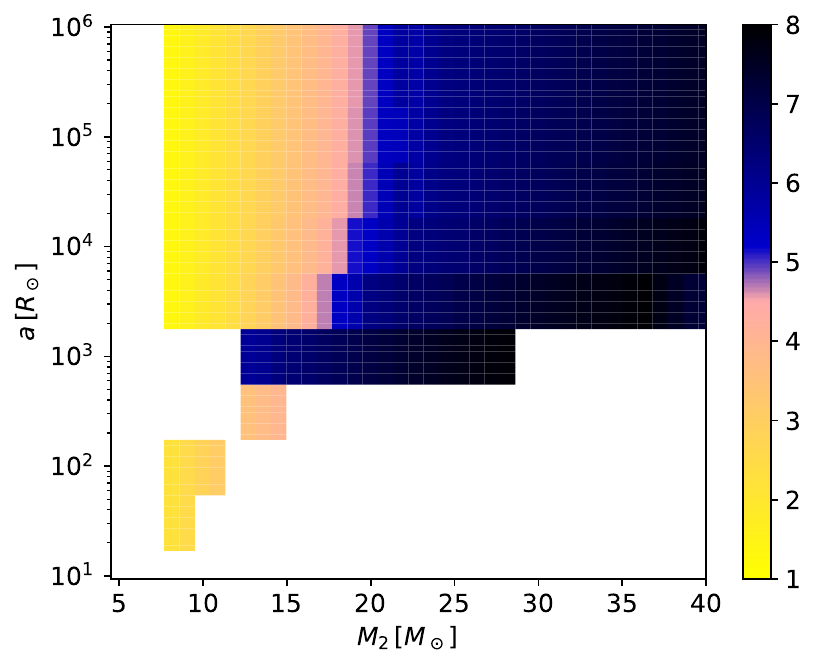}
    \caption{CO core mass of secondary star in the $q=0.9$ case.}
       \label{fig:Q09_2Mco}
\end{figure}  
  
\begin{figure}[h]
 \centering
          \includegraphics[width=\hsize]{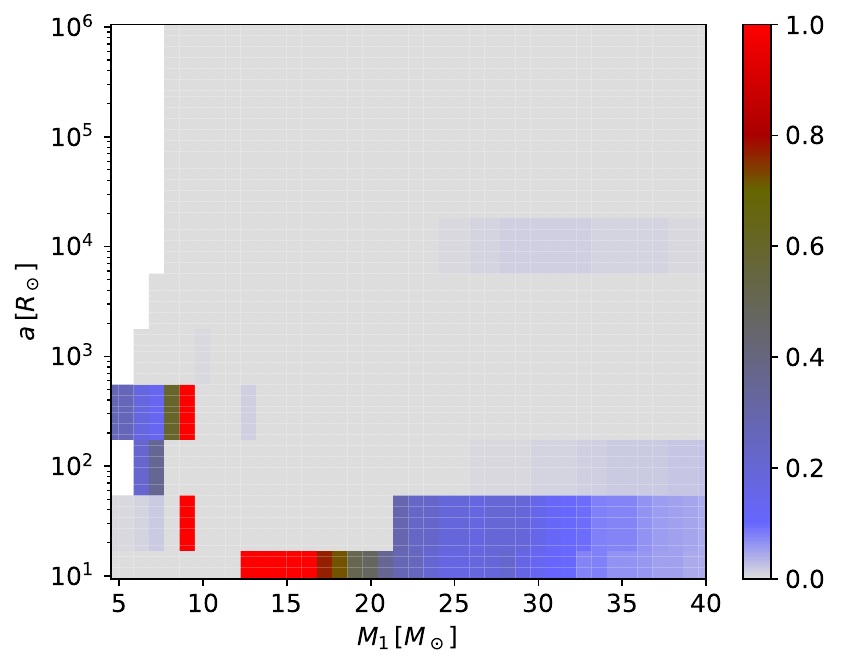}
    \caption{Angular momentum of primary star in the $q=0.9$ case.}
       \label{fig:Q09_am}
  \end{figure}
  
\begin{figure}[h]
 \centering
          \includegraphics[width=\hsize]{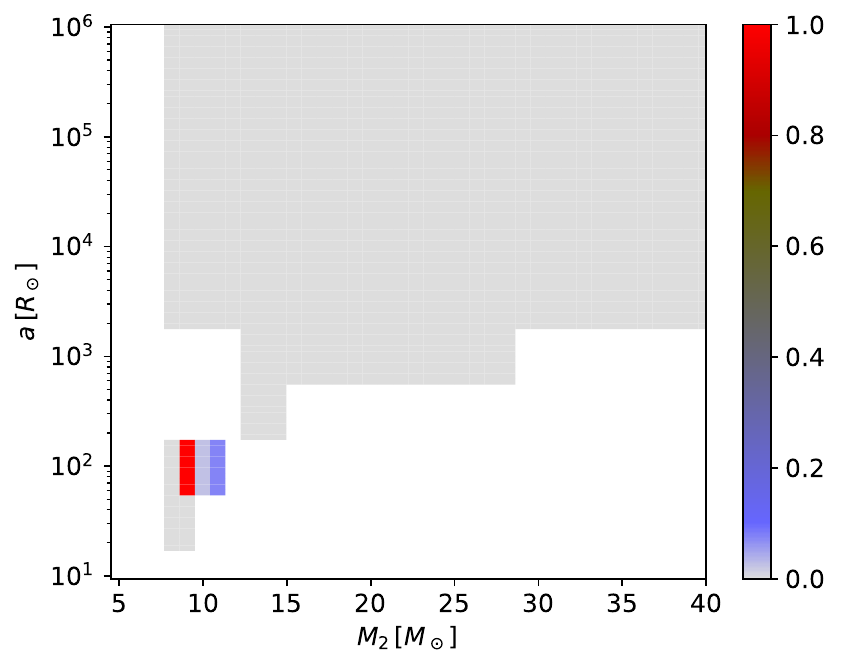}
    \caption{Angular momentum of secondary star in the $q=0.9$ case.}
       \label{fig:Q09_2am}
  \end{figure}

  \begin{figure}
  \begin{center}
    \includegraphics[width=9cm]{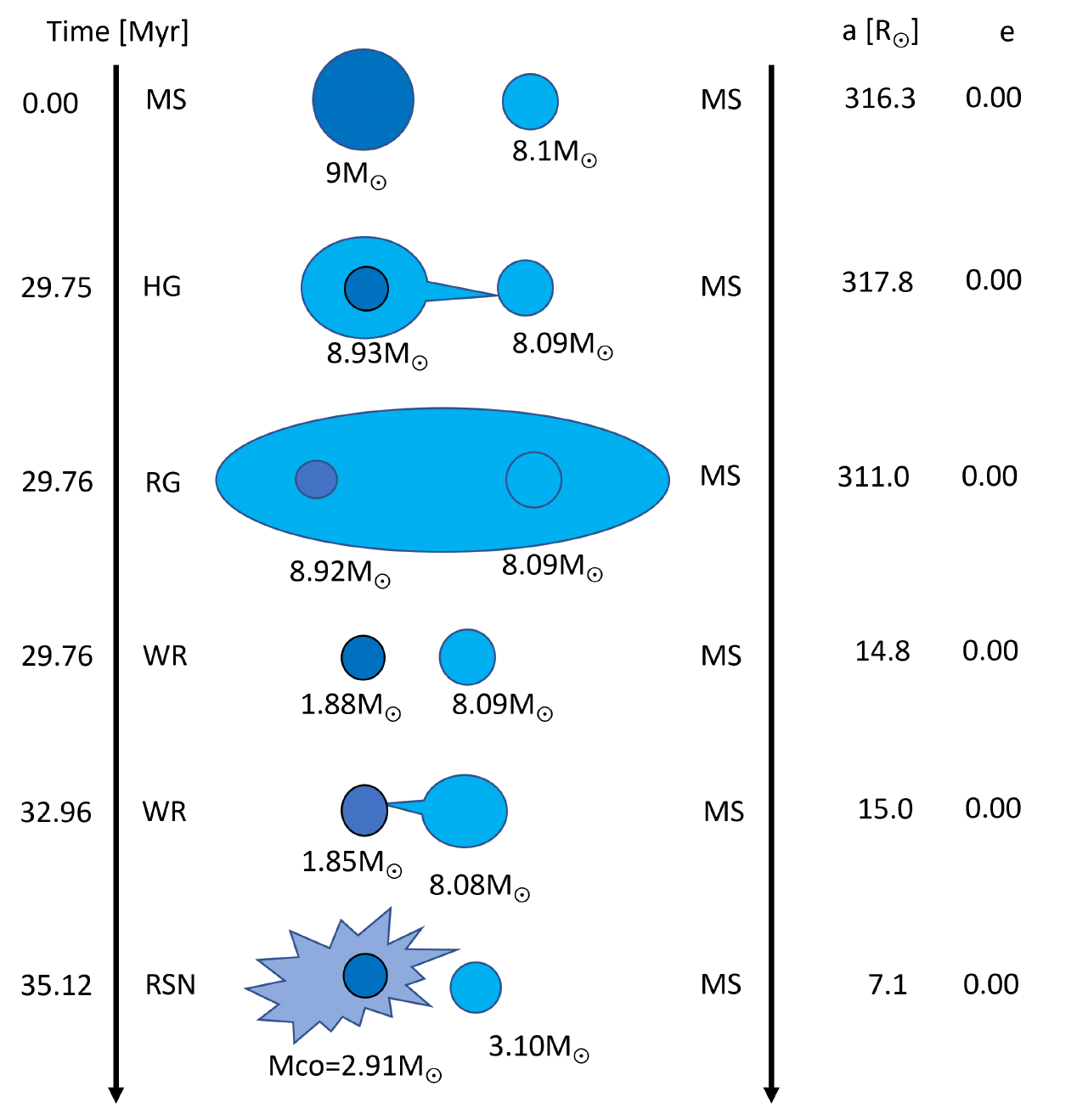}
  \end{center}
  \caption{Example of the RSN progenitor evolutionary path in the case of $q=0.9$. See Fig.~\ref{fig:RSN} for other examples with $q=0.7$.}
  \label{fig:RSN3}
\end{figure}

Figure~\ref{fig:Q09_channel} shows the progenitor of the SN in the mass ratio $q=0.9$.
Figures~\ref{fig:Q09} and \ref{fig:Q09_2} show the fate of the primary star and the secondary star, respectively, as a function of their ZAMS mass and binary separation.
Figures~\ref{fig:Q09_Mco} and \ref{fig:Q09_2Mco} shows CO core mass as a function of their ZAMS mass and binary separation.
Figures.~\ref{fig:Q09_am} and \ref{fig:Q09_2am} show the angular momentum as a function of their ZAMS mass and binary separation.

The main differences of $q=0.9$ case is another evolution path to RSN. 
The point of $(M_1, a)=(9\,\msun, 10^{2.5}\,\rsun)$ in Fig.~\ref{fig:Q09} become RSN.
Figure~\ref{fig:RSN3} shows the evolution path of this progenitor.
The primary star loses the hydrogen envelope by the common envelope phase, but it can get enough mass and angular momentum by the mass transfer from the secondary star to occur a RSN. The secondary star, whose initial mass is enough massive to become a SN, cannot become a SN due to a lot of mass loss.

\section{the fractions of each SN type for 12 models}\label{fraction}
In order to calculate the fractions of each SN type for 12 models We assume the binary fraction $f_b$ as $f_b$=70\% \citep[e.g.,][]{Sana2012}, and $f_b$=50\% \citep[e.g.,][]{Tian2018}.
According to Figure \ref{fig:Q07}, in the case of effectively single stars, relatively light stars ($8\,\msun< M<20\,\msun$) tend to become type II SN, while more massive ones ($>20\,\msun$) tend to become BH.
If $f_b$=70\%, the number of type II SN and the number of BH increase 8163 and 2959, respectively.
If $f_b$=50\%, the number of type II SN and the number of BH increase 19049 and 6903, respectively.
We added these values to our result, in order to calculate the fractions of each SN type.
Tables \ref{tab:SN70}, and \ref{tab:SN50}  show the fractions of each SN type for 12 models with $f_b=$70\%, and $f_b=$50\%, respectively.

\begin{table*}
\caption{The fractions of each SN type for 12 models with the binary fraction $f_b=$70\%}
\label{tab:SN70}
\begin{center}
\begin{tabular}{c|cccccc}
\hline
 model &  AIC & Ibc & II & RSN & BH & GRB\\
 \hline
 Sana\_MT1\_CE01 & 0.000467628&0.102274784&0.615594642&0.001131358&0.266774271&0.013757316\\
  Sana\_MT1\_CE1 & 0.011630121&0.387514464&0.360665583&0.019012465&0.210499334&0.010678033\\
 Sana\_MT1\_CE10 &0.007020693&0.421914822&0.323432766&0.020212544&0.216551525&0.010867649\\
Sana\_MT05\_CE01 &0.000935377&0.111080114&0.631412255&0.000984607&0.240867767&0.014719879\\
 Sana\_MT05\_CE1 &0.003097751&0.330902118&0.422572609&0.024831181&0.207106799&0.011489543\\
Sana\_MT05\_CE10 & 0.002353262&0.419460455&0.334146175&0.033235565&0.199498653&0.01130589\\
 Abt\_MT1\_CE01 &0.003633029&0.044357559&0.669913326&0.00122831&0.269830286&0.011037489\\
Abt\_MT1\_CE1 &0.007645104&0.190229184&0.556380526&0.009255492&0.229251424&0.00723827\\
 Abt\_MT1\_CE10 &0.009395388&0.191088011&0.548203609&0.00915914&0.235066422&0.00708743\\
Abt\_MT05\_CE01 &0.004652061&0.047012148&0.674478618&0.001088407&0.261726705&0.011042062\\
Abt\_MT05\_CE1 &0.004862751&0.166549209&0.580341097&0.010712146&0.230205434&0.007329363\\
 Abt\_MT05\_CE10 &0.00661415&0.184545927&0.557260702&0.011964923&0.232387039&0.007227259\\
 \hline
\end{tabular}
\end{center}
\end{table*}

\begin{table*}
\caption{The fractions of each SN type for 12 models with $f_b$=50\%}
\label{tab:SN50}
\begin{center}
\begin{tabular}{c|cccccc}
\hline
 model &  AIC & Ibc & II & RSN & BH & GRB\\
 \hline
Sana\_MT1\_CE01 &0.00038214&0.083577821&0.637250068&0.000924533&0.266623111&0.011242326\\
Sana\_MT1\_CE1 &0.009554638&0.318359587 &0.42729931&0.015619547&0.22039446&0.008772458\\
Sana\_MT1\_CE10 &0.005672326&0.340883484&0.402294831&0.016330601&0.226038308&0.00878045\\
Sana\_MT05\_CE01 &0.000752296&0.089338507&0.651501953&0.000791891&0.245776581&0.011838771\\
Sana\_MT05\_CE1 &0.002492023&0.266198149&0.483478811&0.019975739&0.218612378&0.0092429\\
Sana\_MT05\_CE10 & 0.001878267&0.334794207&0.414865527&0.026527112&0.212911041&0.009023846\\
Abt\_MT1\_CE01 &0.002891248&0.035300759&0.683009101&0.000977517&0.26903749&0.008783886\\
Abt\_MT1\_CE1 &0.00610929&0.152014305&0.592072824&0.007396169&0.236623229&0.005784184\\
Abt\_MT1\_CE10 &0.007400833&0.15052178&0.587657643&0.007214739&0.241622171&0.005582834\\
Abt\_MT05\_CE01 &0.003691116&0.037301167&0.686784411&0.000863582&0.262598546&0.008761178\\
Abt\_MT05\_CE1 &0.003855395&0.13204727&0.612183606&0.008493044&0.237609655&0.00581103\\
Abt\_MT05\_CE10 &0.005185423&0.144682029&0.595449646&0.009380371&0.239636438&0.005666094\\
 \hline
\end{tabular}
\end{center}
\end{table*}


\bsp	
\label{lastpage}
\end{document}